\documentclass[%
reprint,
superscriptaddress,
nofootinbib,
 amsmath,amssymb,
 aps,
prc
]{revtex4-2}
\usepackage{graphicx}
\graphicspath{{./Images/}}
\usepackage{dcolumn}
\usepackage{bm}
\usepackage{slashed}
\usepackage{physics}
\usepackage{xfrac}
\usepackage{float}
\usepackage{color}
\usepackage{mathrsfs}
\usepackage[colorlinks=true,linkcolor=blue]{hyperref}%
\usepackage{hyperref}

\allowdisplaybreaks[1]

\begin{document}

\preprint{APS/123-QED}

\title{Final State Gluon Emission in Deep-Inelastic Scattering at Next-to-Leading Twist}

\author{Chathuranga Sirimanna}
\affiliation{Department of Physics and Astronomy, Wayne State University, Detroit, MI 48201.}

\author{Shanshan Cao}
\affiliation{Institute of Frontier and Interdisciplinary Science, Shandong University, Qingdao, Shandong, 266237, China}
\affiliation{Department of Physics and Astronomy, Wayne State University, Detroit, MI 48201.}

\author{Abhijit~Majumder} 
\affiliation{Department of Physics and Astronomy, Wayne State University, Detroit, MI 48201.}

\date{\today}

\begin{abstract}
A detailed reanalysis of the single gluon emission rate at next-to-leading twist is carried out. 
As was the case in prior efforts, the problem is cast in the framework of deep-inelastic scattering (DIS) of 
an electron off a large nucleus. The quark produced in the interaction propagates through the remaining nucleus and 
engenders scattering and gluon radiation, which is calculated in the limit of one re-scattering. 
This medium induced single gluon emission rate forms the basis of several energy loss calculations in both DIS and 
heavy-ion collisions. Unlike prior efforts, a complete 
transverse momentum gradient expansion of the hadronic tensor, including $N_c$ suppressed terms, phase terms and finite gluon momentum fraction terms\, ignored 
previously, is carried out. These terms turn out to be surprisingly large. 
In contrast to prior efforts, the full next-to-leading twist gluon emission kernel is found to be positive definite 
and slowly increasing with the exchanged transverse momentum. Phenomenological consequences of these new contributions are discussed.
\end{abstract}


\maketitle



\section{Introduction}

Jet modification in a Quark Gluon Plasma (QGP)~\cite{Majumder:2010qh,Cao:2020wlm} is now understood as a multi-scale process~\cite{Cao:2017zih,Kumar:2020vkx,Tachibana:2018yae,Vujanovic:2020wuk}. 
Jets start as single partons emanating from a hard scattering with a large virtuality, smaller but comparable to their energy $\mu^2 \lesssim E^2$. 
The first emissions are vacuum like~\cite{Caucal:2018dla}, with the cross section for medium induced emission increasing as the virtuality of partons in the shower continues to decrease with successive emissions.
In this early, high-virtuality portion of the shower, partons transition from a purely vacuum like emission to that induced by scattering in a medium. 
As the virtuality of the partons continues to drop, the probability or rate of medium induced emission continues to increase compared to vacuum like emission. 
A transition in the dynamics occurs once the virtuality approaches the medium induced scale $\mu_S^2 \sim \hat{q} \tau$~\cite{Cao:2021rpv}, where $\hat{q}$ is transverse momentum transport coefficient, defined as, 
\begin{eqnarray}
\hat{q} = \frac{\langle \vec{k}^2_\perp \rangle_L}{L},
\end{eqnarray}
the mean transverse momentum square accrued per unit length by a single parton by scattering in the medium~\cite{Baier:2002tc,Majumder:2012sh,JETSCAPE:2021ehl,Kumar:2020wvb}, 
and $\tau$ is the lifetime of a given 
parton in the medium. Beyond this point, the parton transitions to a multiple scattering or transport phase where emissions are dominantly medium induced (the vacuum like emission probability is negligible compared to 
the medium induced probability).

To address this change in the physical picture from high virtuality to lower virtuality, one typically uses two different jet event generators transitioning at an intermediate scale that is comparable to $\mu_S^2$. 
Within the JETSCAPE framework~\cite{Putschke:2019yrg,JETSCAPE:2017eso}, the high virtuality regime is typically dealt with using the MATTER event generator~\cite{Cao:2017qpx,Majumder:2013re}. 
The lower virtuality, transport state is typically dealt with using the LBT~\cite{Cao:2016gvr,He:2015pra} or MARTINI generators~\cite{Schenke:2009gb}. While all three of these generators use a medium induced single gluon emission kernel, 
in both the MATTER and LBT generators the single gluon emission kernel is adopted from the next-to-leading twist single gluon 
emission kernel in Ref.~\cite{Guo:2000nz,Wang:2001ifa} with multiple scattering modifications from Ref.~\cite{Majumder:2009ge}. As a result, the single gluon emission kernel at next-to-leading twist 
plays an essential role in current state-of-the-art simulations. While the MARTINI generator uses the Arnold-Moore-Yaffe single gluon 
emission rate~\cite{Arnold:2001ba,Arnold:2001ms,Arnold:2002ja}, 
the numerical rate at temperatures of $T \gtrsim 200$~MeV for path lengths larger than a few Fermi, 
are very similar. Prior to the use of multi-stage event generators, there has been more than 15 years of development of jet modification where the single gluon emission rate at next-to-leading twist has been used to study a variety of high transverse momentum (high-$p_T$) observables~\cite{Majumder:2004pt,Deng:2009ncl,Chen:2011vt, Wang:2013cia, Cao:2016gvr, Cao:2017hhk,  Majumder:2011uk,Qin:2009gw,Qin:2009uh,Majumder:2004pt} (see Ref.~\cite{Cao:2020wlm} for a complete list).

The evaluation of the single gluon emission rate at next-to-leading twist is thus a matter of some importance. There have already been several attempts to evaluate this, both based on dropping contributions from terms in the phase~\cite{Guo:2000nz,Majumder:2009ge,Wang:2001ifa}, or keeping terms from the phase but dropping terms suppressed in the large $N_c$ limit~\cite{Aurenche:2008hm,Aurenche:2008mq}. In this Paper, we hope to carry out, possibly, the last calculation of the single gluon emission rate at next-to-leading twist, by including all contributions that were previously ignored. The final results obtained show a surprisingly different 
functional dependence on the length from the hard scattering at which the single gluon is emitted. 
However when integrated up to twice the mean formation time, the full calculation yields results numerically similar to the original subset of contributions evaluated in  Ref.~\cite{Guo:2000nz,Wang:2001ifa}.

In Sect.~\ref{sec:Review}, we give an overview of the full framework of DIS on a large nucleus, and highlight salient features about the total cross section and the hadronic tensor. In Subsect.~\ref{subsec:dispute}, we highlight the dispute over the collinear expansion between the original derivations of Guo and Wang (GW)~\cite{Guo:2000nz,Wang:2001ifa} and the later work of Aurenche, Zakharov and Zaraket (AZZ). We demonstrate that the dispute arises because both calculations are, in a sense, incomplete. The complete next-to-leading twist calculations is explained in detail in Sect.~\ref{sec:level3} with three subsections. When the emitted gluon is collinear and soft, the central cut diagrams with scatterings after the emission dominate the process. The calculations of the four central cut diagrams with post-emission scatterings (PES) are given in Subsect.~\ref{subsec:level3.1}.  

If at least one scattering, either in the amplitude or the complex conjugate, happens before the emission, the contributions are expected to be suppressed by the momentum fraction of the gluon. Therefore, these diagrams were neglected in the limit of the soft gluon approximation. There are five such diagrams with at least one pre-emission scattering. It is shown in Subsect.~\ref{subsec:level3.2} that all contributions arising from these diagrams are indeed suppressed by the momentum fraction of the gluon, except for one term.  The calculation is not complete without considering the left and right cut diagrams that isolate both scatterings into either the amplitude or the complex conjugate. In Subsect.~\ref{subsec:level3.3}, we illustrate the calculation of left and right cut diagrams by evaluating two of them, and showing that they also have non-zero contribution to the medium modification. The calculations of all the other left and right cut diagrams are given in Appendix~\ref{sec:appendix1}, which do not contribute to the final result.

The final result for the medium modified kernel can be obtained by combining all the results from the calculated Feynman diagrams. The calculated medium modified kernel and its comparison with the GW and AZZ results are given in Sect.~\ref{sec:level4}. The most obvious difference between this result and the previous once is the momentum fraction $\left( z \right)$ dependence. The momentum fraction dependence of this result is also studied Sect.~\ref{sec:level4}. The significance and the conclusion of this study is given in Sect.~\ref{sec:level5}.

\section{Final state in DIS on a nucleus: a Review}
\label{sec:Review}

In this section, we decompose the general structure of terms that arise in the next-to-leading twist modification of the final-state fragmentation function. Throughout this Paper, we will consider modifications to the fragmentation function, as has been the case in the prior efforts~\cite{Wang:2001ifa,Guo:2000nz,Majumder:2009ge,Majumder:2009zu,Aurenche:2008hm}. However, our results could be immediately applied to a final-state jet function $J(z,Q)$ as well. We will assume that the light-cone momentum of the outgoing quark $q^-$ is sufficiently high such that the preponderance of hadron formation takes place outside the nuclear medium, allowing us to use the vacuum fragmentation function as an input to the medium modified evolution equation. 

After highlighting the general factorized structure of the contributions we will focus on the source of dispute between the earlier calculation of Guo and Wang~\cite{Guo_2000,Wang_2001} and the later calculation of Aurenche et al.~\cite{Aurenche:2008hm}.

\subsection{General structure of terms}
\label{subsec:GeneralStructure}

We consider the deep inelastic scattering (DIS) of an electron off a large nucleus with atomic number $A$. 
The mean momentum of nucleons within this nucleus is $p$ and thus the nucleus has momentum $Ap$. The process is considered in the high energy limit where the nucleus and individual nucleons are moving with a large momentum in the ($+$)-$z$ direction, such that the internal motion of the nucleons within the nucleus can be ignored. In light-cone coordinates, 
\begin{eqnarray}
    p &\equiv& [p^+, p^-, \vec{p}_\perp] = \left[ \frac{p^0 + p^3}{\sqrt{2}}, \frac{ p^0 - p^3  }{\sqrt{2}} , \vec{p}_\perp    \right] \nonumber \\ 
    &\simeq&  \left[ p^+ , \frac{M^2}{2 p^+} , 0 , 0 \right]. 
\end{eqnarray}

\begin{figure}[!htb]
    \includegraphics[width=0.9\linewidth]{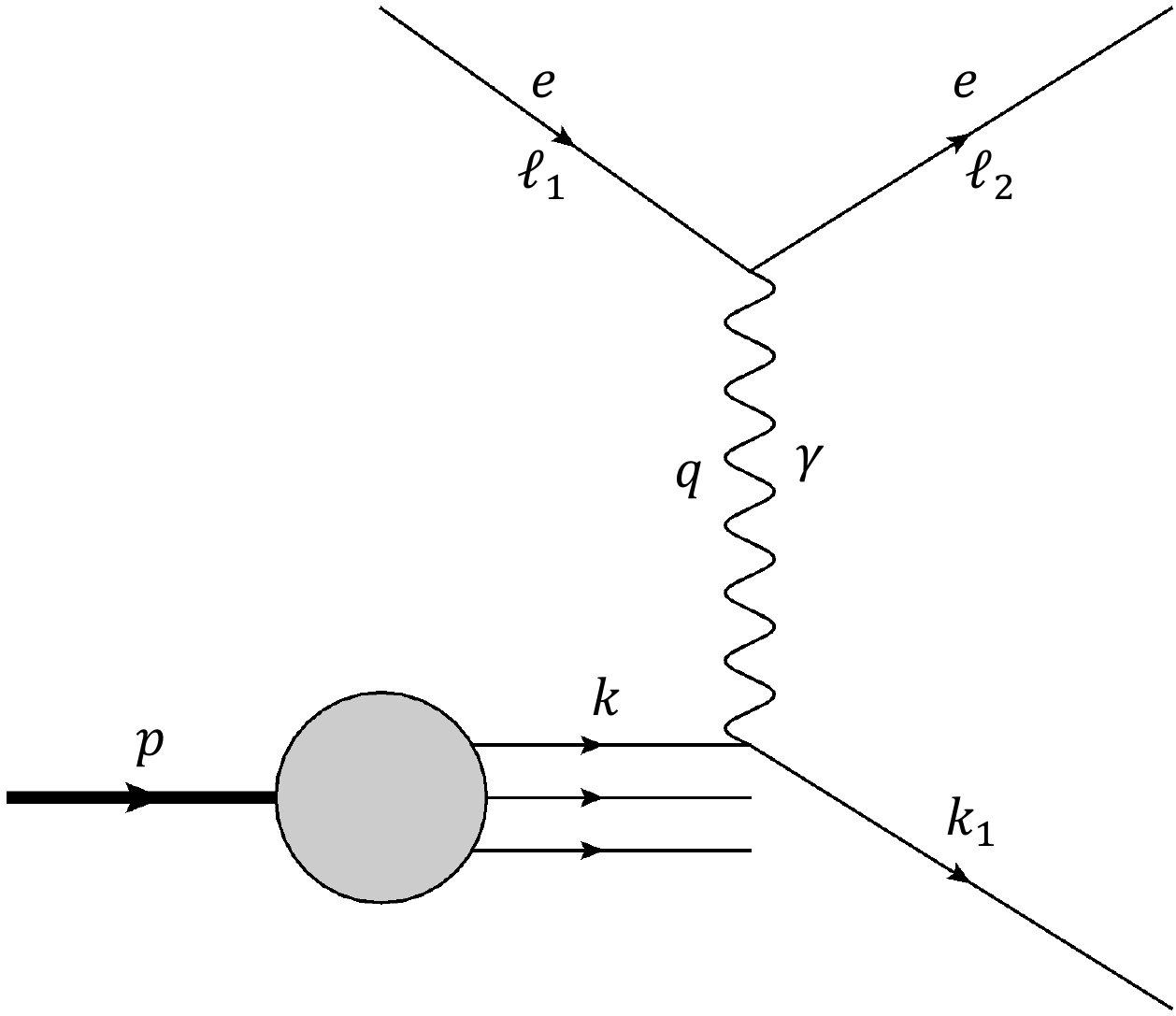}
    \caption{A schematic picture of the deep-inelastic scattering of a high energy electron off a quark in a proton.}
    \label{fig:DIS}
\end{figure}

Shown in Fig.~\ref{fig:DIS} is a diagrammatic illustration of a part of the process: the electron enters the nucleus with momentum $l_1$, and exits with $l_2$ after striking a single nucleon. While further interactions of the electron are possible, these are suppressed by factors of $\alpha_\mathrm{EM}$.
The electron scatters by exchanging a single highly virtual photon (with momentum $q = l_2 - l_1$). At high enough energy and virtuality ($q^2 = - Q^2$) the photon will resolve and scatter off a single quark with momentum $k$ within the nucleon.  
The process is considered in the Breit frame, where in light-cone coordinates,  
\begin{equation}
    q \equiv \left[ q^+ , q^-, \vec{q}_\perp   \right] = \left[ -\frac{Q^2}{2q^-} , q^- , 0, 0 \right].
\end{equation}
In this frame, the struck quark's $(+)$-light-cone momentum is almost balanced by the momentum of the photon, i.e.,
\begin{eqnarray}
k_1^+ = q^+ + k^+ \sim \lambda^2 Q,
\end{eqnarray}
where, $\lambda \rightarrow 0$ is a small dimensionless variable used to identify power suppressed contributions. 
In the limit of an exact cancellation, one can re-scale $k^+ = x_B p^+$ and obtain the definition of the dimensionless Bjorken variable, 
\begin{eqnarray}
\label{Eqn:x_B}
x_B = \frac{Q^2}{2q^- p^+}.
\end{eqnarray}
While the momenta of the virtual photon $Q^2, q^-$ can be chosen by the experiment, quarks from the nucleon  enter the hard scattering with momentum fractions that fluctuate about $x_B$, i.e., $x = x_B + \delta x$. 
For $\delta x > 0$, the final quark is time-like and will radiate, for $\delta x < 0$, the final state quark is space-like and will re-scatter.
The photon transfers its $(-)$-light-cone momentum to the parton, $k_1^- \simeq q^- $.
Thus, in this frame the final quark travels with a large negative $z$-component of momentum. 

The struck quark is virtual $k_1^2 \sim \lambda^2 Q^2$, and propagates through the remaining nucleus, showering and interacting with partons within the other nucleons. While some portion of the shower will excite the nucleus (or nucleons) and be constricted within it, some portion will escape and hadronize into a shower or jet of hadrons. 
A semi-inclusive process with an observed hadron in the final state can be written as
\begin{equation}
\label{eq:semi-inclusive}
e(\ell_1) + A(p) \rightarrow e(\ell_2) + h(\ell_h) +X, 
\end{equation}
where $\ell_h$ is the momentum of the hadron being observed, and $X$ is the observed final state of the remnant nucleus and possible jet that emanates along with $h$. 

The differential cross section can be expressed as
%
%
\begin{eqnarray}
\label{Eqn:DiffCrossSec}
\frac{ d\sigma }{ d^3\ell_2 dz_h } = \frac{ \alpha_{\rm EM}^2 }{ 2\pi \left( p + \ell_1 \right)^2 } \frac{ 1 }{ E_{ \ell_2 } Q^4 } L_{ \mu \nu } \frac{ d \mathcal{W}^{ \mu \nu } }{ dz_h },
\end{eqnarray}
where $\alpha_{\rm EM}$ is the electromagnetic coupling constant, $E_{\ell_2}$ is the energy of the final state electron, $Q^2=-q^2$ denotes the virtuality (or offshellness) of the virtual photon, and $z_h = \ell_h^- / q^-$ is the fractional minus-momentum of the final state hadron taken from the virtual photon. Since the largest momentum component of the 
outgoing quark is the ($-$)-light-cone component, the leading hadron will also possess a dominant ($-$)-light-cone momentum.

In the equation above, $L_{\mu \nu}$ and $\mathcal{W}^{\mu \nu}$ are the corresponding leptonic and hadronic tensors. The leptonic tensor is given by $L_{\mu \nu} = \frac{ 1 }{ 2 } \Tr \left[ \ell_1 \cdot \gamma \gamma_\mu \ell_2 \cdot \gamma \gamma_\nu \right] $. In the case where there is one gluon emission from the hard parton and one secondary scattering with the nucleus, the differential hadronic tensor for the production of a hadron with momentum $\ell_h^- = z_h q^-$, in the fragmentation of the outgoing quark, can be written as,
%
%
\begin{eqnarray}
\label{Eqn:Diff_HadTensor}
\frac{ d \mathcal{ W }^{\mu \nu } }{ d z_h } & = & \mathcal{ W }^{ \mu\nu } \frac{ 1 }{ z } D \left( \frac{ z_h }{ z 	} \right),
\end{eqnarray}
where $ D \left( z \right) $ is the vacuum fragmentation function for the momentum fraction $z$. The assumption of using a vacuum fragmentation function is justified in the case where the energy of the quark is very large, such that a large portion of the hadronization takes place outside. The hadronic tensor $ \mathcal{ W }^{ \mu\nu } $ is given by,
%
%
\begin{eqnarray}
\label{Eqn:HadTensor_General}
\mathcal{ W }^{\mu\nu} & = & \sum_q \int \frac{ dy^- }{ \left( 2 \pi \right) } dy_1^- dy_2^- \frac{ d^2 y_\perp }{ \left( 2 \pi \right)^2 } d^2 k_\perp \int dz \nonumber\\
& \times & \mathcal{ H }^{ \mu \nu }(z) e^{ i \vec{ k }_\perp \cdot \vec{ y }_\perp } \frac{ 1 }{ 2 }
\left\langle A \middle| \bar{ \psi } \left( y^- \right) \gamma^+ A^+ \left( y_1^-, \vec{ y }_\perp \right) \right. \nonumber\\
& \times & \left. A^+ \left( y_2^-, \vec{ 0 }_\perp \right) \psi \left( 0 \right) \middle| A \right\rangle.
\end{eqnarray}
Here $\psi$ represents the hard quark field and $A^+$ is the $(+)$-component of the exchanged gluon field for rescattering. 

All our calculations will be carried out in the $A^-=0$ gauge. 
The energy scaling of the various components of the vector potential can be carried out as in Ref.~\cite{Idilbi:2008vm}. Note that we have assumed the perturbative part (or the partonic hard part) $\mathcal{H}^{\mu\nu}$ can be factorized from the quark fragmentation function $D_q(z_h/z)$ and the soft matrix element. This is straightforwardly demonstrated as in Ref.~\cite{Qiu:1990xxa,Qiu:1990xy}. We can further factorize the soft four-point matrix element as,
\begin{eqnarray}
&& \left\langle A \middle| \bar{ \psi } \left( y^- \right) \gamma^+ F_\alpha^+ \left( y_1^-\right) 
 F^{ +\alpha } \left( y_2^-\right) \psi \left( 0 \right) \middle| A \right\rangle \nonumber \\
&\simeq& C^A_p \left\langle A \middle| \bar{ \psi } \left( y^- \right) \gamma^+\psi \left( 0 \right) \middle| A \right\rangle \nonumber \\
&\times& \left\langle A-1 \middle| F_\alpha^+ \left( y_1^-\right) 
 F^{ +\alpha } \left( y_2^-\right) \middle| A-1 \right\rangle. \label{Eqn:MatrixElementFactorization}
\end{eqnarray}
In the equation above, the factor $C^A_p$ contains the correlation between the nucleon struck by the photon and the nucleon where the quark re-scatters. We use $A-1$ for the second matrix element to indicate that after the first nucleon is struck by the photon, the propagating quark can scatter off any of the remaining $A-1$ nucleons. In this way, we are explicitly ignoring the case where the rescattering occurs in the same nucleon. The first term will be absorbed into the nuclear parton distribution function and the second term will be absorbed into the jet quenching parameter $\hat{q}$ later in Eqs.~\eqref{Eqn:PDF} and~\eqref{eq:q-hat} for calculating jet energy loss. The partonic hard part $\mathcal{H}^{\mu\nu}$ will be the only quantity that varies with different rescattering processes. During detailed calculation in the next section, it will be shown that the leading twist partonic hard part $\mathcal{ H }_0^{ \mu\nu }$ can be extracted from the next-to-leading twist $\mathcal{H}^{\mu\nu}$ as follows,
%
%
\begin{eqnarray}
\label{Eqn:NLO_H}
\mathcal{ H }^{ \mu\nu } & = & \mathcal{ H }_0^{ \mu\nu } \int d \ell_\perp^2 \frac{ \alpha_S }{ 2 \pi } \frac{ 1 + z^2 }{ 1 - z } \frac{ 2 \pi \alpha_S }{ N_C } \nonumber\\
& \times &\mathcal{ H } \left( y^-, y_1^-, y_2^-, \vec{ \ell }_\perp, \vec{ k }_\perp, z \right),
\end{eqnarray}
where $\mathcal{H}$ is the $\vec{ k }_\perp$ dependent part of the partonic hard part, and the leading twist hard part is given by
%
%
\begin{eqnarray}
\label{Eqn:H0}
\mathcal{ H }_0^{ \mu\nu } & = & \frac{ 1 }{ 2 } e_q^2 \left( 2 \pi \right) \delta \left[ \left(  q + xp \right)^2 \right] \nonumber\\
& \times & \Tr \left[ p \cdot \gamma \gamma^\mu \left( q + xp \right) \cdot \gamma \gamma^\nu \right].
\end{eqnarray}

The hadronic tensor $ \mathcal{ W }^{\mu \nu} $ can be calculated from the $ T $-matrix using the contributing Feynman diagrams. They are related by,
%
%
\begin{eqnarray}
\label{Eqn:WfromT}
\mathcal{ W }^{ \mu\nu } = \frac{ 1 }{ \left( 2 \pi \right) } \mathrm{Disc} \left[ T^{ \mu\nu } \right].
\end{eqnarray}

To obtain the next-to-leading twist contribution of the hard part, one may use the collinear expansion around $\vec{ k }_\perp = 0$ as follows,

\begin{eqnarray}
\label{Eqn:ColExpansion}
\mathcal{ H }^{\mu\nu} & = & \mathcal{ H }^{\mu\nu} \left( k_\perp = 0 \right) 
+ \frac{ \partial \mathcal{ H }^{\mu\nu} }{ \partial k_\perp^\rho } \bigg|_{ k_\perp = 0 } k_\perp^\rho \nonumber\\
& + & \frac { \partial^2 \mathcal{ H }^{\mu\nu} }{ \partial k_\perp^\rho \partial k_\perp^\sigma } \bigg|_{ k_\perp = 0 } k_\perp^\rho k_\perp^\sigma + ... ~~,
\end{eqnarray}
where the first non-zero and non-gauge contribution to the double scattering process is from the double derivative term~\cite{Guo:2000nz}. By substituting this double derivative term into Eq.~\eqref{Eqn:HadTensor_General},
%
\begin{eqnarray}
\label{Eqn:Diff_HadTensorSimplify}
\mathcal{ W }^{\mu \nu } & = & \sum_q \int dz \int  d\xi^-  f_q^A \left( x \right) \nonumber\\
%
& \times & \hat{ q } \left( \xi^- \right) \frac{1}{4} \nabla_{ k_\perp }^2 \mathcal{ H }^{ \mu \nu } (z)|_{k_\perp=0},
\end{eqnarray}
Here we have made a variable change $ \xi^- =( y_1^- + y_2^- ) / 2  $ and $ \delta y^- = ( y_1^- - y_2^- ) /  2  $. The $ \delta y^- $ integration is absorbed into $ \hat{ q } $ since its contribution is negligible in the partonic hard part. In the above equation, $f$ is the nuclear quark distribution function defined as~\cite{Majumder:2009ge},
%
%
\begin{eqnarray}
\label{Eqn:PDF}
f_q^A \left( x  \right) & = & A \int \frac{ dy^- }{ 2 \pi } e^{ -i  x  p^+ y^- } \nonumber\\
& \times & \left\langle P \middle| \bar{ \psi } \left( y^- \right) \frac{ \gamma^+ }{ 2 } \psi \left( 0 \right) \middle| P \right\rangle,
\end{eqnarray}
and the gluon jet transport coefficient is defined as \cite{Kumar:2017des,Bianchi:2017wpt},
%
%
\begin{align}
\label{eq:q-hat}
&\hat{ q } \left( \xi^- \right)  =  \frac{ 4 \pi^2 \alpha_S C_A}{ N_C^2 - 1 } \int \frac{ d \delta y^- }{ \left( 2 \pi \right) } 
\frac{ d^2 k_\perp d^2 y_\perp }{ \left( 2 \pi \right)^2 }  \\
& \times e^{ -i \frac{ k_\perp^2 }{ 2 q^- } \delta y^- + i k_\perp \cdot y_\perp } \nonumber \\
& \times \left \langle P \middle| F_\alpha^+ \left( \xi^- \!\!\!+ \!\!\frac{\delta y^-}{2} , y_\perp\right) F^{\alpha+} \left( \xi^- \!\!\!-\!\! \frac{\delta y^-}{2} , 0_\perp \right) \middle| P \right \rangle, \nonumber
\end{align}
where $k_\perp^\rho A^+ k_\perp^\sigma A^+$ is converted into the field strength $F^{\rho+}F^{\sigma+}$ via partial integration. 
See Ref.~\cite{Idilbi:2008vm} for details on this conversion. Unlike statements made in prior efforts, the termination of the Taylor expansion at $k_\perp^2$  is not because $k_\perp \ll l_\perp$, but rather because of the vanishing size of the expectation $\langle P | F_\alpha^+ \left( y_1^-,y_\perp\right) F^{\alpha+} \left( y_2^-, 0_\perp \right) | P \rangle $ for $k_\perp \gg \Lambda_\mathrm{QCD}$. See Ref.~\cite{Kumar:2019uvu} for further details. 

Conservation of color charge inside a nucleon requires the parton scattering in the amplitude term and the corresponding scattering in the complex conjugate term emerge from the same nucleon. Here we assume $y \simeq 0$ and $y_1^- \simeq y_2^- = \xi^-$, considering that the size of a nucleon is negligible compared to that of a large nucleus. 
This helps further simplify the hadronic tensor as 
%
%
\begin{eqnarray}
\label{Eqn:HadTensor_Final}
\mathcal{ W }^{\mu\nu} & = & \sum_q \int dz f_q^A (x) 
%
\mathcal{ K } \left( q^-, z \right) \mathcal{ H }_0^{ \mu\nu }, 
\end{eqnarray} 
where $\mathcal{K}$ will be referred to as the ``medium modified kernel" in the rest of this work. 

An arbitrary term in the medium modified kernel can be symbolically decomposed into a complex amplitude (which only depends on factors of momentum) times a space-time dependent phase factor, times a non-perturbative matrix element which only depends on position as follows,
\begin{eqnarray}
\mathcal{K}_{ab} &=& \int \rho_a^* (\{p_a\}) e^{- i f_a ( \{p_a\} \cdot y_a ) }  \rho_b(\{p_b\}) e^{i f_b( \{p_b\}\cdot y_b )} \nonumber \\
&\times & \langle P | A (y_a) A (y_b) | P \rangle.  \label{general_form}
\end{eqnarray}
In the equation above, $\{p_{a/b}\}$ refers to the collection of momenta active at location $y_{a/b}$. Integration over all positions and internal momenta is implied. In the phase factors, $f_{a/b} (\{ p_{a/b} \})$ refer to different possible functions of the momenta that may appear. The factors of $A$ in the matrix element represent gluon vector potentials with suppressed indices.

The entire medium modification kernel is obtained by summing over all diagrammatic contributions, 
\begin{equation}
\mathcal{K} = \sum_{a,b} \mathcal{K}_{a,b}.    \label{k-sum}
\end{equation}
In the equation above, $\mathcal{K}$ represents the square of an entire matrix element. As such $\mathcal{K}$ should be positive definite, as is the case for a probability density. Taking the first non-vanishing derivative with respect to $k_\perp$ should yield a positive contribution, the matrix element becomes larger as the $k_\perp^2$ grows from zero, i.e., 
$\nabla_{k_\perp}^2 \mathcal{K}(k_\perp = 0) \geq 0$. In the limit of an average over $k_\perp$ weighted with a medium matrix element which disfavors large values of $k_\perp^2$, both the matrix element and the first non-vanishing derivative should grow with distance traversed. In fact, the location where the first non-vanishing derivative with respect to $k_\perp^2$ matches the term with $\mathcal{K} (k_\perp  = 0)$, is where the assumption of single re-scattering, and expansion up to the first non-vanishing derivative should break down.  

As will be explained in the subsequent subsection, the primary dispute in the evaluation of the next-to-leading twist contribution to single gluon emission arises from either not including derivatives of the space-time and momentum dependent phase terms~\cite{Guo:2000nz,Wang:2001ifa}, or including only a subset of the terms in the sum of Eq.~\eqref{k-sum}~\cite{Aurenche:2008hm,Aurenche:2008mq}.

\subsection{Dispute over collinear expansion in the single gluon emission rate}
\label{subsec:dispute}

In the evaluation of the medium induced gluon emission cross section from single rescattering, there are 19 Feynman diagrams that can potentially contribute to the double scattering process, most of the them have a vanishing contribution at the next-to-leading twist (or at first power correction). The first calculation of the medium modification kernel (referred to as the GW kernel) was provided in Refs.~\cite{Wang:2001ifa, Guo:2000nz} as
%
%
\begin{align}
\label{Eqn:GW_kernel}
 \mathcal{K}_\mathrm{GW} & \left( q^-, z \right) = \frac{\alpha_S}{ 2 \pi } \int \frac{ d \mu^2 }{ \mu^4 } 
\int d \xi^- C_F \frac{ 1 + z^2 }{ 1 - z } \nonumber\\
 \times & \frac{ 1 }{ z \left( 1 - z \right) }\hat{ q } \left( \xi^- \right) \left\{ 2 - 2 \cos \left( \frac{ \mu^2 \xi^- }{ 2 q^- } \right) \right\}.
\end{align}
In this work, the authors invoke both the soft and collinear approximation on the emitted gluon. The expectation of the  transverse momentum exchange in the secondary scattering is considered small compared to the transverse momentum of the emitted gluon in the high energy and high virtuality regime where the higher-twist formalism is applicable. Therefore, $\expval{ k_\perp^2 } \ll \ell_\perp^2 $ is assumed.  This implies $\langle ( \vec{ \ell }_\perp - \vec{ k }_\perp )^2 \rangle \simeq \ell_\perp^2 $, or 
$ \ell_\perp^2 \gg \langle k_\perp^2 \rangle - 2 \langle \vec{ \ell }_\perp \cdot \vec{ k }_\perp \rangle $. The averages are carried out over different configurations of the medium, and thus for media with rotational symmetry $\langle \vec{\ell}_\perp \cdot \vec{k}_\perp  \rangle \simeq 0$, for a fixed $\vec{\ell}_\perp$.

To obtain the equation above, the authors of Refs.~\cite{Guo:2000nz,Wang:2001ifa} neglect the $ k _\perp $-dependence in the phase factors before applying the double derivative with respect to $k_\perp$ in Eq.~\eqref{Eqn:Diff_HadTensorSimplify}, leading to the GW kernel Eq.~\eqref{Eqn:GW_kernel}. The argument to ignore these contributions is predicated on the limit of very high energies $q^-$ and medium lengths shorter than a formation time $L^- \leq 2q^- z(1-z)/l_\perp^2$.
Critical to the analysis in Refs.~\cite{Guo:2000nz,Wang:2001ifa} is the assumption that power corrections to the phase factors can be ignored. This along with the retention of only the next-to-leading twist or power correction, leads to only one contributing diagram topology (shown in Fig.~\ref{fig:DoubleggScat1}).

Later it was suggested by Aurenche, Zakharov and Zaraket (AZZ)~\cite{Aurenche:2008hm} that the $ \vec{ k }_\perp $ dependence of the phase factors in the hadronic tensor cannot be ignored in the derivative expansion, specifically for the case of longer media. By taking into account the phase factor for \emph{only} the most dominant diagram of the GW kernel (as illustrated in Fig.~\ref{fig:DoubleggScat1}), the single gluon emission kernel was re-calculated in Ref.~\cite{Aurenche:2008hm} as, 
%
%
\begin{align}
\label{Eqn:AZZ_kernel}
 \mathcal{K}&_\mathrm{AZZ}  \left( q^-, z \right) = \frac{\alpha_S}{ 2 \pi } \int \frac{ d \mu^2 }{ \mu^4 } \int d \xi^-  C_F \frac{ 1 + z^2 }{ 1 - z } \nonumber\\
 &\times  \frac{ 1 }{ z \left( 1 - z \right) } \hat{ q } \left( \xi^- \right) \left\{ 2 - 2 \cos \left( \frac{ \mu^2 \xi^- }{ 2 q^- } \right) \right. \\
 &-  \left. \left( \frac{ \mu^2 \xi^- }{ 2 q^- } \right) 
\sin \left( \frac{ \mu^2 \xi^- }{ 2 q^- } \right) + \left( \frac{ \mu^2 \xi^- }{ 2 q^- } \right)^2
\cos \left( \frac{ \mu^2 \xi^- }{ 2 q^- } \right) \right\}. \nonumber
\end{align}
This will be referred to as the AZZ kernel in the remainder of this paper. One observes two additional terms in the AZZ kernel compared to the GW kernel. As will be shown in Fig.~\ref{fig:Kernel_Comp}, these two terms drive the medium-induced gluon kernel to become oscillatory at large path length $\xi^-$, breaking the positive definiteness of the gluon spectrum. This leads to the long dispute over the validity of collinear expansion in the calculation of medium-induced gluon emission.

In this work we are about to address two questions related to the medium-induced gluon radiation. The first is the role played by the soft gluon approximation that has been assumed in both GW and AZZ. With the recent increase of the colliding energy in heavy-ion experiments, wide angle radiation is expected to become important in understanding jet showers, especially when the parton energy and virtuality are both high. Jet observables are no longer confined to leading hadrons. To understand the modification to the entire jet, one needs to compute the medium modified distribution of splits beyond the soft gluon approximation. Therefore, the soft gluon approximation will be abandoned in this work, and a more general result will be provided. The second question is the significance of the contribution from the phase factor in the hadronic tensor. In addition to the diagram considered by AZZ, we will conduct detailed calculations of all the remaining diagrams for double scattering. We will show that after taking into account the complete set of diagrams, the positive definiteness of the medium-induced gluon spectrum is resurrected. In the case of only post-emission scattering (PES) diagrams, and in combination with the soft gluon approximation, our results are  consistent with both GW and AZZ kernels at small $\xi^-$.

\subsection{Soft gluon approximation and post-emission scattering}
\label{subsec:level2.1}

In most calculations of single gluon emission, either in medium or in vacuum, the soft gluon approximation is invoked  to simplify the calculation. In this approximation, emitted gluons are assumed to be soft or in other words have significantly less longitudinal momentum (momentum in the direction of the parent parton) compared to the parent parton. This is true even for a gluon splitting into two gluons, where the momentum fraction of one of the emitted gluons is typically very small.  If we ignore the $\vec{k}_\perp$ dependent terms in the phase, and in addition invoke the soft gluon approximation, it allows us to ignore the contributions from several Feynman diagrams where there are at least one pre-emission scattering. This simplification is possible because all such contributions after the collinear expansion are suppressed by the momentum fraction of the gluon.

In Subsect.~\ref{subsec:level3.2}, we demonstrate that retaining $k_\perp$ dependent terms that originate from the 
phase of terms with scattering before emission do not always lead to terms suppressed by gluon momentum fraction. 
In fact, terms with one scattering before and one after emission are shown to contain contributions that are not suppressed by momentum fraction of the emitted gluon. These highlight the importance of carrying out the $k_\perp$ expansion on the momentum and space-time dependent phase terms [$f_{a/b}$ in Eq.~\eqref{general_form}] as well as the complex amplitude and the complex conjugate [$\rho_{a/b}$ in Eq.~\eqref{general_form}].

In the subsequent section we will carry out the full calculation of the single gluon emission rate at the  next-to-leading twist. We will explicitly demonstrate the approximations under which one can recover the GW and AZZ kernels. The expansion of contributions from the phase will demonstrate that these calculations are not complete. The final results and numerical comparisons will be presented in Sect.~\ref{sec:level4}.

\section{\label{sec:level3}Full next-to-leading twist calculation}

In this section we carry out the full analytic calculation of the single gluon emission kernel at the next-to-leading twist. 
The Taylor expansion of the result of each diagram in terms of the exchanged transverse momentum $k_\perp$, is carried out both on the momentum dependent amplitude and complex conjugate, as well as on the space-time and momentum dependent phase factors. The presence of $k_\perp$ contributions from the phase leads to contributions from several diagram topologies. We also demonstrate later in the section that most of the non-central cut diagrams continue to have a vanishing contribution to the single gluon emission at the next-to-leading twist. Only two non-central cut diagrams contribute to the final result. 

\subsection{\label{subsec:level3.1} Central cut diagrams for post-emission scattering }
\begin{figure}[tbp]
    \addtolength{\abovecaptionskip}{-3mm}
%
	\centering
	\includegraphics[width=\linewidth]{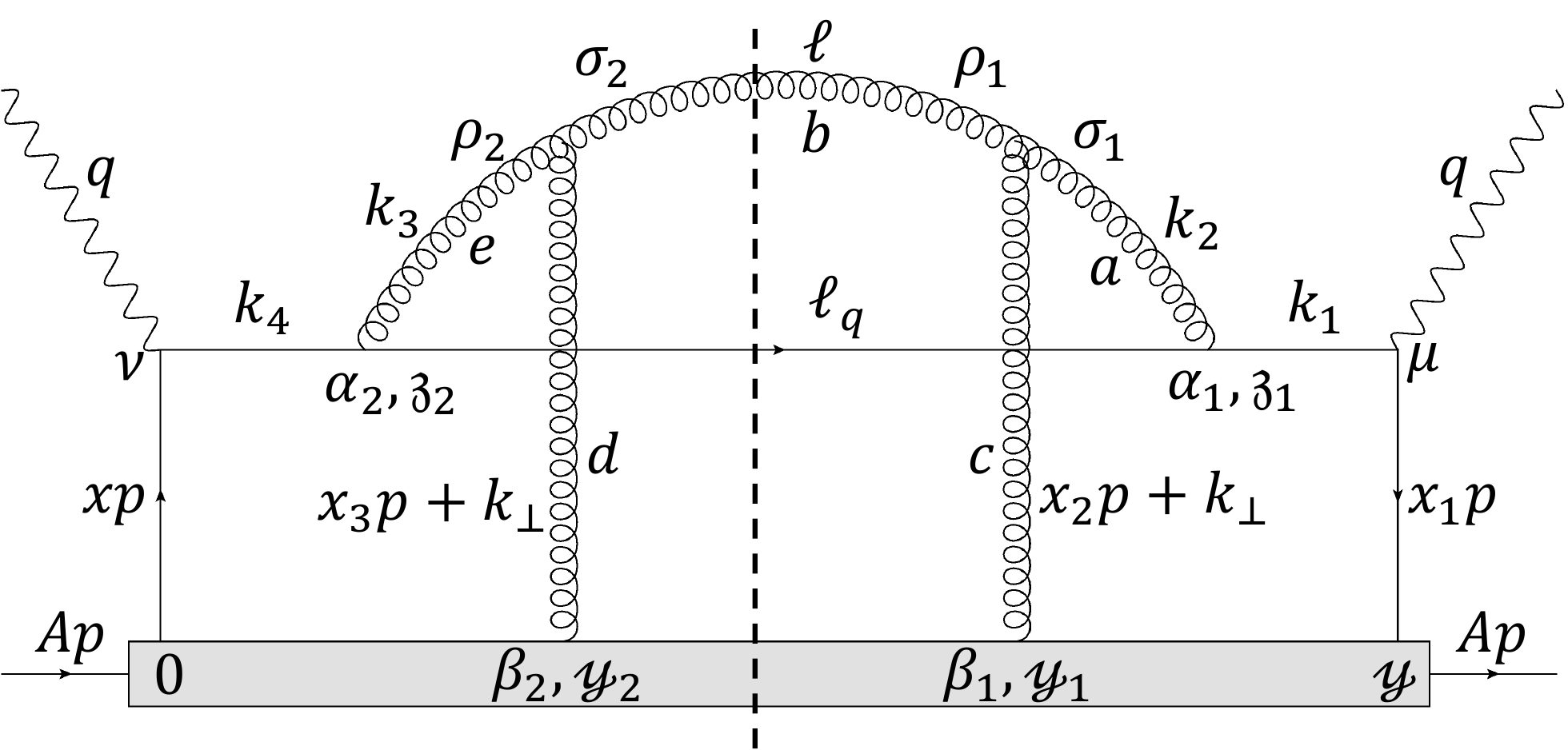}
	\caption{Feynman diagram for the quark-gluon double rescattering process.}
	\label{fig:DoubleggScat1}
\end{figure}

In this subsection, we first consider contributions from diagrams in which both scatterings happen after gluon emission. Even though all the Landau–Pomeranchuk–Migdal (LPM)~\cite{Landau:1953um, Landau:1953gr,Migdal:1956tc} interference diagrams were considered in the GW calculation, by neglecting the phase factors, the only contributing diagram is the double gluon-gluon re-scattering diagram with the discontinuity taken at the center of the diagram as shown in Fig.~\ref{fig:DoubleggScat1}. We first review the contribution from this digram in detail, with the phase factors taken into account. Light cone coordinates are used here to simplify the calculation. In this calculation the nucleus travels in the positive light cone direction with momentum $p=[p^+, 0^-, \vec{0}_\perp]$, and the virtual photon has both negative and positive components of the light cone momentum as $q=[ x_B p^+,  q^-, \vec{ 0 }_\perp]$.  The Bjorken variable $x_B$ is given by Eq.~\eqref{Eqn:x_B}.
%
%

The $\mathcal{T}$-matrix associated with this diagram can be calculated using the QCD Feynman rules as
%
%
\begin{eqnarray}
\label{Eqn:DoubleggScat1Tmatrix}
\mathcal{T}^{\mu\nu}_{\ref{fig:DoubleggScat1}} & = & - \int d^4y d^4z_1 d^4z_2 d^4y_1 d^4y_2 
\int \frac{d^4k_1}{\left( 2\pi \right)^4} \frac{d^4k_2}{\left( 2\pi \right)^4} \frac{d^4k_3}{\left( 2\pi \right)^4} \nonumber\\
& \times & \int \frac{d^4k_4}{\left( 2\pi \right)^4} \frac{d^4\ell}{\left( 2\pi \right)^4} \frac{d^4\ell_q}{\left( 2\pi \right)^4} e^{iqy} 
e_q^2 \left\langle A \middle| \bar{\psi} \left( y \right) \gamma^\mu \right.\nonumber\\
& \times & \left. \frac{i \slashed{k}_1 e^{-i k_1 \left( y - z_1 \right)}}{ k_1^2 + i \epsilon} 
 igt^a \gamma^{\alpha_1} \frac{i d_{\alpha_1 \sigma_1} e^{-i k_2 \left( z_1 - y_1 \right)}}{ k_2^2 + i \epsilon} \right. \nonumber\\
& \times & \left. ig \left( -k_2 - \ell \right)^{\beta_1} f^{abc} A^c_{\beta_1} \left( y_1 \right) g^{\rho_1 \sigma_1} \right. \nonumber\\
& \times & \left. \frac{ i d_{\rho_1 \sigma_2} e^{-i \ell \left( y_1 - y_2 \right)}}{ \ell^2 + i \epsilon} 
\frac{i \slashed{\ell}_q e^{-i \ell_q \left( z_1 - z_2 \right)}}{ \ell_q^2 + i \epsilon} 
i g f^{bed} g^{\rho_2 \sigma_2} \right. \nonumber\\
& \times & \left. A^d_{\beta_2} \left( y_2 \right) \left( - \ell - k_3 \right)^{\beta_2}  
\frac{i d_{\rho_2 \alpha_2} e^{-i k_3 \left( y_2 - z_2 \right)}}{ k_3^2 + i \epsilon} igt^e \right. \nonumber\\
& \times & \left. \gamma^{ \alpha_2 }  \frac{i \slashed{k}_4 e^{-i k_4 z_2 }}{ k_4^2 + i \epsilon} \gamma^\nu \psi \left( 0 \right) \middle| A \right\rangle .
\end{eqnarray}

To conveniently map the $\mathcal{T}$-matrix to its corresponding Feynman diagram, we label it with the figure number (``2" in this case) as its subscript. This notation will be applied through the rest of this work. The negative light cone gauge is used in this calculation, where $n\cdot A = A^- = 0$ with $n = \left[ 1, 0, 0_\perp \right]$ in the light cone coordinates. The polarization tensor of the gluon propagator, $d_{ \alpha \beta } \left( \ell \right)$, in the light cone gauge is given by 
%
%
\begin{eqnarray}
\label{Eqn:PolarizationMatrix}
d_{ \alpha \beta } \left( \ell \right) = -g_{ \alpha \beta } + \frac{ n_\alpha \ell_\beta + n_\beta \ell_\alpha }{ n \cdot \ell } .
\end{eqnarray}

The hadronic tensor is then obtained by taking the discontinuity at the center as shown in Fig.~\ref{fig:DoubleggScat1}. Integrating over $z_1$ and $z_2$ followed by $k_2$ and $k_4$ integrals introduce two relations of momentum conservation as $k_2 = k_1 - \ell_q$ and $k_3 = k_4 - \ell_q$. Using these constrains, we can set $k_1 = \left[x_1 p^+ + q^+, q^-, 0_\perp\right]$, $k_4 = \left[x p^+ + q^+, q^-, 0_\perp\right]$, and $\ell_q = \left[x_1 p^+ + x_2 p^+ + q^+ - \ell^+, q^- - \ell^-, k_\perp - \ell_\perp \right]$. Therefore we can simplify the hadronic tensor by performing a set of integrals with respect to the above 4-momentum components and 4-position components. The remaining integrals over the ($+$)-components of the momenta and the remaining transverse component can be expressed as $dk_1^+ = p^+ dx_1$, $dk_4^+ = p^+ dx$, $d\ell_q^+ = p^+ dx_2$, and $d^2 {\ell_q}_\perp = d^2 k_\perp$. An additional $\delta$-function, $\int dz \delta \left( 1 - z -  \ell^- / q^- \right) = 1$, is enforced to introduce the momentum fraction ($z$) taken by the final quark.  
All the color-related terms can be combined to give a ${ C_A }/{ 2 N_C }$ factor, where $N_C$ is the number of colors, and $C_A = N_C$.  
With these simplifications, the hadronic tensor reads
%
%
\begin{eqnarray}
\label{Eqn:DoubleggScat1_W}
\mathcal{W}^{\mu\nu}_{\ref{fig:DoubleggScat1}} & = & \frac{1}{2\pi} \int dy^- dy_1^- dy_2^- d^2y_\perp \int \frac{dx_1}{ \left( 2\pi \right) } \frac{dx_2}{ \left( 2\pi \right) } 
\frac{dx}{ \left( 2\pi \right) } \frac{d^2k_\perp }{ \left( 2\pi \right)^2 } \nonumber\\
& \times & e_q^2 g^4 \left( p^+ \right)^2  e^{ -ix_1 p^+ y^- - ix_2 p^+ y_1^- - i \left( x - x_1 - x_2 \right) p^+ y_2^- } \nonumber\\
& \times & e^{ i\vec{ k }_\perp \cdot \vec{ y }_\perp } \int \frac{ d^4 \ell }{ \left( 2\pi \right)^4 } \frac{ C_A }{ 2 N_C } \int dz \delta \left( 1 - z - \frac{ \ell^- }{ q^- } \right) \frac{ 1 }{ 2 } \nonumber\\
& \times & \left\langle A \middle| \bar{ \psi } \left( y^- \right) \gamma^+ A^+ \left( y_1^-, \vec{ y }_\perp \right) 
 A^+ \left( y_2^-, \vec{ 0 }_\perp \right) \psi \left( 0 \right) \middle| A \right\rangle  \nonumber\\
& \times & \frac{1}{2} \Tr \left[ p . \gamma \gamma^\mu \frac{ k_1 . \gamma }{ k_1^2 - i \epsilon } \gamma^{\alpha_1}
\frac{d_{\alpha_1 \sigma_1}}{ \left( k_1 - \ell_q \right)^2 - i\epsilon } g^{\perp}_{\sigma_1 \rho_1} \right. \nonumber\\
& \times & \left( -2 \ell^- \right) \left. \left( 2\pi \right) \delta \left( \ell^2 \right) d_{\rho_1 \sigma_2}  
\left( 2\pi \right) \delta \left( \ell_q^2 \right) \ell_q . \gamma g^\perp_{\sigma_2 \rho_2} \right. \nonumber\\
& \times & \left. \left( -2 \ell^- \right) \frac{ d_{ \rho_2 \alpha_2 }}{ \left( k_4 - \ell_q \right)^2 + i\epsilon } \gamma^{\alpha_2 }
\frac{ k_4 . \gamma }{ k_4^2 + i\epsilon } \gamma^\nu \right]  . \label{Eqn:W2_general_form}
\end{eqnarray}
Here we have assumed that $A_\alpha \left( y \right) \simeq	p_\alpha \frac{ A^+ \left( y \right) }{ p^+ } $. This approximation together with the collinear approximation at the leading twist are used to separate the matrix element in Eq.~\eqref{Eqn:DoubleggScat1Tmatrix} into a non-perturbative matrix element and a trace part as shown in Eq.~\eqref{Eqn:DoubleggScat1_W}. 

To further perform contour integration over $x_1$ and $x_2$, we apply the following simplifications to the $\delta$-functions and the terms in the denominator. The $\delta$-functions can be written as
%
%
\begin{eqnarray}
\label{Eqn:Deltal2}
\delta \left( \ell^2 \right) = \frac{ 1 }{ 2 \ell^- } \delta \left( \ell^+ - \frac{ \ell_\perp^2 }{ 2 \ell^- } \right) ,
\end{eqnarray}
%
%
\begin{eqnarray}
\label{Eqn:Deltalq2}
\delta \left( \ell_q^2 \right) = \frac{ 1 }{ 2 p^+ q^- z } \delta \left( x_1 + x_2 - x_L - x_D -x_B \right),
\end{eqnarray}
where the Bjorken variable $x_B$ is defined in Eq.~\eqref{Eqn:x_B}, and the fractional momenta $x_L$ and $x_D$ are defined as
%
\begin{eqnarray}
x_L = \frac{ \ell_\perp^2 }{ 2p^+q^-z \left( 1 - z \right)},
\label{Eqn:x_L}
\end{eqnarray}
%
%
\begin{eqnarray}
x_D = \frac{ k_\perp^2 - 2 \vec{\ell}_\perp \cdot \vec{k}_\perp }{ 2p^+q^-z} .
\label{Eqn:x_D}
\end{eqnarray}
With these definitions, the four terms in the denominator can be rewritten as
%
%
\begin{align}
\label{Eqn: k12}
&k_1^2 = 2p^+ q^- \left( x_1 - x_B \right) ,\\
\label{Eqn:k42}
&k_4^2 = 2 p^+ q^- \left( x - x_B \right) ,\\
\label{Eqn:k22}
&\left( k_1 - \ell_q \right)^2  = 2p^+ q^- \left( 1 - z \right) \nonumber\\
&\quad\quad\quad \times \left( x_1 - x_L - x_B - \frac{ x_D }{ \left( 1 - z \right) } \right) ,\\
\label{Eqn:k32}
&\left( k_4 - \ell_q \right)^2 = 2p^+ q^- \left( 1 - z \right) \nonumber\\
&\quad\quad\quad \times \left( x - x_L - x_B - \frac{ x_D }{ \left( 1 - z \right) } \right) .
\end{align}
The contour integration part is then extracted from Eq.~\eqref{Eqn:DoubleggScat1_W} and calculated separately as follows,
%
%
\begin{align}
\label{Eqn:ContInt}
\mathcal{I}^C_{\ref{fig:DoubleggScat1}} & =  \int dx \frac{ d x_1 }{ \left( 2\pi \right) } \frac{ dx_2 }{ \left( 2\pi \right) } 
\frac{ 1 }{ z \left( 1 - z \right)^2 \left( 2 p^+ q^- \right)^5 } \nonumber\\
& \quad\times  \frac{ e^{ -ix_1 p^+ y^- - ix_2 p^+ y_1^- - i \left( x - x_1 - x_2 \right) p^+ y_2^- }}
{ \left( x_1 - x_B - i\epsilon \right) \left( x_1 - x_L - x_B - \frac{ x_D }{ \left( 1 -z \right) } - i\epsilon \right) } \nonumber\\
& \quad\times  \frac{ \delta \left( x_1 + x_2 - x_B - x_L - x_D \right) }{ \left( x - x_B + i\epsilon \right) 
\left( x - x_L - x_B - \frac{ x_D }{ \left( 1 -z \right) } + i\epsilon \right) } \nonumber\\
& =  \int dx \frac{ \delta \left( x - x_B \right) }{ \left( \vec{ \ell }_\perp - \vec{ k }_\perp \right)^4 }
 \theta \left( y_2^- \right) \theta \left( y_1^- - y^- \right) \nonumber\\
& \quad\times  e^{ -i \left( x_B + x_L \right) p^+ y^- - i x_D p^+ \left( y_1^- - y_2^- \right)} \frac{ z }{ \left( 2 p^+ q^- \right)^3 } \nonumber\\
& \quad\times  \left[ e^{ -i \frac{ x_D }{ \left( 1 - z \right) }p^+ \left( y - y_1^- \right) } - e^{ i x_L p^+ \left( y - y_1^- \right) } \right] \nonumber\\
& \quad\times  \left[ e^{ -i \frac{ x_D }{ \left( 1 - z \right) }p^+  y_2^- } - e^{ i x_L p^+ y_2^- } \right] .
\end{align}

In this work we consider the kinematics satisfies $ q^- \sim Q$, $\vec{ \ell }_\perp \sim \lambda Q$ and $\vec{ k }_\perp \sim \lambda Q$, where $Q$ is the scale of the hard process and $\lambda$ is a small number ($\lambda\ll1$).  These can be used to show that $\ell^- = \left( 1 - z \right) q^- \sim Q$, and $\ell^+ \sim \lambda^3 Q$. In the end, the trace part is simplified by using the trace identities and Dirac matrix identities with power counting. By expanding to the leading order and neglecting terms suppressed by higher powers of $\lambda$, we obtain the numerator of the last 3 lines in Eq.~\eqref{Eqn:W2_general_form} as, 
%
%
\begin{eqnarray}
\label{Eqn:Trace}
\mathrm{Tr}\bigg[\ldots\bigg] & = & \Tr \left[ p . \gamma \gamma^\mu \left( q + xp \right) . \gamma \gamma^\nu \right] 8 \left( q^- \right)^2 \nonumber\\
& \times & \frac{ 1 + z^2 }{ z } \left( \vec{ \ell }_\perp - \vec{ k }_\perp \right)^2 .
\end{eqnarray}

By substituting Eqs.~\eqref{Eqn:ContInt} and~\eqref{Eqn:Trace} into Eq.~\eqref{Eqn:DoubleggScat1_W}, one obtains the hadronic tensor as
%
%
\begin{eqnarray}
\label{Eqn:DoubleggScat1_WFinal}
\mathcal{W}^{\mu\nu}_{\ref{fig:DoubleggScat1}} & = & \frac{1}{2\pi} \int \frac{dy^-}{2\pi} dy_1^- dy_2^- \frac{ d^2y_\perp }{ \left( 2\pi \right)^2 } 
d^2k_\perp e^{ i \vec{ k }_\perp \cdot \vec{ y }_\perp } \int dz \int dx \nonumber\\
& \times &  \left( 2\pi \right) \delta \left[ \left( q + xp \right)^2 \right]  \frac{ e_q^2 }{2} 
\Tr \left[ p . \gamma \gamma^\mu \left( q + xp \right) . \gamma \gamma^\nu \right]  \frac{ 1 }{ 2 } \nonumber\\
& \times & \left\langle A \middle| \bar{ \psi } \left( y^- \right) \gamma^+ A^+ \left( y_1^-, \vec{ y }_\perp \right) 
 A^+ \left( y_2^-, \vec{ 0 }_\perp \right) \psi \left( 0 \right) \middle| A \right\rangle  \nonumber\\
& \times & \int \frac{ d \ell_\perp^2 }{ \left( \vec{ \ell }_\perp - \vec{ k }_\perp \right)^2 } \frac{ \alpha_s }{ \left( 2\pi \right) } 
C_A \frac{ 1 + z^2 }{ 1 - z } \frac{ 2 \pi \alpha_s }{ N_C } \theta \left( y_2^- \right)  \nonumber\\
& \times &  \theta \left( y_1^- - y^- \right) e^{ -i \left( x_B + x_L \right) p^+ y^- - i  x_D p^+ \left( y_1^- - y_2^- \right)} \nonumber\\
& \times & \left[ e^{ -i \frac{ x_D }{ \left( 1 - z \right) }p^+ \left( y - y_1^- \right) } - e^{ i x_L p^+ \left( y - y_1^- \right) } \right] \nonumber\\
& \times & \left[ e^{ -i \frac{ x_D }{ \left( 1 - z \right) }p^+  y_2^- } - e^{ i x_L p^+ y_2^- } \right] .
\end{eqnarray}

By comparing the above equation to Eqs.~\eqref{Eqn:HadTensor_General}$\sim$\eqref{Eqn:H0}, one can extract the $\vec{ k }_\perp$-dependent part of the partonic hard part as, 
%
%
\begin{eqnarray}
H_{\ref{fig:DoubleggScat1}} & = & \frac{ C_A } { \left( l_\perp - k_\perp \right)^2 } 
\theta \left( y_2^- \right) \theta \left( y_1^- - y^- \right). \nonumber \\
& \times &  \left[ e^{ -i \frac{ x_D }{ \left( 1 - z \right) }p^+ \left( y - y_1^- \right) } - e^{ i x_L p^+ \left( y - y_1^- \right) } \right] \nonumber \\
& \times & \left[ e^{ -i \frac{ x_D }{ \left( 1 - z \right) }p^+  y_2^- } - e^{ i x_L p^+ y_2^- } \right].
\label{Eqn:DoubleggScat1_H}
\end{eqnarray}
The double derivative of this $\vec{ k }_\perp$-dependent part at $k_\perp = 0$ then reads
%
%
\begin{align}
\label{Eqn:DoubleggScat1_d2H}
\nabla_{ k_\perp }^2 H_{\ref{fig:DoubleggScat1}} & \Big|_{ k_\perp = 0 }  =  \frac{ 4 C_A }{ \ell_\perp^4 } \left[ 2 
 - 2 \cos \left( \frac{ \ell_\perp^2 \xi^- }{ 2 q^- z \left( 1 - z \right) } \right) \right. \nonumber\\ 
& -  \left. 2 \left( \frac{ \ell_\perp^2 \xi^- }{ 2 q^- z \left( 1 - z \right) } \right) \sin \left( \frac{ \ell_\perp^2 \xi^- }{ 2 q^- z \left( 1 - z \right) } \right) \right. \\
& +  2  \left. \left( \frac{ \ell_\perp^2 \xi^- }{ 2 q^- z \left( 1 - z \right) } \right)^2 \cos \left( \frac{ \ell_\perp^2 \xi^- }{ 2 q^- z \left( 1 - z \right) } \right) \right]. \nonumber
\end{align}
Here $y = 0$ and $y_1^- = y_2^- = \xi^-$ are used as discussed earlier. The result we obtained here for diagram Fig.~\ref{fig:DoubleggScat1} is identical to the result by AZZ~\cite{Aurenche:2008hm, Aurenche:2008mq}.  However, as other contributing diagrams are included in the rest of this work, our final result will deviate from AZZ. 

%
%
\begin{figure}[tbp]
    \addtolength{\abovecaptionskip}{-3mm}
%
	\centering
	\includegraphics[width=\linewidth]{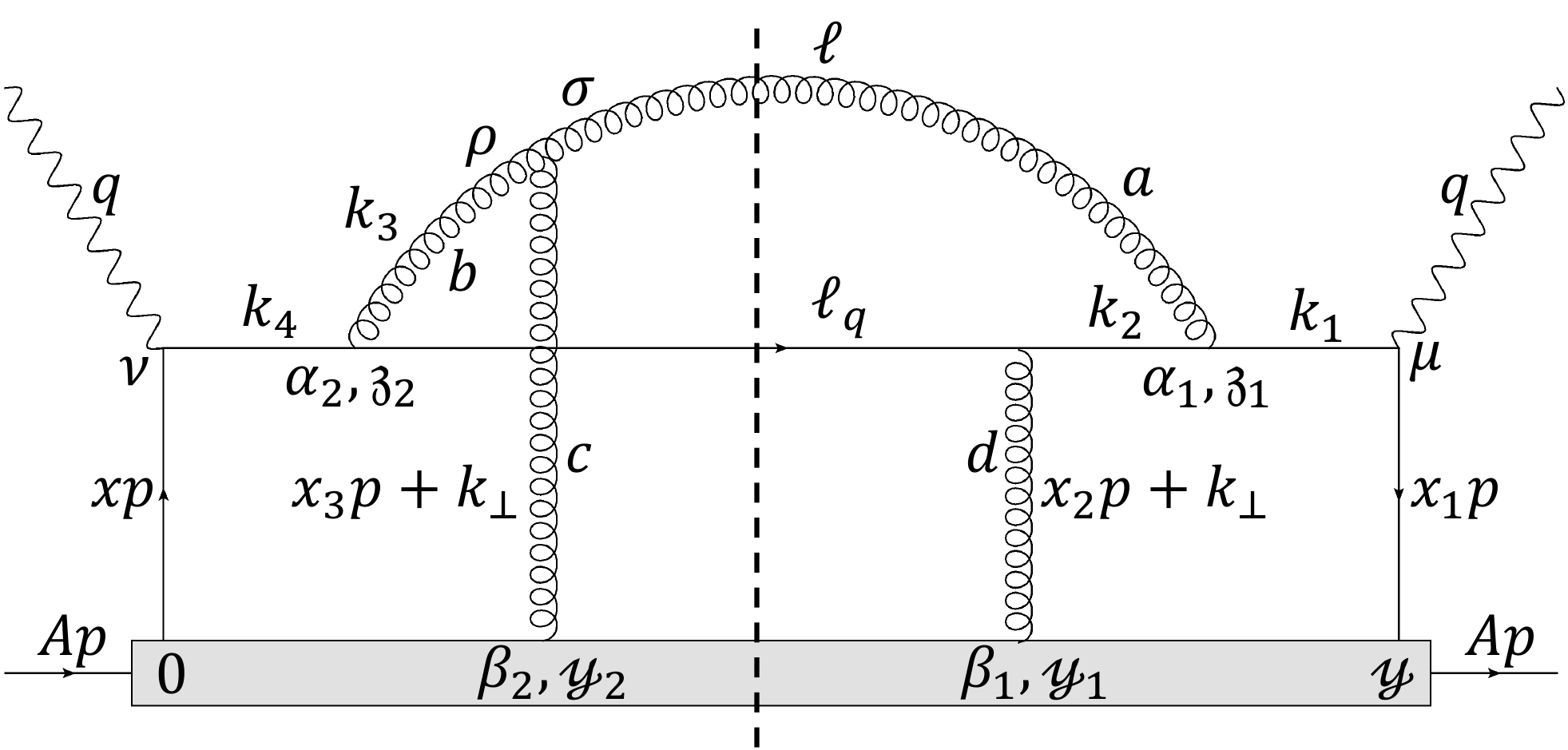}
	\caption{Feynman diagram for the first of two possible quark-gluon and gluon-gluon rescattering process.}
	\label{fig:qgAndggScat1}
\end{figure}

\begin{figure}[tbp]
    \addtolength{\abovecaptionskip}{-3mm}
%
	\centering
	\includegraphics[width=\linewidth]{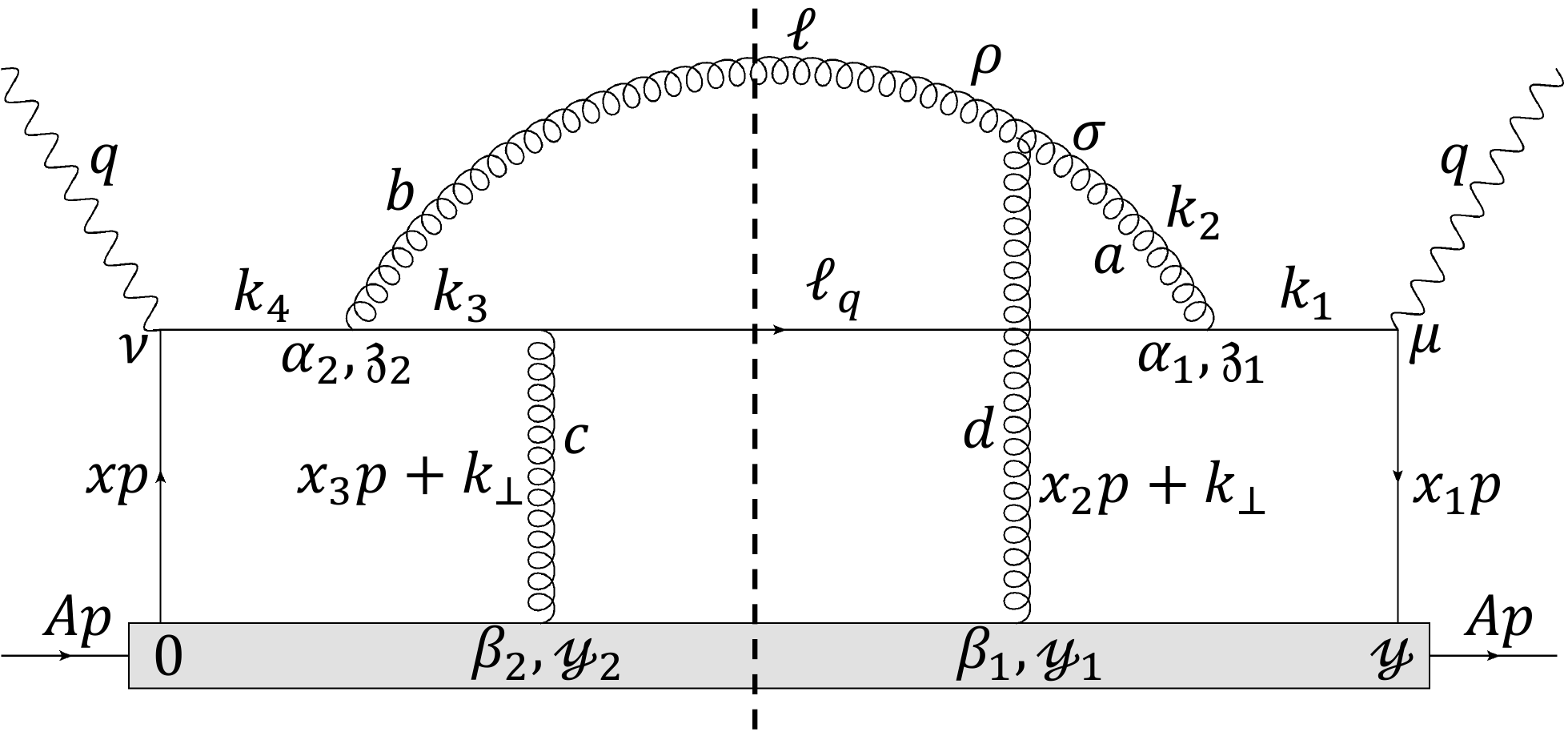}
	\caption{Feynman diagram for the second of two possible quark-gluon and gluon-gluon rescattering process.}
	\label{fig:qgAndggScat2}
\end{figure}

There are two diagrams with quark-gluon and gluon-gluon rescattering, as shown in Figs.~\ref{fig:qgAndggScat1} and~\ref{fig:qgAndggScat2}. Following similar steps used for the gluon-gluon double rescattering calculation above, one may first obtain the hadronic tensor for Fig.~\ref{fig:qgAndggScat1} as follows,
%
%
\begin{eqnarray}
\label{Eqn: qgAndggScat1_W}
\mathcal{W}^{\mu\nu}_{\ref{fig:qgAndggScat1}} & = & - \frac{1}{2\pi} \int \frac{dy^-}{2\pi} dy_1^- dy_2^- \frac{ d^2y_\perp }{ \left( 2\pi \right)^2 } 
d^2k_\perp \int dz \int dx \nonumber\\
& \times &  \left( 2\pi \right) \delta \left[ \left( q + xp \right)^2 \right]  \frac{ e_q^2 }{2} 
\Tr \left[ p . \gamma \gamma^\mu \left( q + xp \right) . \gamma \gamma^\nu \right]  \frac{ 1 }{ 2 } \nonumber\\
& \times & \left\langle A \middle| \bar{ \psi } \left( y^- \right) \gamma^+ A^+ \left( y_1^-, \vec{ y }_\perp \right) 
 A^+ \left( y_2^-, \vec{ 0 }_\perp \right) \psi \left( 0 \right) \middle| A \right\rangle  \nonumber\\
& \times & e^{ i \vec{ k }_\perp \cdot \vec{ y }_\perp } \int d \ell_\perp^2 \frac{ \vec{ \ell }_\perp \cdot \left( \vec{ \ell }_\perp - \vec{ k }_\perp \right) }{ \ell_\perp^2 \left( \vec{ \ell }_\perp - \vec{ k }_\perp \right)^2 } 
\frac{ \alpha_s }{ \left( 2\pi \right) } \frac{C_A}{2} \frac{ 1 + z^2 }{ 1 - z } \nonumber\\
& \times &  \frac{ 2 \pi \alpha_s }{ N_C }  e^{ -i \left( x_B + x_L \right) p^+ y^- - i  x_D p^+ \left( y_1^- - y_2^- \right)} \nonumber\\
& \times & \theta \left( y_1^- - y^- \right) \theta \left( y_2^- \right) \left[ 1 - e^{ i x_L p^+ \left( y - y_1^- \right) } \right] \nonumber\\
& \times & \left[ e^{ -i \frac{ x_D }{ \left( 1 - z \right) }p^+  y_2^- } - e^{ i x_L p^+ y_2^- } \right] .
\end{eqnarray}
The corresponding $\vec{k}_\perp$-dependent part of the partonic hard part can be then obtained as,
%
%
\begin{eqnarray}
\label{Eqn:qgAndggScat1_H}
H_{\ref{fig:qgAndggScat1}} & = & - \frac{C_A}{2} \frac{ \vec{ \ell }_\perp \cdot \left( \vec{ \ell }_\perp - \vec{ k }_\perp \right) }{ \ell_\perp^2 \left( \vec{ \ell }_\perp - \vec{ k }_\perp \right)^2 } \theta \left( y_2^- \right) \theta \left( y_1^- - y^- \right)  \nonumber\\
& \times & \left[ 1 - e^{ i x_L p^+ \left( y - y_1^- \right) } \right] \nonumber\\
& \times & \left[ e^{ -i \frac{ x_D }{ \left( 1 - z \right) }p^+  y_2^- } - e^{ i x_L p^+ y_2^- } \right].
   \end{eqnarray}

The collinear expansion of the above hard part can be evaluated right away. However, the expansion can be more convenient if it is implemented when results of the Figs.~\ref{fig:qgAndggScat1} and~\ref{fig:qgAndggScat2} are combined together. Note that these two diagrams are complex conjugate of each other, thus closely related.

The hadronic tensor for Fig.~\ref{fig:qgAndggScat2} reads,
\begin{eqnarray}
\label{Eqn:qgAndggScat2_W}
\mathcal{W}^{\mu\nu}_{\ref{fig:qgAndggScat2}} & = & - \frac{1}{2\pi} \int \frac{dy^-}{2\pi} dy_1^- dy_2^- \frac{ d^2y_\perp }{ \left( 2\pi \right)^2 } 
d^2k_\perp \int dz \int dx \nonumber\\
& \times &  \left( 2\pi \right) \delta \left[ \left( q + xp \right)^2 \right]  \frac{ e_q^2 }{2} 
\Tr \left[ p . \gamma \gamma^\mu \left( q + xp \right) . \gamma \gamma^\nu \right]  \frac{ 1 }{ 2 } \nonumber\\
& \times & \left\langle A \middle| \bar{ \psi } \left( y^- \right) \gamma^+ A^+ \left( y_1^-, \vec{y}_\perp \right) 
 A^+ \left( y_2^-, \vec{0}_\perp \right) \psi \left( 0 \right) \middle| A \right\rangle  \nonumber\\
& \times & e^{ i \vec{ k }_\perp \cdot \vec{ y }_\perp } \int d \ell_\perp^2 \frac{ \vec{ \ell }_\perp \cdot \left( \vec{ \ell }_\perp - \vec{ k }_\perp \right) }{ \ell_\perp^2 \left( \vec{ \ell }_\perp - \vec{ k }_\perp \right)^2 }  
\frac{ \alpha_s }{ \left( 2\pi \right) } \frac{ C_A }{ 2 } \frac{ 1 + z^2 }{ 1 - z }  \nonumber\\
& \times &  \frac{ 2 \pi \alpha_s }{ N_C } e^{ -i \left( x_B + x_L \right) p^+ y^- - i  x_D p^+ \left( y_1^- - y_2^- \right)}  \theta \left( y_1^- - y^- \right) \nonumber\\
& \times & \theta \left( y_2^- \right) \left[ e^{ -i \frac{ x_D }{ \left( 1 - z \right) }p^+ \left( y - y_1^- \right) } - e^{ i x_L p^+ \left( y - y_1^- \right) } \right] \nonumber\\
& \times & \left[ 1 - e^{ i x_L p^+ y_2^- } \right],
\end{eqnarray}
with its corresponding $\vec{k}_\perp$-dependent part of the partonic hard part as,
%
%
\begin{eqnarray}
\label{Eqn:qgAndggScat2_H}
H_{\ref{fig:qgAndggScat2}} & = & - \frac{ C_A }{ 2 } \frac{ \vec{ \ell }_\perp \cdot \left( \vec{ \ell }_\perp - \vec{ k }_\perp \right) }{ \ell_\perp^2 \left( \vec{ \ell }_\perp - \vec{ k }_\perp \right)^2 } 
\theta \left( y_2^- \right) \theta \left( y_1^- - y^- \right) \nonumber\\
 & \times & \left[ 1 - e^{ i x_L p^+ y_2^- } \right] \nonumber\\
 & \times & \left[ e^{ -i \frac{ x_D }{ \left( 1 - z \right) }p^+ \left( y - y_1^- \right) } - e^{ i x_L p^+ \left( y - y_1^- \right) } \right]. 
\end{eqnarray}

By taking $y = 0$ and $y_1^- = y_2^- = \xi^-$, we obtain the combination Eqs.~\eqref{Eqn:qgAndggScat1_H} and~\eqref{Eqn:qgAndggScat2_H} as
%
\begin{eqnarray}
\label{Eqn:qgAndggScat1plus2_H}
H_{\ref{fig:qgAndggScat1},\ref{fig:qgAndggScat2}} & = & \frac{ C_A }{ 2 } \frac{ \vec{ \ell }_\perp \cdot \left( \vec{ \ell }_\perp - \vec{ k }_\perp \right) }{ \ell_\perp^2 \left( \vec{ \ell }_\perp - \vec{ k }_\perp \right)^2 } 
\left[ 2 \cos \left( \frac{ \xi^- \left( \vec{ \ell }_\perp - \vec{ k }_\perp \right)^2 }{ 2 q^- z \left( 1 - z \right) } \right) \right.\nonumber\\
& - & \left. 2 - 2 \cos \left( \frac{ \xi^- \left( k_\perp^2 - 2 \vec{ k }_\perp \cdot \vec{ \ell }_\perp \right) }{ 2 q^- z \left( 1 - z \right) } \right) \right. \nonumber\\
& + & \left. 2 \cos \left( \frac{ \xi^- \ell_\perp^2 }{ 2 q^- z \left( 1 - z \right) } \right) \right] .
\end{eqnarray}
The double derivative of this combined term is then
%
%
\begin{align}
\label{Eqn:qgAndggScat1plus2_d2H}
\nabla_{ k_\perp }^2 \left( \mathcal{ H }_{ \ref{fig:qgAndggScat1}, \ref{fig:qgAndggScat2} } \right)& \Big|_{ k_\perp = 0 } = 2 C_A 
	\left( \frac{ \xi^- }{ 2 q^- z \left( 1 - z \right) } \right)^2 \nonumber\\
	& \times \left[ 2 - 2 \cos  \left( \frac{ \xi^- \ell_{\perp}^2 }{ 2 q^- z \left( 1 - z \right) } \right) \right] .
\end{align}

%
\begin{figure}[tbp]
    \addtolength{\abovecaptionskip}{-3mm}
%
	\centering
	\includegraphics[width=\linewidth]{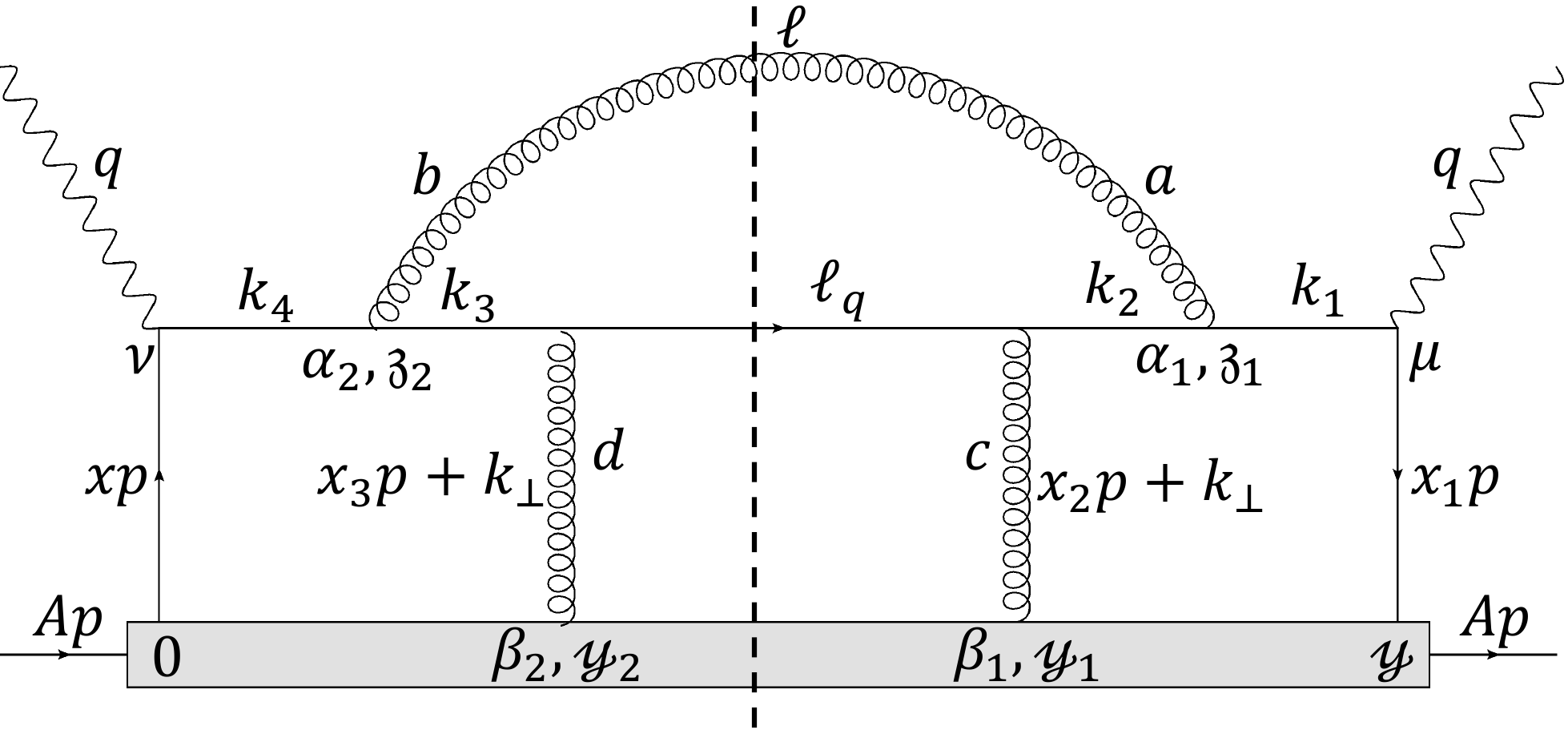}
	\caption{Feynman diagram for the double quark-gluon rescattering process.}
	\label{fig:qgDoubleScat1}
\end{figure}

The last diagram for this subsection is for double quark-gluon rescattering, as shown in Fig.~\ref{fig:qgDoubleScat1}. Its hadronic tensor reads
%
%
\begin{eqnarray}
\label{Eqn:qgDoubleScat1_W}
\mathcal{W}^{\mu\nu}_{\ref{fig:qgDoubleScat1}} & = & \frac{1}{2\pi} \int \frac{dy^-}{2\pi} dy_1^- dy_2^- \frac{ d^2y_\perp }{ \left( 2\pi \right)^2 } 
d^2k_\perp e^{ i \vec{ k }_\perp \cdot \vec{ y }_\perp } \int dz \int dx \nonumber\\
& \times &  \left( 2\pi \right) \delta \left[ \left( q + xp \right)^2 \right]  \frac{ e_q^2 }{2} 
\Tr \left[ p . \gamma \gamma^\mu \left( q + xp \right) . \gamma \gamma^\nu \right]  \frac{ 1 }{ 2 } \nonumber\\
& \times & \left\langle A \middle| \bar{ \psi } \left( y^- \right) \gamma^+ A^+ \left( y_1^-, \vec{ y }_\perp \right) 
 A^+ \left( y_2^-, \vec{ 0 }_\perp \right) \psi \left( 0 \right) \middle| A \right\rangle  \nonumber\\
& \times & \int \frac{ d \ell_\perp^2 }{ \left( \vec{ \ell }_\perp - \vec{ k }_\perp \right)^2 } \frac{ \alpha_s }{ \left( 2\pi \right) } 
C_A \frac{ 1 + z^2 }{ 1 - z } \frac{ 2 \pi \alpha_s }{ N_C } \theta \left( y_2^- \right)  \nonumber\\
& \times &  \theta \left( y_1^- - y^- \right) e^{ -i \left( x_B + x_L \right) p^+ y^- - i  x_D p^+ \left( y_1^- - y_2^- \right)} \nonumber\\
& \times & \left[ 1 - e^{ i x_L p^+ \left( y - y_1^- \right) } \right] \left[ 1 - e^{ i x_L p^+ y_2^- } \right],
\end{eqnarray}
and the corresponding $\vec{k}_\perp$-dependent part of the partonic hard part is
%
%
\begin{eqnarray}
\label{Eqn:qgDoubleScat1_H}
H_{\ref{fig:qgDoubleScat1}} & = & \frac{ C_F }{  \ell_\perp^2 } \left[ 1 - e^{ i x_L p^+ \left( y - y_1^- \right) } \right] \left[ 1 - e^{ i x_L p^+ y_2^- } \right] \nonumber\\
& \times & \theta \left( y_2^- \right) \theta \left( y_1^- - y^- \right) .
\end{eqnarray}

As there is no $\vec{k}_\perp$ dependence in this hard part, the derivative with respect to $\vec{k}_\perp$ is zero. Therefore, there is no contribution from this diagram to the medium modification kernel in the end.
%
%
\begin{eqnarray}
\label{Eqn:qgDoubleScat1_d2H}
\nabla_{ k_\perp }^2 H_{\ref{fig:qgDoubleScat1}}  \Big|_{ k_\perp = 0 } & = & 0 .
\end{eqnarray}

Figures~\ref{fig:DoubleggScat1}$-$\ref{fig:qgAndggScat2} are the only three contributing diagrams for the single scattering single emission DIS when scattering happens after emission. Here we only present calculations for central cut diagrams. Diagrams with left and right cuts will be discussed in Sect.~\ref{subsec:level3.3} and Appendix~\ref{sec:appendix1}. However, we will see that only two non-central cut diagrams have contribution to our final result at the next-to-leading twist.

\subsection{\label{subsec:level3.2} Central cut diagrams including pre-emission scattering }

In the previous subsection, we concentrated on diagrams in which both rescatterings happen after the gluon emission. In this subsection, we investigate contributions from central cut diagrams in which one or both rescatterings occur before the gluon emission. If one ignores the $\vec{k}_\perp$-dependent phase factors in the hadronic tensor as the GW work did, one would find these diagrams including pre-emission scattering are suppressed by the momentum fraction of the emitted gluon compared to those post-emission scattering diagrams. 
As a result, in the limit of small gluon momentum (or soft gluon approximation), these pre-emission contributions were considered negligible. However, in this subsection, we will show that after taking into account the contributions from the $\vec{k}_\perp$-dependent phase factors before the collinear expansion, diagrams with pre-emission scattering will no longer be entirely suppressed by the emitted gluon momentum fraction. Therefore, these diagrams cannot be neglected even with the soft gluon approximation. This is also the main source for our results differing from GW and AZZ kernels even at small path length ($\xi^-/\tau_F$) in the end.

\begin{figure}[tbp]
    \addtolength{\abovecaptionskip}{-3mm}
%
	\centering
	\includegraphics[width=\linewidth]{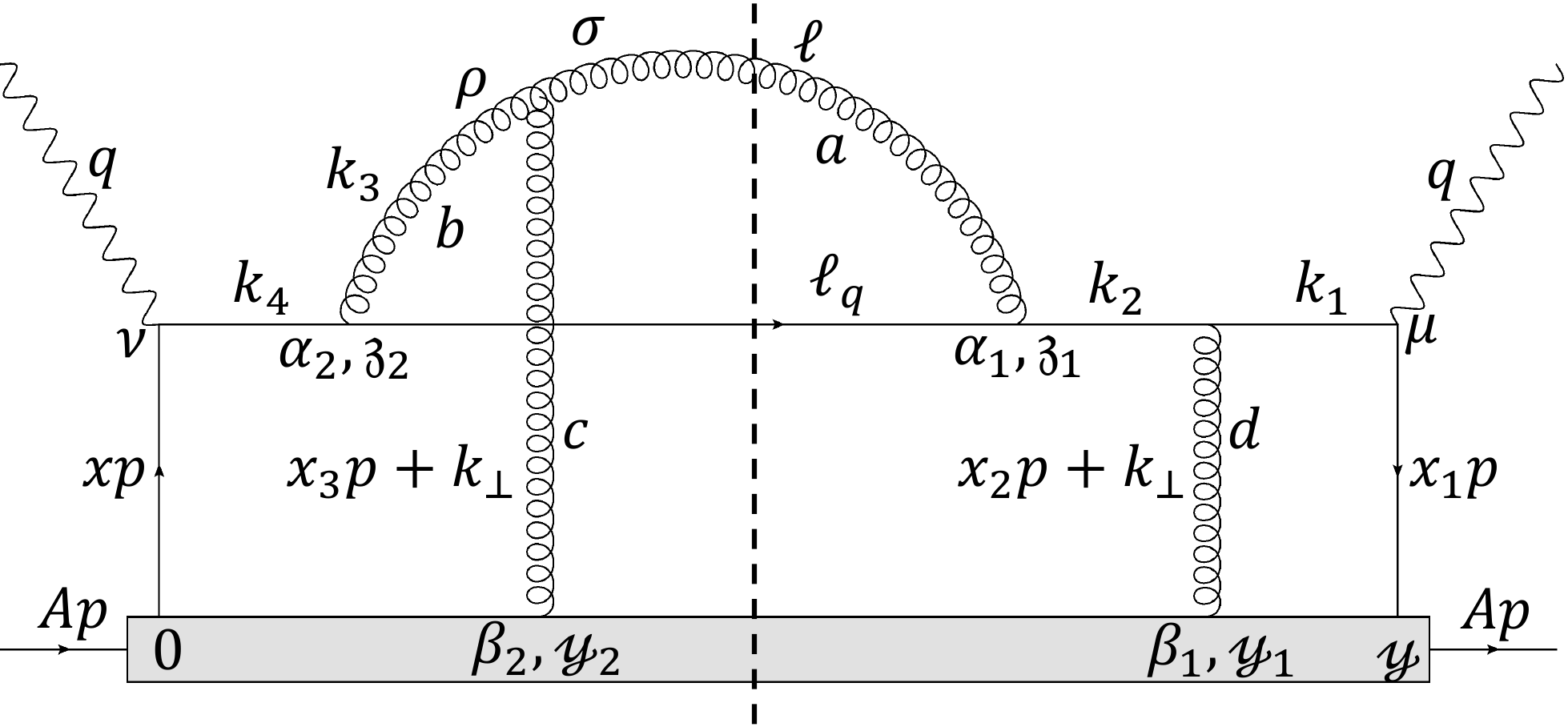}
	\caption{Feynman diagram for the first of two possible quark-gluon and gluon-gluon rescattering process with scattering on quark happens before the emission.}
	\label{fig:qgAndggScat3}
\end{figure}

\begin{figure}[tbp]
    \addtolength{\abovecaptionskip}{-3mm}
%
	\centering
	\includegraphics[width=\linewidth]{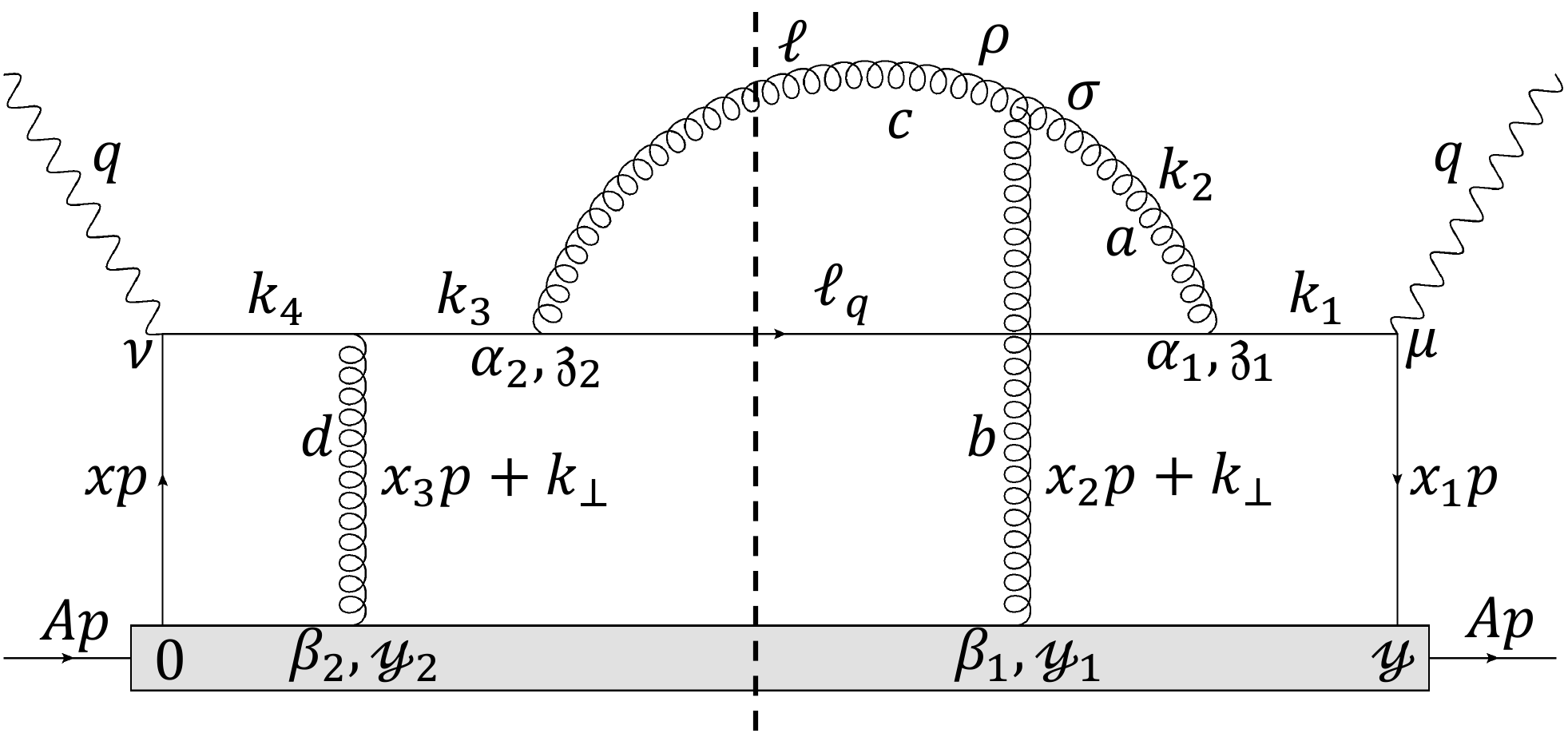}
	\caption{Feynman diagram for the second of two possible quark-gluon and gluon-gluon rescattering process with scattering on quark happens before the emission.}
	\label{fig:qgAndggScat4}
\end{figure}

There are a total of five central cut diagrams that include at least one secondary scattering happening before the gluon emission. 
There are two diagrams, as shown in Figs.~\ref{fig:qgAndggScat3} and~\ref{fig:qgAndggScat4}, corresponding to quark-gluon+gluon-gluon rescattering process with the quark-gluon rescattering occuring before the gluon emission. They will be shown to be the only two contributing diagrams among the five. Following the routines constructed in the previous subsection, one may first express the hadronic tensor for Fig.~\ref{fig:qgAndggScat3} as follows,
%
%
%
%
\begin{eqnarray}
\label{Eqn:qgAndggScat3_W}
\mathcal{W}^{\mu\nu}_{\ref{fig:qgAndggScat3}}  & = & - \frac{1}{2\pi} \int \frac{dy^-}{2\pi} dy_1^- dy_2^- \frac{ d^2y_\perp }{ \left( 2\pi \right)^2 } 
d^2k_\perp  \int dz \int dx \nonumber\\
& \times & \left( 2\pi \right) \delta \left[ \left( q + xp \right)^2 \right] \frac{ e_q^2 }{2} 
\Tr \left[ p . \gamma \gamma^\mu \left( q + xp \right) . \gamma \gamma^\nu \right] \frac{ 1 }{ 2 } \nonumber\\
& \times & \left\langle A \middle| \bar{ \psi } \left( y^- \right) \gamma^+ A^+ \left( y_1^-, \vec{ y }_\perp \right) 
 A^+ \left( y_2^-, \vec{ 0 }_\perp \right) \psi \left( 0 \right) \middle| A \right\rangle  \nonumber\\
& \times & e^{ i \vec{ k }_\perp \cdot \vec{ y }_\perp } \int d \ell_\perp^2 \frac{ \left( \vec{ \ell }_\perp - \vec{ k }_\perp \right) \cdot \left( \vec{ \ell }_\perp - \left( 1 - z \right) \vec{ k }_\perp \right) }
{ \left( \vec{ \ell }_\perp - \vec{ k }_\perp \right)^2 \left( \vec{ \ell }_\perp - \left( 1 - z \right) \vec{ k }_\perp \right)^2 } \nonumber\\
& \times & \frac{ \alpha_s }{ \left( 2\pi \right) } \frac{ C_A }{ 2 } \frac{ 1 + z^2 }{ 1 - z } \frac{ 2 \pi \alpha_s }{ N_C } \theta \left( y_2^- \right) \theta \left( y_1^- - y^- \right) \nonumber\\
& \times & e^{ -i \left( x_B + x_L \right) p^+ y^- - i  x_D p^+ \left( y_1^- - y_2^- \right)} e^{ i x_L p^+ \left( y^- - y_1^- + y_2^- \right) } \nonumber\\
& \times & \left[ 1 - e^{ -i \left( x_L + \frac{ x_D }{ \left( 1 - z \right) } \right) p^+ y_2^- } \right] .
\end{eqnarray}
The corresponding $\vec{k}_\perp$-dependent part of the partonic hard part is then extracted as
%
%
\begin{eqnarray}
\label{Eqn:qgAndggScat3_H}
H_{\ref{fig:qgAndggScat3}} & = & - \frac{ C_A }{ 2 } \frac{ \left( \vec{ \ell }_\perp - \vec{ k }_\perp \right) . \left( \vec{ \ell }_\perp - \left( 1 - z \right) \vec{ k }_\perp \right) }
{ \left( \vec{ \ell }_\perp - \vec{ k }_\perp \right)^2 \left( \vec{ \ell }_\perp - \left( 1 - z \right) \vec{ k }_\perp \right)^2 } \nonumber\\
& \times & e^{ i x_L p^+ \left( y^- - y_1^- + y_2^- \right) } \left[ 1 - e^{ -i \left( x_L + \frac{ x_D }{ \left( 1 - z \right) } \right) p^+ y_2^- } \right] \nonumber\\
& \times & \theta \left( y_2^- \right) \theta \left( y_1^- - y^- \right) . 
\end{eqnarray}
Similar to our earlier calculation of Figs.~\ref{fig:qgAndggScat1} and~\ref{fig:qgAndggScat2}, collinear expansion of the above partonic hard part for Fig.~\ref{fig:qgAndggScat3} can be postponed till it is combined with result from Fig.~\ref{fig:qgAndggScat4} later, since they are complex conjugate of each other. 

%

The hadronic tensor of the diagram shown in the Fig.~\ref{fig:qgAndggScat4} is given by
%
%
\begin{eqnarray}
\label{Eqn:qgAndggScat4_W}
\mathcal{W}^{\mu\nu}_{\ref{fig:qgAndggScat4}} & = & - \frac{1}{2\pi} \int \frac{dy^-}{2\pi} dy_1^- dy_2^- \frac{ d^2y_\perp }{ \left( 2\pi \right)^2 } 
d^2k_\perp \int dz \int dx \nonumber\\
& \times & \left( 2\pi \right) \delta \left[ \left( q + xp \right)^2 \right] \frac{ e_q^2 }{2} 
\Tr \left[ p . \gamma \gamma^\mu \left( q + xp \right) . \gamma \gamma^\nu \right]  \frac{ 1 }{ 2 } \nonumber\\
& \times & \left\langle A \middle| \bar{ \psi } \left( y^- \right) \gamma^+ A^+ \left( y_1^-, \vec{ y }_\perp \right) 
 A^+ \left( y_2^-, \vec{ 0 }_\perp \right) \psi \left( 0 \right) \middle| A \right\rangle  \nonumber\\
& \times & e^{ i\vec{ k }_\perp \cdot \vec{ y }_\perp } \int d \ell_\perp^2 \frac{ \left( \vec{ \ell }_\perp - \vec{ k }_\perp \right) \cdot \left( \vec{ \ell }_\perp - \left( 1 - z \right) \vec{ k }_\perp \right) }
{ \left( \vec{ \ell }_\perp - \vec{ k }_\perp \right)^2 \left( \vec{ \ell }_\perp - \left( 1 - z \right) \vec{ k }_\perp \right)^2 } \nonumber\\
& \times & \frac{ \alpha_s }{ \left( 2\pi \right) } \frac{ C_A }{ 2 } \frac{ 1 + z^2 }{ 1 - z } \frac{ 2 \pi \alpha_s }{ N_C } \theta \left( y_2^- \right) \theta \left( y_1^- - y^- \right) \nonumber\\
& \times & e^{ -i x_B p^+ y^- - i  \left( x_L + x_D \right) p^+ \left( y_1^- - y_2^- \right)} e^{ i x_L p^+ \left( y^- - y_1^- + y_2^- \right) } \nonumber\\
& \times & \left[ 1 - e^{ -i \left( x_L + \frac{ x_D }{ \left( 1 - z \right) } \right) p^+ \left( y^- - y_1^- \right) } \right],
\end{eqnarray}
and its corresponding $\vec{k}_\perp$-dependent part of the partonic hard part is 
%
%
\begin{eqnarray}
\label{Eqn:qgAndggScat4_H}
H_{\ref{fig:qgAndggScat4}} & = & - \frac{ C_A }{ 2 } \frac{ \left( \vec{ \ell }_\perp - \vec{ k }_\perp \right) \cdot \left( \vec{ \ell }_\perp - \left( 1 - z \right) \vec{ k }_\perp \right) }
{ \left( \vec{ \ell }_\perp - \vec{ k }_\perp \right)^2 \left( \vec{ \ell }_\perp - \left( 1 - z \right) \vec{ k }_\perp \right)^2 } \nonumber\\
& \times & e^{ i x_L p^+ \left( y^- - y_1^- + y_2^- \right) }
\left[ 1 - e^{ -i \left( x_L + \frac{ x_D }{ \left( 1 - z \right) } \right) p^+ \left( y^- - y_1^- \right) } \right] \nonumber\\
& \times & \theta \left( y_2^- \right) \theta \left( y_1^- - y^- \right).
\end{eqnarray}
By combining Eqs.~\eqref{Eqn:qgAndggScat3_H} and~\eqref{Eqn:qgAndggScat4_H} and then taking the second derivative with respect to $\vec{k}_\perp$ at $\vec{k}_\perp = 0$, we have
%
%
\begin{align}
\label{Eqn:qgAndggScat4_d2H}
\nabla_{ k_\perp }^2 \mathcal{H}_{ \ref{fig:qgAndggScat3}, \ref{fig:qgAndggScat4} } &\Big |_{ k_\perp = 0} 
 = - \frac{ 4 C_A }{ \ell_\perp^4 } \bigg\{ \left( 1 - z \right)  \\
& \times \left[ 1 - \cos \left( \frac{ \ell_\perp^2 \xi^- }{ 2 q^- z \left( 1 - z \right) } \right) \right. \nonumber\\
& - \left. \left( \frac{ \ell_\perp^2 \xi^- }{ 2 q^- z \left( 1 - z \right) } \right) 
\sin \left( \frac{ \ell_\perp^2 \xi^- }{ 2 q^- z \left( 1 - z \right) } \right) \right] \nonumber\\
& + \left( \frac{ \ell_\perp^2 \xi^- }{ 2 q^- z \left( 1 - z \right) } \right)^2 
\cos \left( \frac{ \ell_\perp^2 \xi^- }{ 2 q^- z \left( 1 - z \right) } \right) \bigg\}, \nonumber
\end{align}
in which $y = 0$ and $y_1^- = y_2^- = \xi^-$ have been taken as before.

Note that in the above equation, the first part inside $\{ \ldots \}$ is suppressed by the gluon momentum fraction $(1-z)$, while the second part is not. The latter makes contribution from such diagrams with pre-emission scattering non-negligible even in the soft gluon limit. This is one of the main findings of this work after taking into account the contribution from the $\vec{k}_\perp$-dependent phase factors in the hadronic tensor.

\begin{figure}[tbp]
    \addtolength{\abovecaptionskip}{-3mm}
%
	\centering
	\includegraphics[width=\linewidth]{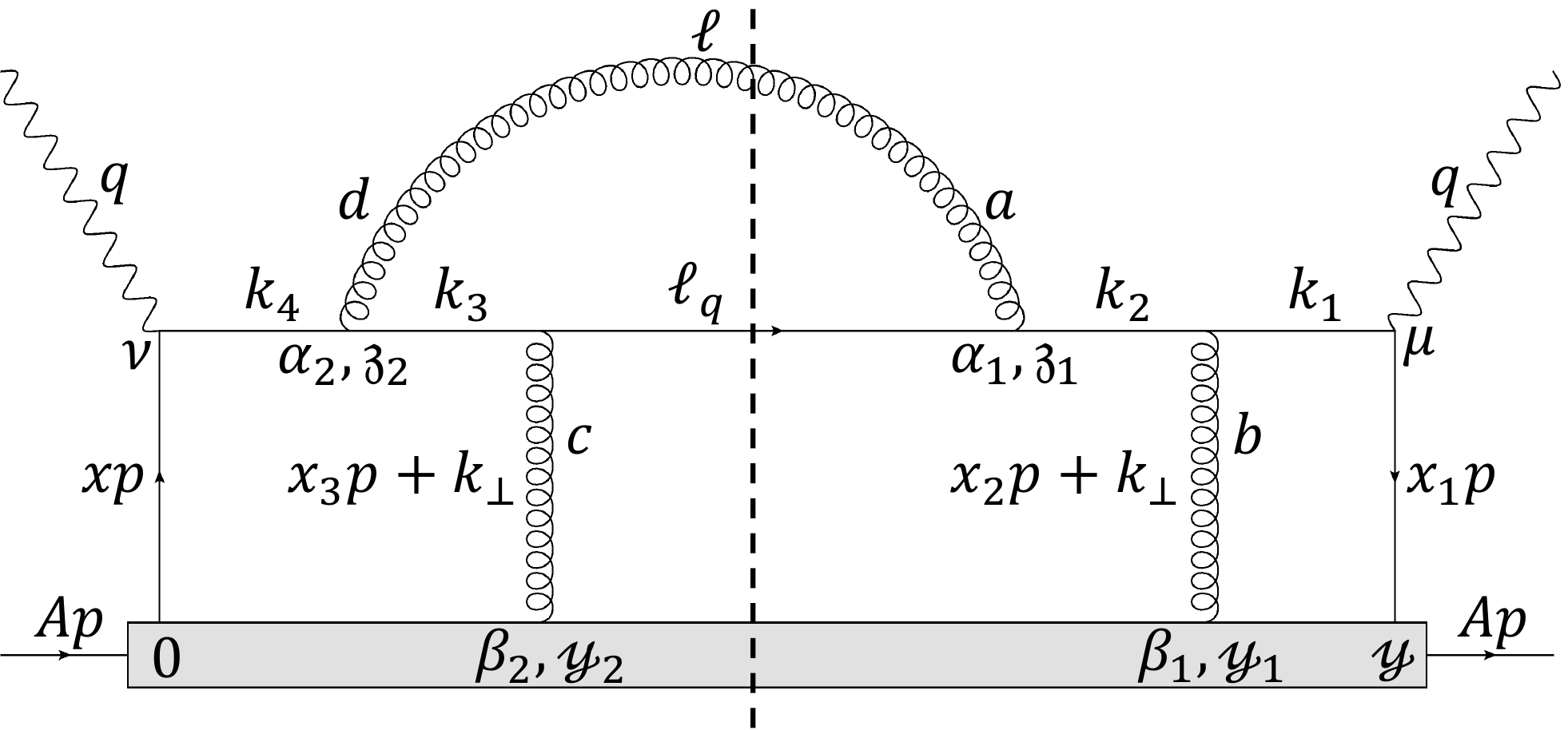}
	\caption{Feynman diagram for the first of two possible double quark-gluon rescattering process with only one scattering happens before emission.}
	\label{fig:qgDoubleScat2}
\end{figure}

\begin{figure}[tbp]
    \addtolength{\abovecaptionskip}{-3mm}
%
	\centering
	\includegraphics[width=\linewidth]{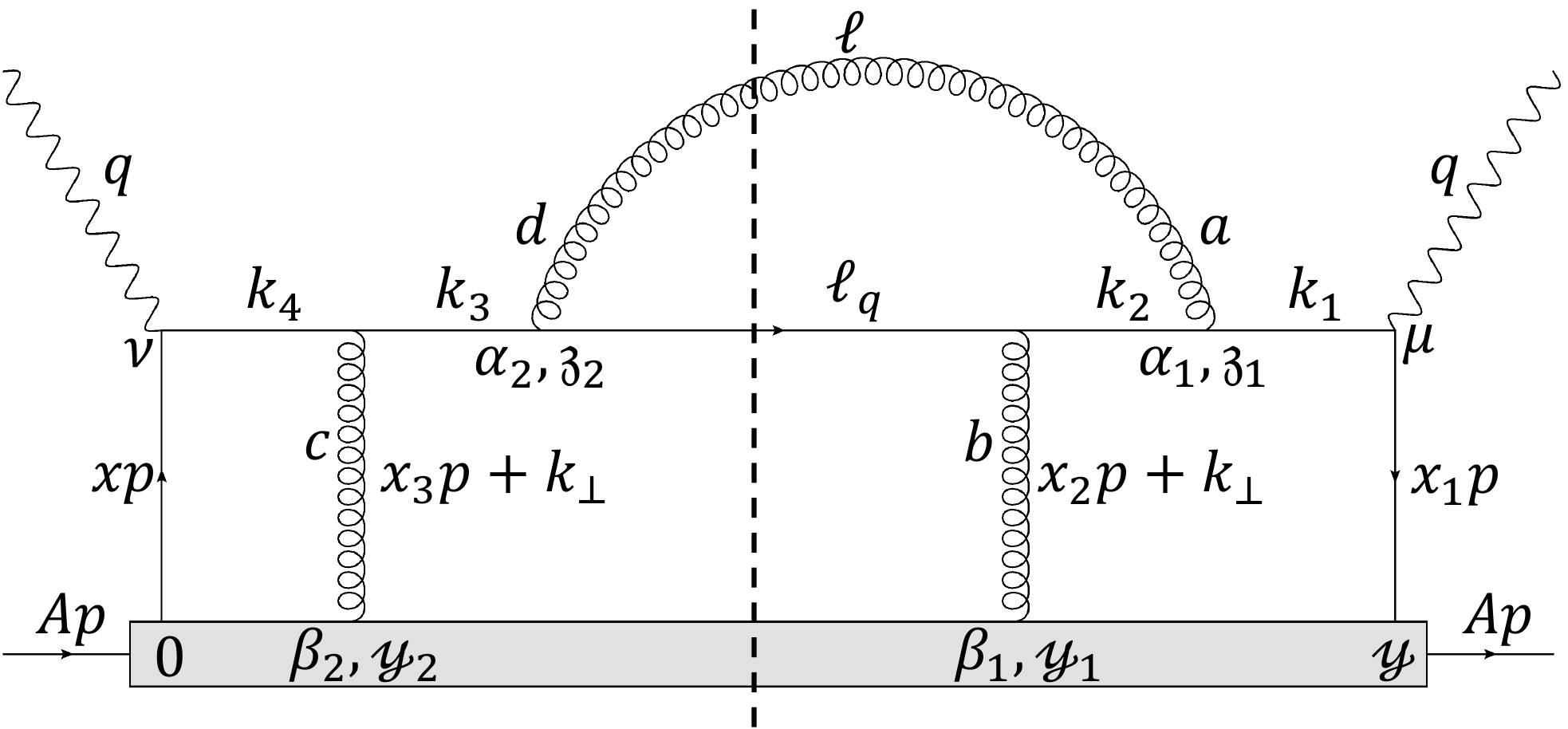}
	\caption{Feynman diagram for the second of two possible double quark-gluon rescattering process with only one scattering happens before emission.}
	\label{fig:qgDoubleScat3}
\end{figure}

There are three quark-gluon double rescattering diagrams, two with one scattering (Figs.~\ref{fig:qgDoubleScat2} and~\ref{fig:qgDoubleScat3}), one with both scatterings (Fig.~\ref{fig:qgDoubleScat4}), occurring before the gluon emission. However, as will be shown below, none of them contributes to the final result of our medium modification kernel at the next-to-leading twist. 

%
%

First, the hadronic tensor corresponding to Fig.~\ref{fig:qgDoubleScat2} can be written as
\begin{eqnarray}
\label{Eqn:qgDoubleScat2_W}
\mathcal{W}^{\mu\nu}_{\ref{fig:qgDoubleScat2}} & = & - \frac{1}{2\pi} \int \frac{dy^-}{2\pi} dy_1^- dy_2^- \frac{ d^2y_\perp }{ \left( 2\pi \right)^2 } 
d^2k_\perp e^{ i\vec{ k }_\perp \vec{ y }_\perp } \int dz \int dx \nonumber\\
& \times & \left( 2\pi \right) \delta \left[ \left( q + xp \right)^2 \right] \frac{ e_q^2 }{2} 
\Tr \left[ p . \gamma \gamma^\mu \left( q + xp \right) . \gamma \gamma^\nu \right]  \frac{ 1 }{ 2 } \nonumber\\
& \times & \left\langle A \middle| \bar{ \psi } \left( y^- \right) \gamma^+ A^+ \left( y_1^-, \vec{ y }_\perp \right) 
 A^+ \left( y_2^-, \vec{ 0 }_\perp \right) \psi \left( 0 \right) \middle| A \right\rangle  \nonumber\\
& \times & \int d \ell_\perp^2 \frac{ \vec{ \ell }_\perp \cdot \left( \vec{ \ell }_\perp - \left( 1 - z \right) \vec{ k }_\perp \right) }
{ \ell_\perp^2 \left( \vec{ \ell }_\perp - \left( 1 - z \right) \vec{ k }_\perp \right)^2 } \frac{ \alpha_s }{ \left( 2\pi \right) } C_F  \frac{ 1 + z^2 }{ 1 - z } \nonumber\\
& \times & \frac{ 2 \pi \alpha_s }{ N_C } e^{ -i x_B p^+ y^- - i  \left( x_L + x_D \right) p^+ \left( y_1^- - y_2^- \right)} \theta \left( y_2^- \right) \nonumber\\
& \times & \theta \left( y_1^- - y^- \right)  e^{ i x_L p^+ \left( y^- - y_1^- \right) } \left[ 1 - e^{ i x_L p^+ y_2^-} \right],
\end{eqnarray}
from which the $\vec{k}_\perp$-dependent part of the partonic hard part can be extracted as
%
%
\begin{eqnarray}
\label{Eqn:qgDoubleScat2_H}
H_{\ref{fig:qgDoubleScat2}} & = & C_F \frac{ \vec{ \ell }_\perp \cdot \left( \vec{ \ell }_\perp - \left( 1 - z \right) \vec{ k }_\perp \right) }
{ \ell_\perp^2 \left( \vec{ \ell }_\perp - \left( 1 - z \right) \vec{ k }_\perp \right)^2 } e^{ i x_L p^+ \left( y^- - y_1^- \right) } \nonumber\\
& \times & \left[ 1 - e^{ -i x_L p^+  y_2^- } \right] \theta \left( y_2^- \right) \theta \left( y_1^- - y^- \right).
\end{eqnarray}
%
%

%

Similarly, the hadronic tensor associated with Fig.~\ref{fig:qgDoubleScat3} can be expressed as
\begin{eqnarray}
\label{Eqn:qgDoubleScat3_W}
\mathcal{W}^{\mu\nu}_{\ref{fig:qgDoubleScat3}} & = & - \frac{1}{2\pi} \int \frac{dy^-}{2\pi} dy_1^- dy_2^- \frac{ d^2y_\perp }{ \left( 2\pi \right)^2 } 
d^2k_\perp e^{ i\vec{ k }_\perp \vec{ y }_\perp } \int dz \int dx \nonumber\\
& \times & \left( 2\pi \right) \delta \left[ \left( q + xp \right)^2 \right] \frac{ e_q^2 }{2} 
\Tr \left[ p . \gamma \gamma^\mu \left( q + xp \right) . \gamma \gamma^\nu \right]  \frac{ 1 }{ 2 } \nonumber\\
& \times & \left\langle A \middle| \bar{ \psi } \left( y^- \right) \gamma^+ A^+ \left( y_1^-, \vec{ y }_\perp \right) 
 A^+ \left( y_2^-, \vec{ 0 }_\perp \right) \psi \left( 0 \right) \middle| A \right\rangle  \nonumber\\
& \times & \int d \ell_\perp^2 \frac{ \vec{ \ell }_\perp \cdot \left( \vec{ \ell }_\perp - \left( 1 - z \right) \vec{ k }_\perp \right) }
{ \ell_\perp^2 \left( \vec{ \ell }_\perp - \left( 1 - z \right) \vec{ k }_\perp \right)^2 } \frac{ \alpha_s }{ \left( 2\pi \right) } C_F \frac{ 1 + z^2 }{ 1 - z } \nonumber\\
& \times & \frac{ 2 \pi \alpha_s }{ N_C } e^{ -i x_B p^+ y^- - i  \left( x_L + x_D \right) p^+ \left( y_1^- - y_2^- \right)} \theta \left( y_2^- \right) \nonumber\\
& \times & \theta \left( y_1^- - y^- \right) e^{ i x_L p^+ y_2^- } \left[ 1 - e^{ i x_L p^+ \left( y^- - y_1^- \right) } \right],
\end{eqnarray}
and its $\vec{k}_\perp$-dependent part of the partonic hard part is given by
%
%
\begin{eqnarray}
\label{Eqn:qgDoubleScat3_H}
H_{\ref{fig:qgDoubleScat3}} & = & C_F \frac{ \vec{ \ell }_\perp \cdot \left( \vec{ \ell }_\perp - \left( 1 - z \right) \vec{ k }_\perp \right) }
{ \ell_\perp^2 \left( \vec{ \ell }_\perp - \left( 1 - z \right) \vec{ k }_\perp \right)^2 } e^{ i x_L p^+  y_2^- } \nonumber\\
& \times & \left[ 1 - e^{ -i x_L p^+ \left( y^- - y_1^- \right) } \right] \theta \left( y_2^- \right) \theta \left( y_1^- - y^- \right).
\end{eqnarray}

Since Figs.~\ref{fig:qgDoubleScat2} and~\ref{fig:qgDoubleScat3} are complex conjugate of each other, one can naturally add Eqs.~\eqref{Eqn:qgDoubleScat2_H} and~\eqref{Eqn:qgDoubleScat3_H} together and take $y = 0$ and $y_1^- = y_2^- = \xi^-$. Although the combination $\nabla_{ k_\perp }^2 \mathcal{ H }_{ \ref{fig:qgDoubleScat2}, \ref{fig:qgDoubleScat3} }$ still has $\vec{k}_\perp$ dependence, one may easily verify that its second order derivative with respect to $\vec{k}_\perp$ at $\vec{k}_\perp=0$ is zero:
%
%
\begin{eqnarray}
\label{Eqn:qgDoubleScat2plus3_d2H}
\nabla_{ k_\perp }^2 \mathcal{ H }_{ \ref{fig:qgDoubleScat2}, \ref{fig:qgDoubleScat3} }  \Big|_{ k_\perp = 0 } & = & 0 .
\end{eqnarray}

%
\begin{figure}[tbp]
    \addtolength{\abovecaptionskip}{-3mm}
%
	\centering
	\includegraphics[width=\linewidth]{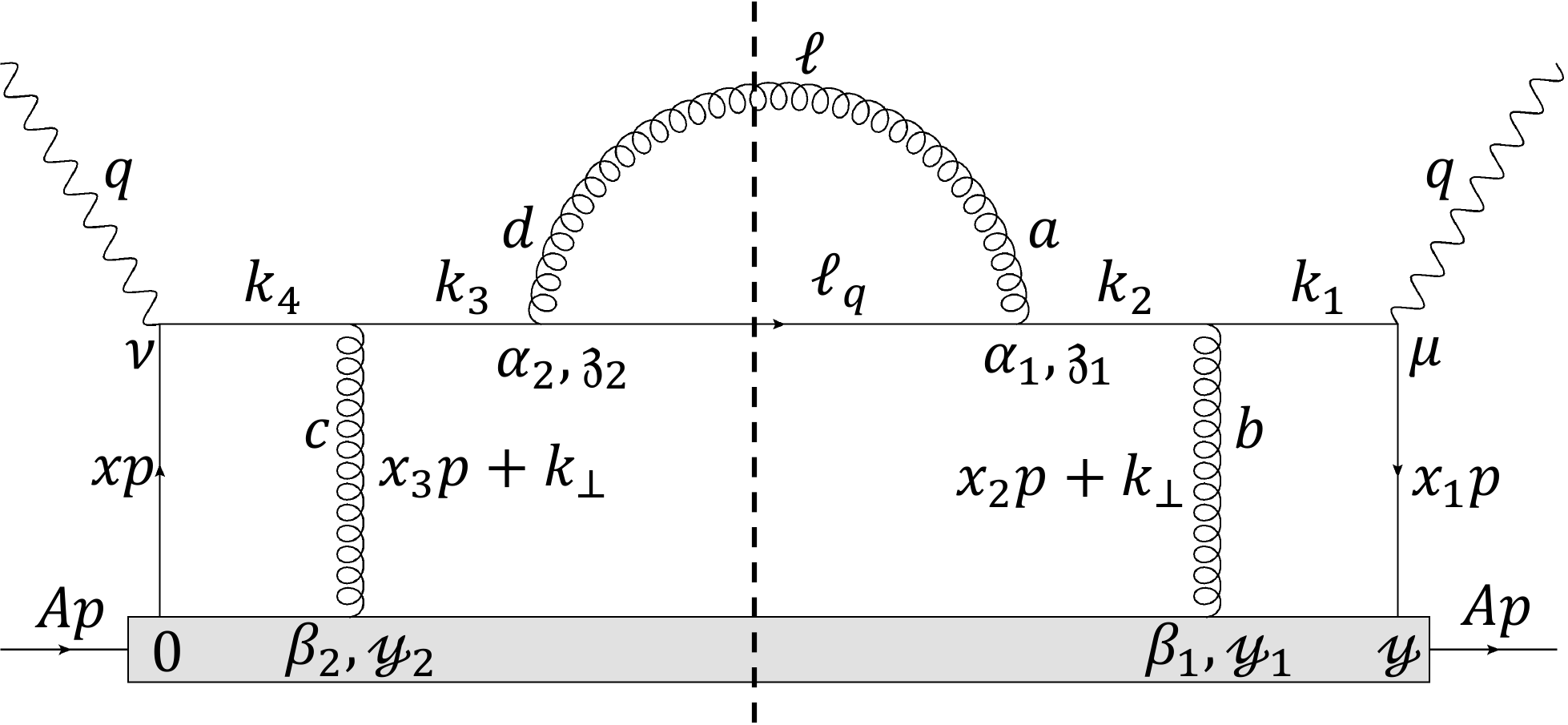}
	\caption{Feynman diagram for the double quark-gluon rescattering process with both scatterings happens before emission.}
	\label{fig:qgDoubleScat4}
\end{figure}

The last central cut diagram of this subsection is for the double quark-gluon rescattering where both scatterings happen before the gluon emission. The hadronic tensor associated with Fig.~\ref{fig:qgDoubleScat4} is given by
\begin{eqnarray}
\label{Eqn:qgDoubleScat4_W}
\mathcal{W}^{\mu\nu}_{\ref{fig:qgDoubleScat4}} & = & - \frac{1}{2\pi} \int \frac{dy^-}{2\pi} dy_1^- dy_2^- \frac{ d^2y_\perp }{ \left( 2\pi \right)^2 } 
d^2k_\perp e^{ i\vec{ k }_\perp \vec{ y }_\perp } \int dz \int dx \nonumber\\
& \times & \left( 2\pi \right) \delta \left[ \left( q + xp \right)^2 \right] \frac{ e_q^2 }{2} 
\Tr \left[ p . \gamma \gamma^\mu \left( q + xp \right) . \gamma \gamma^\nu \right]  \frac{ 1 }{ 2 } \nonumber\\
& \times & \left\langle A \middle| \bar{ \psi } \left( y^- \right) \gamma^+ A^+ \left( y_1^-, \vec{ y }_\perp \right) 
 A^+ \left( y_2^-, \vec{ 0 }_\perp \right) \psi \left( 0 \right) \middle| A \right\rangle  \nonumber\\
& \times & \int \frac{ d \ell_\perp^2 } { \left( \vec{ \ell }_\perp - \left( 1 - z \right) \vec{ k }_\perp \right)^2 }
\frac{ \alpha_s }{ \left( 2\pi \right) } C_F \frac{ 1 + z^2 }{ 1 - z } \frac{ 2 \pi \alpha_s }{ N_C } \nonumber\\
& \times & e^{ -i x_B p^+ y^- - i  \left( x_L + x_D \right) p^+ \left( y_1^- - y_2^- \right)} \nonumber\\
& \times & \theta \left( y_2^- \right) \theta \left( y_1^- - y^- \right) .
\end{eqnarray}

Even though this diagram belongs to the single scattering single emission calculation, this looks more like a higher order correction to the diagram of single emission with no scattering. For this reason, there is no $\vec{k}_\perp$-dependent phase term associated with this diagram. The corresponding $\vec{k}_\perp$-dependent part of the partonic hard part can be extracted as
%
%
\begin{eqnarray}
\label{Eqn:qgDoubleScat4_H}
H_{\ref{fig:qgDoubleScat4}} & = & \frac{ C_F } { \left( \vec{ \ell }_\perp - \left( 1 - z \right) \vec{ k }_\perp \right)^2 } \theta \left( y_2^- \right) \theta \left( y_1^- - y^- \right).
\end{eqnarray}

Here we can ignore the two $\theta$-functions since there is no other $y_1^-$ or $y_2^-$ dependence in Eq.~\eqref{Eqn:qgDoubleScat4_H}. Taking the collinear expansion of this hard part is non-trivial since taking double derivative at $\vec{k}_\perp =0$ directly would yield divergence. To avoid such divergence, one can perform the $\ell_\perp^2$ integral of the hard part before taking the $\vec{k}_\perp$ derivative:
%
%
\begin{eqnarray}
\int_0^{ Q^2 } d\ell_\perp^2 H_{ \ref{fig:qgDoubleScat4} } = \int_0^{ Q^2 } d \ell_\perp^2 \frac{ C_F } { \left( \vec{ \ell }_\perp - \left( 1 - z \right) \vec{ k }_\perp \right)^2 } .
\end{eqnarray}
The above integral can be performed with the change of variable $ \vec{ p }_\perp = \vec{\ell}_\perp - \left( 1 - z \right) \vec{k}_\perp$. This also changes the limits of the integral from $0 \rightarrow Q^2$ to $ \left( 1 - z \right)^2 k_\perp^2 \rightarrow Q^2 - 2 \left( 1 - z \right) \vec{ \ell }_\perp \cdot \vec{ k }_\perp + \left( 1 - z \right)^2 k_\perp^2$. One may still use $Q^2$ as the upper limit considering that it is a large number. For generality, we keep the full form of this upper limit and can show that the final result is zero either with or without this large $Q^2$ assumption. The integral is then re-written as 
%
%
\begin{align}
\int_0^{ Q^2 } & d\ell_\perp^2 H_{ \ref{fig:qgDoubleScat4} } = \int_{ \left( 1 - z \right)^2 k_\perp^2 }^{ Q^2 
- 2 \left( 1 - z \right) \vec{ \ell }_\perp \cdot \vec{ k }_\perp + \left( 1 - z \right)^2 k_\perp^2 } d p_\perp^2 \frac{ C_F } { p_\perp^2 }\\
=\; & C_F \log \left( \frac{ Q^2 - 2 \left( 1 - z \right) \vec{ \ell }_\perp \cdot \vec{ k }_\perp + \left( 1 - z \right)^2 k_\perp^2 }{ \left( 1 - z \right)^2 k_\perp^2  } \right) . \nonumber
\end{align}
One may easily observe that the second derivative of this expression with respect to $\vec{ k }_\perp$ at $\vec{ k }_\perp = 0$ is zero no matter whether the above mentioned large $Q^2$ assumption is used or not. Therefore, there is no contribution to the final medium modification kernel from this double pre-emission quark-gluon rescattering diagram.

We have now completed calculations for all the nine central cut diagrams. Five of them have non-zero contributions to the medium modification kernel at the next-to-leading twist. If one only considers post-emission rescattering and assumes diagrams including pre-emission scattering is suppressed in the soft gluon limit, the number of contributing diagrams is reduced to five, including the two contributing non-central cut diagrams. However, we have shown that such suppression no longer exists when we take into account the $\vec{k}_\perp$-dependent phase factors in our calculations. The contributions from post-emission and pre-emission rescatterings will be compared in detail later in Sec.~\ref{sec:level4}.

\subsection{\label{subsec:level3.3} Non-central cut diagrams }

There are eight double-rescattering-single-emission diagrams that can be cut asymmetrically, leading to ten left or right cut diagrams in total. However, only two of them have contribution to our medium modification kernel at the next-to-leading twist. In this subsection, we present detailed calculations for the two contributing non-central cut diagrams. The remaining eight diagrams will be summarized in Appendix~\ref{sec:appendix2}.

%
%
\begin{figure}[tbp]
    \addtolength{\abovecaptionskip}{-3mm}
%
	\centering
	\includegraphics[width=\linewidth]{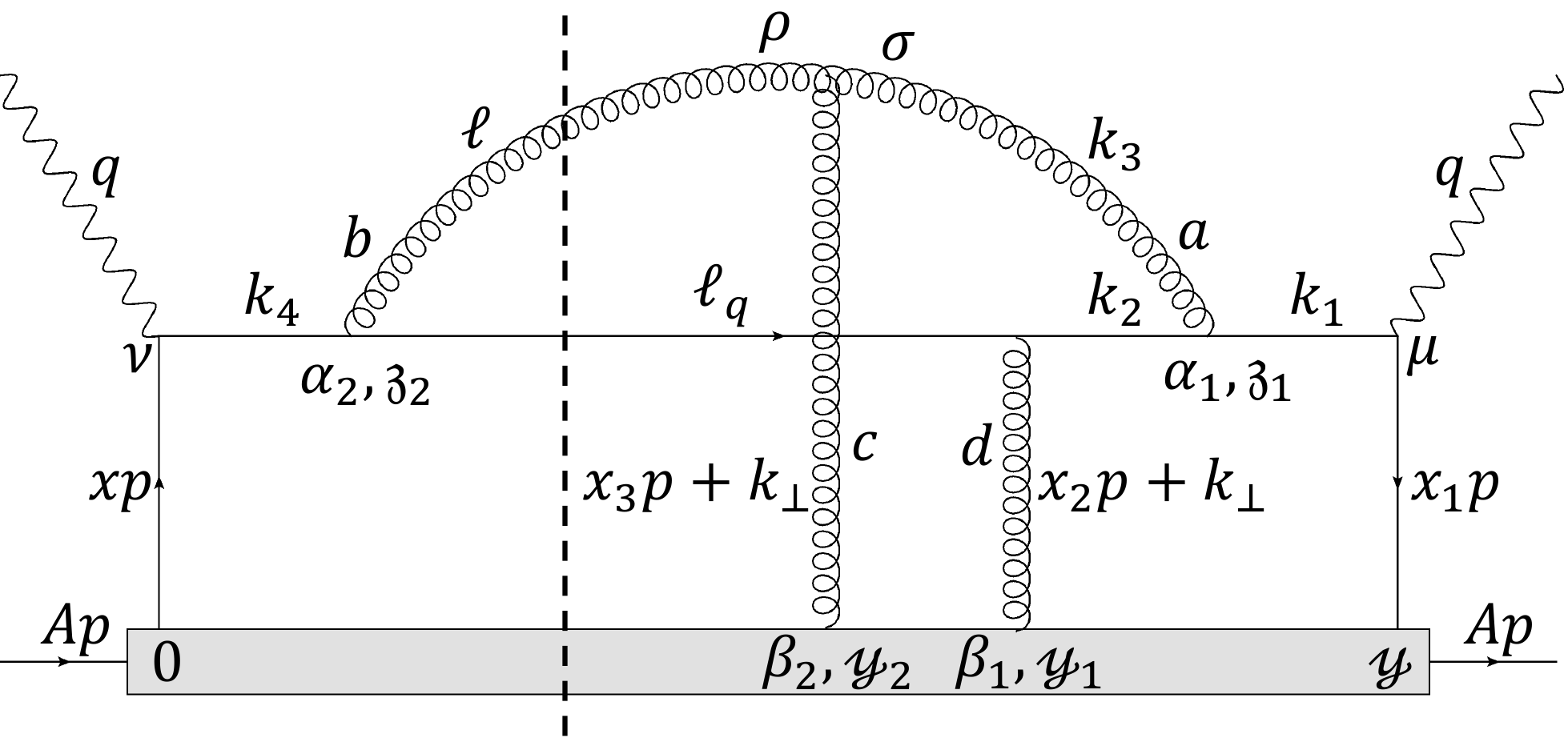}
	\caption{Feynman diagram for the quark-gluon and gluon-gluon rescattering process with left cut.}
	\label{fig:qgAndggScatLeft}
\end{figure}

The hadronic tensor for the left cut diagram illustrated in Fig.~\ref{fig:qgAndggScatLeft} can be expressed as follows,
\begin{eqnarray}
\label{Eqn:qgAndggScatLeft_W}
\mathcal{W}^{\mu\nu}_{\ref{fig:qgAndggScatLeft}}  & = & - \frac{1}{2\pi} \int \frac{dy^-}{2\pi} dy_1^- dy_2^- \frac{ d^2y_\perp }{ \left( 2\pi \right)^2 } 
d^2k_\perp \int dz \int dx \nonumber\\
& \times & \left( 2\pi \right) \delta \left[ \left( q + xp \right)^2 \right] \frac{ e_q^2 }{2} 
\Tr \left[ p . \gamma \gamma^\mu \left( q + xp \right) . \gamma \gamma^\nu \right] \frac{ n }{ 2 } \nonumber\\
& \times & \left\langle A \middle| \bar{ \psi } \left( y^- \right) \gamma^+ A^+ \left( y_1^-, \vec{ y }_\perp \right) 
 A^+ \left( y_2^-, \vec{ 0 }_\perp \right) \psi \left( 0 \right) \middle| A \right\rangle  \nonumber\\
& \times & e^{ i\vec{ k }_\perp \cdot \vec{ y }_\perp } \int d \ell_\perp^2 \frac{ \vec{ \ell }_\perp \cdot \left( \vec{ \ell }_\perp + \vec{ k }_\perp \right) }{ \ell_\perp^2 \left( \vec{ \ell }_\perp + \vec{ k }_\perp \right)^2 }  
\frac{ \alpha_s }{ \left( 2\pi \right) } \frac{ C_A }{ 2 }  \frac{ 1 + z^2 }{ 1 - z } \nonumber\\
& \times & \frac{ 2 \pi \alpha_s }{ N_C } e^{ -i \left( x_B + x_L \right) p^+ y^- + i  x_D^+ p^+ \left( y_1^- - y_2^- \right)} \nonumber\\
& \times & \left\{ \left[ \theta \left( y_2^- - y_1^- \right) e^{ -i \left( x_L + \frac{ x_D^+ }{ \left( 1 - z \right) } \right) p^+ \left( y_1^- - y_2^- \right) } \right. \right. \nonumber\\
& + & \left. \left. \theta \left( y_1^- - y_2^- \right) \right]  e^{ - i \left( x_L + \frac{ x_D^+ }{ \left( 1 - z \right) } \right) p^+ \left( y_2^- - y^- \right) } \right. \nonumber\\
& \times & \left. \theta \left( y_2^- - y^- \right) + \theta \left( y_1^- - y^- \right) \theta \left( y_2^- - y^- \right) \right. \nonumber\\
& \times & \left.  \left[ e^{ -i \left( x_L + \frac{ x_D^+ }{ \left( 1 - z \right) } \right) p^+ \left( y_1^- - y^- \right) } - 1 \right] \right\} \nonumber\\
& \times & e^{ -i \frac{x_D^+ p^+ \left( y^- - y_2^- \right) }{ \left( 1 - z \right) }} .
\end{eqnarray}
Note that to be consistent with the convention, the four momentum of the outgoing gluon should be $ \ell $. As a result, the definition of momentum in this diagram appears slightly different from its corresponding central cut diagram (Fig.~\ref{fig:qgAndggScat1}). And because of these adjustments in the definition of momentum, the related momentum fraction $x_D$ previously defined in Eq.~\eqref{Eqn:x_D} has also been adjusted to $ x_D^+ $ in this diagram as follows, 
%
%
\begin{eqnarray}
x_D^+ = \frac{ k_\perp^2 + 2 \vec{\ell}_\perp\cdot\vec{k}_\perp }{ 2p^+q^-z} .
\label{Eqn:x_D+}
\end{eqnarray}

Additional factor of $n$ associated with the soft four-point matrix element is due to phase space constraint explained in Appendix~\ref{sec:appendix1}. Here $0 \leq n \leq 1/2$ depend on the phase space constraint of the transverse distance between two scatterings.

The $\vec{k}_\perp$-dependent part of the partonic hard part can be then extracted from the hadronic tensor as
%
%
\begin{eqnarray}
\label{Eqn:qgAndggScatLeft_H}
H_{\ref{fig:qgAndggScatLeft}} & = & - \frac{ n C_A }{ 2 } \frac{ \vec{ \ell }_\perp \cdot \left( \vec{ \ell }_\perp + \vec{ k }_\perp \right) } 
{ \ell_\perp^2 \left( \vec{ \ell }_\perp + \vec{ k }_\perp \right)^2 } e^{ -i \frac{x_D^+ p^+ \left( y^- - y_2^- \right) }{ \left( 1 - z \right) }} \nonumber\\
& \times & \left\{ \left[ \theta \left( y_2^- - y_1^- \right) e^{ -i \left( x_L + \frac{ x_D^+ }{ \left( 1 - z \right) } \right) p^+ \left( y_1^- - y_2^- \right) } \right. \right. \nonumber\\
& + & \left. \left. \theta \left( y_1^- - y_2^- \right) \right]  e^{ - i \left( x_L + \frac{ x_D^+ }{ \left( 1 - z \right) } \right) p^+ \left( y_2^- - y^- \right) } \right. \nonumber\\
& \times & \left. \theta \left( y_2^- - y^- \right) + \theta \left( y_1^- - y^- \right) \theta \left( y_2^- - y^- \right) \right. \nonumber\\
& \times & \left.  \left[ e^{ -i \left( x_L + \frac{ x_D^+ }{ \left( 1 - z \right) } \right) p^+ \left( y_1^- - y^- \right) } - 1 \right] \right\} .
\end{eqnarray}
There are two possible orientations of this diagram depending on the location of the secondary scattering. Both quark-gluon rescattering followed by gluon-gluon rescattering and gluon-gluon rescattering followed by quark-quark rescattering are included in the above expression, as characterize by the two $\theta$-functions $\theta \left( y_2^- - y_1^- \right) $ and $\theta \left( y_1^- - y_2^- \right) $ respectively.

%
%
\begin{figure}[tbp]
    \addtolength{\abovecaptionskip}{-3mm}
%
	\centering
	\includegraphics[width=\linewidth]{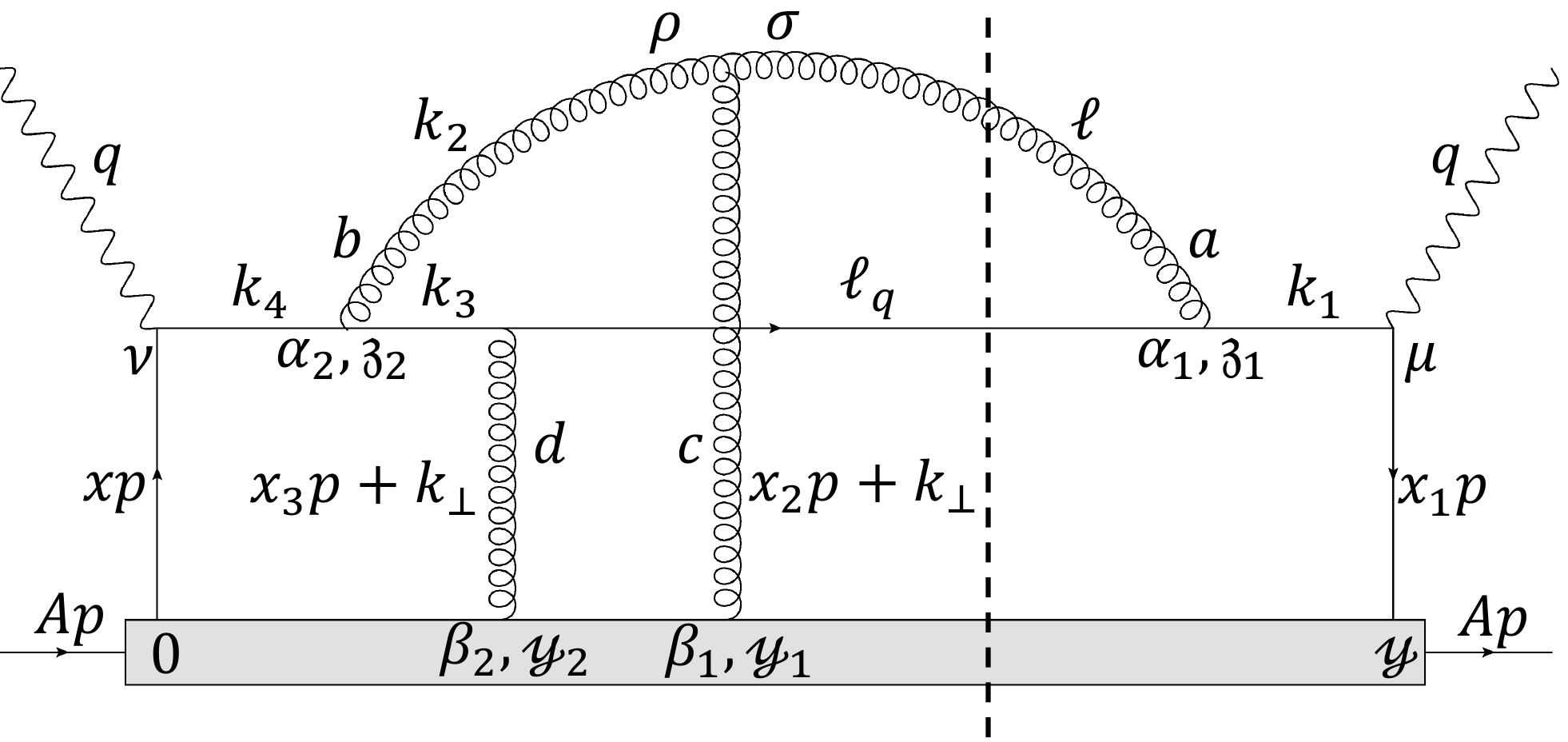}
	\caption{Feynman diagram for the quark-gluon and gluon-gluon rescattering process with right cut.}
	\label{fig:qgAndggScatRight}
\end{figure}

The complex conjugate of this left cut diagram (Fig.~\ref{fig:qgAndggScatLeft}) is shown in Fig.~\ref{fig:qgAndggScatRight} as a right cut diagram. We can combine their hadronic tensors to simplify the further collinear expansion as we did in previous subsections. The hadronic tensor associated with Fig.~\ref{fig:qgAndggScatRight} reads
\begin{eqnarray}
\label{Eqn:qgAndggScatRight_W}
\mathcal{W}^{\mu\nu}_{\ref{fig:qgAndggScatRight}}  & = & - \frac{1}{2\pi} \int \frac{dy^-}{2\pi} dy_1^- dy_2^- \frac{ d^2y_\perp }{ \left( 2\pi \right)^2 } 
d^2k_\perp \int dz \int dx \nonumber\\
& \times & \left( 2\pi \right) \delta \left[ \left( q + xp \right)^2 \right] \frac{ e_q^2 }{2} 
\Tr \left[ p . \gamma \gamma^\mu \left( q + xp \right) . \gamma \gamma^\nu \right] \frac{ n }{ 2 } \nonumber\\
& \times & \left\langle A \middle| \bar{ \psi } \left( y^- \right) \gamma^+ A^+ \left( y_1^-, \vec{ y }_\perp \right) 
 A^+ \left( y_2^-, \vec{ 0 }_\perp \right) \psi \left( 0 \right) \middle| A \right\rangle  \nonumber\\
& \times & e^{ i \vec{ k }_\perp \cdot \vec{ y }_\perp } \int d \ell_\perp^2 \frac{ \vec{ \ell }_\perp \cdot \left( \vec{ \ell }_\perp + \vec{ k }_\perp \right) } { \ell_\perp^2 \left( \vec{ \ell }_\perp + \vec{ k }_\perp \right)^2 }  
\frac{ \alpha_s }{ \left( 2\pi \right) } \frac{ C_A }{ 2 } \frac{ 1 + z^2 }{ 1 - z } \nonumber\\
& \times & \frac{ 2 \pi \alpha_s }{ N_C } e^{ -i \left( x_B + x_L \right) p^+ y^- + i  x_D^+ p^+ \left( y_1^- - y_2^- \right)} e^{ -i \frac{x_D^+ p^+ y_1^-  }{ \left( 1 - z \right) }} \nonumber\\
& \times & \left\{ \left[ \theta \left( y_2^- - y_1^- \right) e^{ -i \left( x_L + \frac{ x_D^+ }{ \left( 1 - z \right) } \right) p^+ \left( y_1^- - y_2^- \right) } \right. \right. \nonumber\\
& + & \left. \left. \theta \left( y_1^- - y_2^- \right) \right] \theta \left( y_2^- \right) e^{ i \left( x_L + \frac{ x_D^+ }{ \left( 1 - z \right) } \right) p^+ y_2^- }  \right. \nonumber\\
& + & \left. \theta \left( y_1^- \right) \theta \left( y_2^- \right) \left[ e^{ -i \left( x_L + \frac{ x_D^+ }{ \left( 1 - z \right) } \right) p^+ y_1^- } - 1 \right] \right\}.
\end{eqnarray}
And its $\vec{k}_\perp$-dependent part of the partonic hard part is given by
%
%
\begin{eqnarray}
\label{Eqn:qgAndggScatRight_H}
H_{\ref{fig:qgAndggScatRight}} & = & - \frac{ n C_A }{ 2 } \frac{ \vec{ \ell }_\perp \cdot \left( \vec{ \ell }_\perp + \vec{ k }_\perp \right) } 
{ \ell_\perp^2 \left( \vec{ \ell }_\perp + \vec{ k }_\perp \right)^2 } e^{ -i \frac{x_D^+ p^+ y_1^-  }{ \left( 1 - z \right) }} \nonumber\\
& \times & \left\{ \left[ \theta \left( y_2^- - y_1^- \right) e^{ -i \left( x_L + \frac{ x_D^+ }{ \left( 1 - z \right) } \right) p^+ \left( y_1^- - y_2^- \right) } \right. \right. \nonumber\\
& + & \left. \left. \theta \left( y_1^- - y_2^- \right) \right] \theta \left( y_2^- \right) e^{ i \left( x_L + \frac{ x_D^+ }{ \left( 1 - z \right) } \right) p^+ y_2^- }  \right. \\
& + & \left. \theta \left( y_1^- \right) \theta \left( y_2^- \right) \left[ e^{ -i \left( x_L + \frac{ x_D^+ }{ \left( 1 - z \right) } \right) p^+ y_1^- } - 1 \right] \right\} . \nonumber
\end{eqnarray}
After combining Eqs.~\eqref{Eqn:qgAndggScatLeft_H} and~\eqref{Eqn:qgAndggScatRight_H} and taking $y^- = 0$ and $y_1^- = y_2^- = \xi^-$ as before, we obtain
%
%
\begin{eqnarray}
\label{Eqn:qgAndggScatNonCentral_d2H}
\mathcal{H}_{ \ref{fig:qgAndggScatLeft}, \ref{fig:qgAndggScatRight} } & = & - \frac{ n C_A }{ 2 }
\frac{ \vec{ \ell }_\perp \cdot \left( \vec{ \ell }_\perp + \vec{ k }_\perp \right) }{ \ell_\perp^2 \left( \vec{ \ell }_\perp + \vec{ k }_\perp \right)^2 }
\left[ 2 \cos \left\{ \frac{ \ell_\perp^2 \xi^- }{ 2 z \left( 1 - z \right) q^- } \right\} \right. \nonumber\\
& - & \left. 2 \cos \left\{ \frac{ \left( k_\perp^2 - 2 \vec{ k }_\perp \cdot \vec{ \ell }_\perp \right) \xi^- }{ 2 z \left( 1 - z \right) q^- } \right\} \right].
\label{Eqn: H_NCqgAndgg}
\end{eqnarray}
One may easily show that the second derivative of this combined hard part with respect to $ \vec{k}_\perp $ at $ \vec{k}_\perp = 0 $ gives non-negligible contribution to the final medium modification:
%
%
\begin{eqnarray}
\label{Eqn:qgAndggScatNonCentral_d2H}
\nabla_{ k_\perp }^2 \mathcal{H}_{ \ref{fig:qgAndggScatLeft}, \ref{fig:qgAndggScatRight} } \Big |_{ k_\perp = 0} & = & -4 n \frac{ C_A }{ \ell_\perp^4 }\left( \frac{ \ell_\perp^2 \xi^- }{ 2 q^- z \left( 1 - z \right) } \right)^2.
\end{eqnarray}
The factor of $n$ in the equation above lies between 0 and 1/2 ($0<n<1/2$) and depends on assumptions made regarding the matrix element $\langle F^{+\mu} F^+_\mu \rangle$ in the left and right cut diagrams. These are described in Appendix~\ref{sec:appendix1}. 
The other eight left and right cut diagrams will be summarized in Appendix~\ref{sec:appendix1}, which will have zero contribution to our final result.

\section{\label{sec:level4} Medium modification kernel}

In the previous section, a complete calculation of the single re-scattering induced single gluon emission process has been carried out. The calculation has taken into account contributions from the momentum-dependent, complex amplitude and conjugate [Eq.~\eqref{general_form}], as well as the $\vec{k}_\perp$-dependent phase factors during the collinear (or $\vec{k}_\perp$) expansion. 
While several new and large contributions were discovered, the basic physical interpretation of the diagrams is unchanged from that in GW~\cite{Guo:2000nz, Wang:2001ifa}. Thus the main result of our work is a quantitative, and not a qualitative, shift from that of GW. 
We recapitulate this physical interpretation in this section, present our final results and carry out numerical comparisons between the results of our work and that of GW and AZZ.

\subsection{Physical interpretation of pole structure}

\begin{figure}[htbp!]
%
\includegraphics[width=\linewidth]{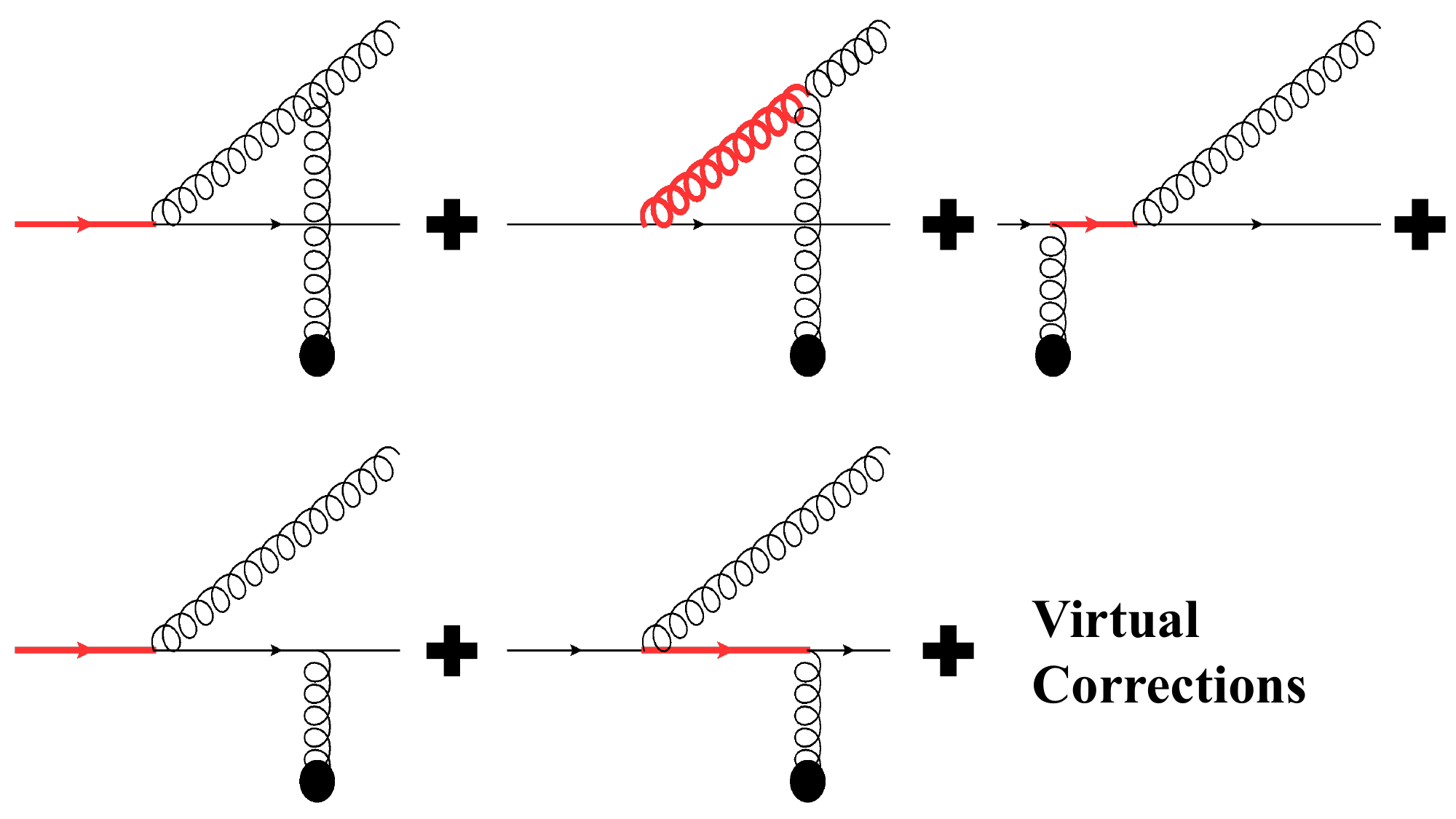}
%
\caption{Splitting processes included in medium splitting function. The thick red line represents the most virtual propagator in the diagram. The dark circle attached to the gluons indicates that those gluons are part of the medium.}
	\label{fig:LPM_Interference_Diagrams}
\end{figure}

Consider all the possible Feynman diagram amplitudes, i.e., the left hand sides of the cut line (for central cuts only).
In the evaluation of contributions to the hadronic tensor from Eqs.~\eqref{Eqn:DoubleggScat1_W}-\eqref{Eqn:qgDoubleScat4_W}, one carries out a series of contour integrations over the momenta in the denominator of the propagators, using the method of residues. However in each case, for central cut diagrams, there is one propagator in the amplitude and one in the complex conjugate which cannot be integrated out by the method of residues. This propagator remains off-shell, while we integrate over the momenta in the other propagators. This is indicated by the red line in Fig.~\ref{fig:LPM_Interference_Diagrams}. In GW, the cases where the off-shell propagator emanated directly from the hard vertex were referred to as soft (S) contributions (the two left diagrams in Fig.~\ref{fig:LPM_Interference_Diagrams}). The cases where the off-shell propagators were separated from the hard vertex by propagator whose momentum had been integrated out is referred to as double hard (H), e.g., the quark after hard scattering off a photon goes on-shell and is knocked off-shell by a second hard scattering in the medium leading to the radiation of a hard gluon (the remaining 3 diagrams in Fig.~\ref{fig:LPM_Interference_Diagrams}). 

Based on the phase factors, these diagrams can be re-organized into five different types of diagrams as illustrated in Fig.~\ref{fig:LPM_Interference_Diagrams}. In each of these five diagrams, the phase varies depending on the most virtual parton, indicated by the thick red line. For cases where the off-shell parton is time-like (the two left diagrams, and top right diagram in Fig.~\ref{fig:LPM_Interference_Diagrams}), the overall sign is positive. For diagrams where the off-shell line is space-like, e.g., the two middle diagrams in Fig.~\ref{fig:LPM_Interference_Diagrams}, the overall sign is negative. 

\subsection{Comparing the full kernel with GW and AZZ}

In this subsection, we gather all contributions to the Taylor expansion in terms of $k_\perp$ and 
reconstitute the hadronic tensor at next-to-leading order and next-to-leading twist. 
Combining all contributions from both the amplitude and complex conjugate, at the next-to-leading twist, i.e., Eqs.~\eqref{Eqn:DoubleggScat1_d2H}, \eqref{Eqn:qgAndggScat1plus2_d2H}, \eqref{Eqn:qgAndggScat4_d2H} and~\eqref{Eqn:qgAndggScatNonCentral_d2H}, we have 
%
%
\begin{eqnarray}
\label{Eqn:d2H_Combined}
\nabla_{ k_\perp }^2 H \Big|_{ k_\perp = 0 } & = & \frac{ 4 C_A }{ \ell_\perp^4 } \left[ \frac{ \left( 1 + z \right) }{ 2 }
\left\{ 2  - 2 \cos \left( \frac{ \ell_\perp^2 \xi^- }{ 2 q^- z \left( 1 - z \right) } \right) \right. \right. \nonumber\\ 
& - & \left. \left. 2 \left( \frac{ \ell_\perp^2 \xi^- }{ 2 q^- z \left( 1 - z \right) } \right) \sin \left( \frac{ \ell_\perp^2 \xi^- }{ 2 q^- z \left( 1 - z \right) } \right) \right\} \right. \nonumber\\
& + &  \left. \left(1 - n \right) \left( \frac{ \ell_\perp^2 \xi^- }{ 2 q^- z \left( 1 - z \right) } \right)^2  \right] . 
\end{eqnarray}
By using the full partonic hard part at next-to-leading twist after collinear expansion given in Eq.~\eqref{Eqn:d2H_Combined}, the hadronic tensor can be written as,
%
\begin{eqnarray}
\label{Eqn:SCM_full}
\mathcal{W}^{\mu\nu}  & = & \int \frac{ d y^- }{ 2 \pi }  d \xi^- \frac{ d \delta y^- }{ 2 \pi } dz dx
 \frac{ d^2 k_\perp d^2 y_\perp }{ \left( 2 \pi \right)^2 } e^{ i k_\perp \cdot y_\perp } \nonumber\\
& \times & \left( 2\pi \right) \delta \left[ \left( q + xp \right)^2 \right] \frac{ e_q^2 }{2} \Tr \left[ p . \gamma \gamma^\mu \left( q + xp \right) . \gamma \gamma^\nu \right] \nonumber\\
& \times & \frac{ 1 }{ 2 } \left\langle A \middle| \bar{ \psi } \left( y^- \right) \gamma^+ F_\alpha^+ \left( \xi^- + \frac{ \delta y^- }{ 2 }\right) \right. \nonumber\\
& \times & \left. F^{ +\alpha } \left( \xi^- - \frac{ \delta y^- }{ 2 }\right) \psi \left( 0 \right) \middle| A \right\rangle 
\frac{ \alpha_s }{ \left( 2\pi \right) } \nonumber\\
& \times & \int \frac{ d \ell_\perp^2 }{ \ell_\perp^4 } C_A \frac{ 1 + z^2 }{ 1 - z }
\frac{ 2 \pi \alpha_s }{ N_C } e^{ -i \left( x_B + x_L \right) p^+ y^- } e^{ -i x_D^+ p^+ \delta y^-}
\nonumber\\
& \times & \left[ \frac{ \left( 1 + z \right) }{ 2 }  \left\{ 2 - 2 \cos \left( \frac{ \ell_\perp^2 \xi^- }{ 2 q^- z \left( 1 - z \right) } \right) \right. \right. \nonumber\\ 
& - & \left. \left. 2 \left( \frac{ \ell_\perp^2 \xi^- }{ 2 q^- z \left( 1 - z \right) } \right) \sin \left( \frac{ \ell_\perp^2 \xi^- }{ 2 q^- z \left( 1 - z \right) } \right) \right\} \right. \nonumber\\
& + &  \left. \left( 1 - n \right) \left( \frac{ \ell_\perp^2 \xi^- }{ 2 q^- z \left( 1 - z \right) } \right)^2 \right] .
\end{eqnarray}

In Eq.~\eqref{Eqn:SCM_full}, the gluon fields in soft hadronic tensor is converted to field strength via partial integration to get the jet transport coefficient, defined in Eq.~\eqref{eq:q-hat}, after factorization. Here a variable change, $ \ell_\perp^2 = z \left( 1 - z \right) \mu^2 $, can be made to remove the $z$ dependence of sine and cosine terms. This allows one to factor the $\mu$ and $\xi^-$ integrations from the $z$ integration. But even with this variable change, the overall $ \left( 1 + z \right) / [2z(1-z)] $ factor remains. Thus the in-medium splitting function is \emph{not} of the form $P(z)\times K$, i.e., the vacuum splitting function times a medium dependent factor that does not depend on $z$. Indeed the medium induced splitting function is different from that in vacuum.

By considering Eq.~\eqref{Eqn:Diff_HadTensor} and Eq.~\eqref{Eqn:SCM_full}, we can reintroduce a fragmentation function on the outgoing quark or gluon. Also here we use the definitions in Eq.~\eqref{Eqn:H0}, Eq.~\eqref{eq:q-hat}, and Eq.~\eqref{Eqn:PDF} for $ \mathcal{ H }_0^{ \mu\nu } $, $ \hat{ q } \left( \xi^- \right) $, and $ f_q^A \left( x \right) $ respectively, to simplify the expression. For the case where the detected hadron emanates from the quark with momentum $p_h^- = z_h q^-$,  the differential hadronic tensor which includes medium induced single gluon emission from the quark at next-to-leading twist (or in the one rescattering approximation) can be expressed as,
%
\begin{eqnarray}
\label{Eqn:SCM_full_mu2}
\frac{ d \mathcal{W}^{\mu\nu} }{ d z_h }  & = & \int d \xi^-
 \int \frac{ dz }{ z } \mathcal{ H }_0^{ \mu \nu } f_q^A \left( x_B + x_L \right) \nonumber\\
& \times & \hat{ q } \left( \xi^- \right) \int \frac{ d \mu^2 }{ \mu^4 } \frac{ \alpha_s }{ \left( 2\pi \right) } C_F \frac{ 1 + z^2 }{ 1 - z } \nonumber\\
& \times & \frac{ 1 }{ z \left( 1 - z \right) }\left[ \frac{ \left( 1 + z \right) }{ 2 }  \left\{ 2 - 2 \cos \left( \frac{ \mu^2 \xi^- }{ 2 q^- } \right) \right. \right. \nonumber\\ 
& - & \left. \left. 2 \left( \frac{ \mu^2 \xi^- }{ 2 q^- } \right) \sin \left( \frac{ \mu^2 \xi^- }{ 2 q^- } \right) \right\} \right. \nonumber\\
& + & \left.  \left( 1 - n \right) \left( \frac{ \mu^2 \xi^- }{ 2 q^- } \right)^2 \right]
D_q^h \left( \frac{ z_h }{ z } \right) .
\end{eqnarray}
In the equation above $C_F = (N_C^2 - 1)/(2N_C)$ is the Casimir factor for a quark.
We will factorize the above equation in the form, 
\begin{eqnarray}
\frac{d \mathcal{W}^{\mu\nu}}{d z_h} = f_q^A \otimes \mathcal{H}_0^{\mu \nu} \otimes \Tilde{D}^h_q,
\end{eqnarray}
where, $\Tilde{D}$ is the medium modified fragmentation function. Thus the next-to-leading twist contribution to the fragmentation function at next-to-leading order is obtained as, 
\begin{eqnarray}
\delta \Tilde{D}_q^h(z_h) &=&  \int \frac{ d \mu^2 }{ \mu^4 } \frac{ \alpha_s }{ \left( 2\pi \right) } C_F  \int \frac{ dz }{ z } \frac{ 1 + z^2 }{ z(1 - z)^2 } \nonumber\\
& \times & \int d \xi^-
\hat{ q } (\xi^-)\left[ \frac{ \left( 1 + z \right) }{ 2 }  \left\{ 2 - 2 \cos \left( \frac{ \mu^2 \xi^- }{ 2 q^- } \right) \right. \right. \nonumber\\ 
& - & \left. \left. 2 \left( \frac{ \mu^2 \xi^- }{ 2 q^- } \right) \sin \left( \frac{ \mu^2 \xi^- }{ 2 q^- } \right) \right\} \right. \nonumber\\
& + & \left.  \left( 1 - n \right) \left( \frac{ \mu^2 \xi^- }{ 2 q^- } \right)^2 \right] 
D_q^h \left( \frac{ z_h }{ z } \right) .
\end{eqnarray}

Adding the above contribution to the vacuum contribution to the fragmentation function at next-to-leading order and leading twist, 
\begin{equation}
\delta D(z) = \int \frac{d \mu^2}{\mu^2}  \frac{\alpha_S}{2 \pi} \int_{z_h}^1 P_{+}(z)  D^h_q \left( \frac{z_h}{z} \right),
\end{equation}
we obtain the full contribution from next-to-leading order up to next-to-leading twist. 
In the equation above, $ P \left( z \right) = C_F \frac{ 1 + z^2 }{ 1 - z }$, and the $(+)$-function, $ P_+ \left( z \right) $ is defined by combining real and virtual contributions. We now simply add next-to-leading twist contributions at each successive order and then differentiate with $\log{\mu^2}$ to obtain the medium modified Dokshitzer-Gribov-Lipatov-Altarelli-Parisi (DGLAP) evolution equation for the medium modified fragmentation function (a more detailed calculation of this resummation using GW kernel can be found in Ref. \cite{Cao:2021rpv}):
%
%
\begin{eqnarray}
\label{Eqn:Med_DGLAP}
&& \frac{\partial D_q^h \left( z_h, \mu^2, q^- \right) \big|_{ t_0^- }^{ t_0^- + \tau } }{ \partial \log \mu^2 }
= \frac{ \alpha_S }{ 2 \pi } \int_{ z_h }^1 \frac{ dz }{ z } \left[ P_+ \left( z \right) \right. \nonumber\\
& + & \left. \left( \frac{ P \left( z \right) }{ z \left( 1 - z \right) } \right)_+ \int_{ t_0^- }^{ t_0^- + \tau } d\xi^- \frac{ 1 }{ \mu^2} \bar{K} \left( q^-, \mu^2, z, \xi^- \right) \right. \nonumber\\
& \times & \left. D_i^h \left( \frac{ z_h }{ z }, \mu^2, zq^- \right) \right],
\end{eqnarray}
where $ \bar{K} \left( q^-, \mu^2, z, \xi^- \right) $ is defined below in Eq.~\eqref{Eqn:GeneralKernel}. In Eq.~\eqref{Eqn:Med_DGLAP}, $i$ can be quark or gluon, and depending on the value of $i$, the splitting function, $P_+ \left( y \right)$ changes accordingly. The medium modification kernel $ \mathcal{K} \left( q^-, z \right) $, which includes the entire contribution from the medium, other than the fragmentation function, is given as,
%
%
\begin{eqnarray}
\label{Eqn:SCM_kernel}
\mbox{}\!\!\!\!& \!\!\!\mathcal{K} & \left( q^-, z \right) = \frac{ \alpha_S }{2\pi} \int \frac{ d \mu^2 }{ \mu^4 } 
C_F \frac{ 1 + z^2 }{ 1 - z } \int d \xi^- \hat{ q } \left( \xi^- \right) \nonumber\\
& \times & \frac{ 1 }{ z \left( 1 - z \right) } \left[ \frac{ \left( 1 + z \right) }{ 2 } \left\{ 2 - 2 \cos \left( \frac{ \mu^2 \xi^- }{ 2 q^- } \right) \right. \right. \nonumber\\
&\!\!\!\! - & \!\!\!\!\left. \left. 2 \left( \frac{ \mu^2 \xi^- }{ 2 q^- } \right) \sin \left( \frac{ \mu^2 \xi^- }{ 2 q^- } \right) \right\} + \left( 1 - n \right) \!\!\left( \frac{ \mu^2 \xi^- }{ 2 q^- } \right)^2 \right]\!\! .
\end{eqnarray}

The general form of the medium modification kernel $\mathcal{ K } \left( q^-, z \right)$ can be given as
%
%
\begin{eqnarray}
\label{Eqn:GeneralKernel}
& \mathcal{K} & \left( q^-, z \right) = \frac{\alpha_S}{ 2 \pi } \int \frac{d \mu^2 }{ \mu^4 } \int d \xi^- C_F \frac{ 1 + z^2 }{ 1 - z } \nonumber\\
& \times & \frac{ 1 }{ z \left( 1 - z \right) }\bar{ K } \left( q^-, \mu^2, z, \xi^- \right),
\end{eqnarray}
in which we use $\bar{K}$ to compare different kernels to visualise the difference between calculations properly.

One can observe obvious difference between our (SCM) result Eq.~\eqref{Eqn:SCM_kernel} and the medium modification kernels provided by either GW Eq.~\eqref{Eqn:GW_kernel}~\cite{Guo:2000nz} or AZZ Eq.~\eqref{Eqn:AZZ_kernel}~\cite{Aurenche:2008mq}. Additional momentum fraction $z$ remains even after the gluon splitting time substitution $\tau_F^-=2q^-z(1-z)/\ell_\perp^2$. Therefore, to compare our result to GW and AZZ kernels, we need to specify the $z$ value taken in our calculation. 

%
%
\begin{figure}[tbp]
    \addtolength{\abovecaptionskip}{-3mm}
%
	\includegraphics[width=0.9\linewidth]{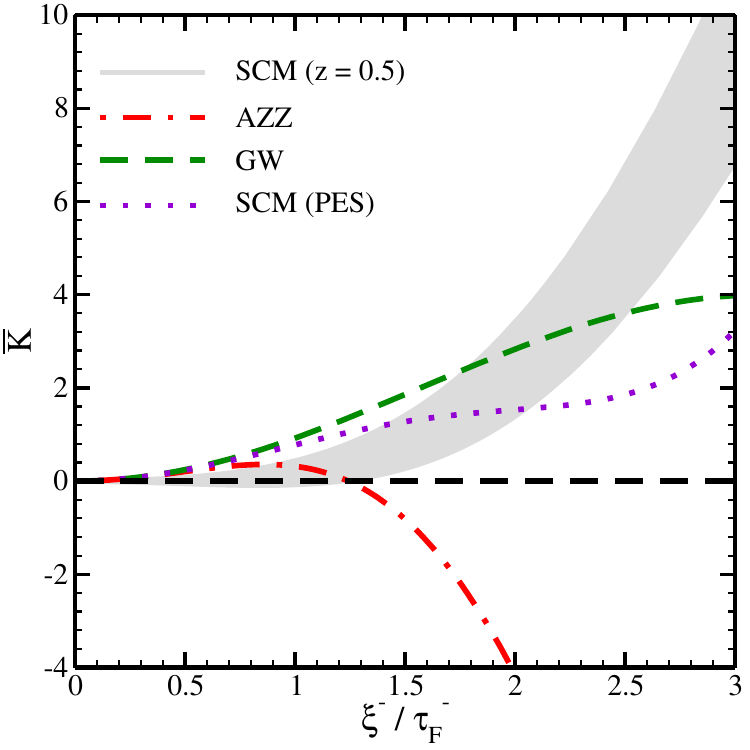}
	\caption{Comparison of the energy loss kernel of this study (labeled as SCM), only using the post-emission scattering (PES), and using the full result at $z=0.5$, to the previous GW and AZZ calculations. The band in SCM results from varying the factor $n$ in Eq.~\eqref{Eqn:SCM_kernel}, between 0 (top of band) to $1/2$ (bottom of band).}
	\label{fig:Kernel_Comp}
\end{figure}

In Fig. \ref{fig:Kernel_Comp}, we compare our new result with both GW and AZZ kernels, as a function of the path-length-formation-time ratio ($\xi^-/\tau_F^-$). The momentum fraction is chosen as $z=0.5$ in this figure. Compared to the GW result, our current calculation predicts a much smaller value of the medium modification kernel up till a length around $1.5 \tau_F$. This is due to the negative contribution from the pre-emission rescattering diagram as given by Eq.~\eqref{Eqn:qgAndggScat4_d2H}, which is absent in the GW work due to its neglect of the $\vec{k}_\perp$-dependent phase factors before collinear expansion. 

The AZZ effort extended this to include only some of the $\vec{k}_\perp$-dependent phase factors, those that emanate from the leading diagram of the GW analysis, 
Fig.~\ref{fig:DoubleggScat1} in the current effort.
As a result, the AZZ kernel coincides with the GW kernel in the small path length limit. Compared to the AZZ kernel, the current effort yields a final result that is positive definite (for $n\rightarrow 1/2$ there is the possibility of a very small negative value, however the integral is positive definite past 1.3$\tau_f$ for any choice of $n$), and steadily growing with length, unlike the AZZ kernel which begins to oscillate at $\xi^- \gtrsim \tau_F^-$, and never becomes positive definite.

To confirm the deviation between our result and the GW kernel indeed comes from those pre-emission rescattering diagrams, we present in Fig.~\ref{fig:Kernel_Comp} our result that only combines contribution from the post-emission rescattering scenarios, i.e., Eqs.~\eqref{Eqn:DoubleggScat1_d2H} and \eqref{Eqn:qgAndggScat1plus2_d2H}:
%
%
\begin{eqnarray}
\label{Eqn:d2H_Combined_SoftGluon}
\nabla_{ k_\perp }^2 H \Big|_{ k_\perp = 0 } & = & \frac{ 4 C_A }{ \ell_\perp^4 } \left[
2  - 2 \cos \left( \frac{ \ell_\perp^2 \xi^- }{ 2 q^- z \left( 1 - z \right) } \right) \right. \nonumber\\ 
& - & \left. 2 \left( \frac{ \ell_\perp^2 \xi^- }{ 2 q^- z \left( 1 - z \right) } \right) \sin \left( \frac{ \ell_\perp^2 \xi^- }{ 2 q^- z \left( 1 - z \right) } \right) \right. \nonumber\\
& + &  \left. \left( \frac{ \ell_\perp^2 \xi^- }{ 2 q^- z \left( 1 - z \right) } \right)^2 \right. \nonumber\\
& \times & \left. \left\{ 1 - n + \cos \left( \frac{ \ell_\perp^2 \xi^- }{ 2 q^- z \left( 1 - z \right) } \right) \right\} \right],
\end{eqnarray}
which yields the following medium modification kernel for only post emission scattering (PES),
%
%
\begin{eqnarray}
\label{SCM_kernel_SoftGluon}
& \mathcal{K}_\mathrm{PES} & \left( q^-, z \right) = \frac{ \alpha_S }{2\pi} \int \frac{ d \mu^2 }{ \mu^4 } 
C_F \frac{ 1 + z^2 }{ 1 - z } \int d \xi^- \frac{ \hat{ q } \left( \xi^- \right) }{ z \left( 1 - z \right) }  \nonumber\\
& \times & \left[ 2  - 2 \cos \left( \frac{ \mu^2 \xi^- }{ 2 q^- } \right) 
- 2 \left( \frac{ \mu^2 \xi^- }{ 2 q^- } \right) \sin \left( \frac{ \mu^2 \xi^- }{ 2 q^- } \right) \right. \nonumber\\
& + &  \left. \left( \frac{ \mu^2 \xi^- }{ 2 q^- } \right)^2 \left\{ 1 - n + \cos \left( \frac{ \mu^2 \xi^- }{ 2 q^- } \right) \right\} \right] . 
\end{eqnarray}
After excluding the pre-emission rescattering contribution, our result no longer has additional $z$ dependence beyond the gluon splitting time $\tau_F^-$ substitution and agrees with the GW kernel up to a path length around $\tau_F^-$.

%
%
\begin{figure}[tbp]
    \addtolength{\abovecaptionskip}{-3mm}
%
	\centering
	\includegraphics[width=0.9\linewidth]{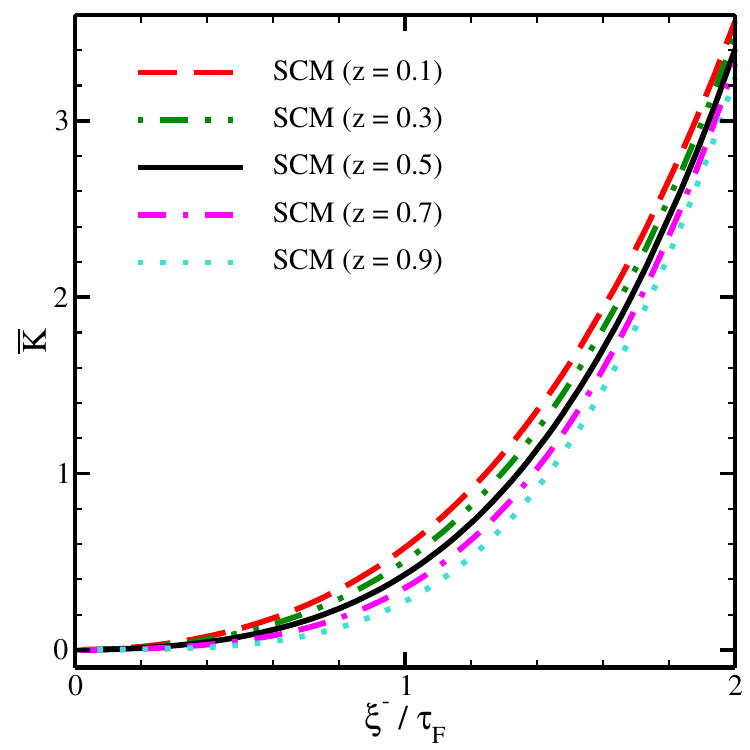}
	\caption{Variation of the energy loss kernel with the momentum fraction variation $z$. Calculated with $n=0$ in Eq.~\eqref{Eqn:SCM_kernel}.}
	\label{fig:Kernel_z_Comp}
\end{figure}

Presented in Fig.~\ref{fig:Kernel_z_Comp} (for $n=0$) is the momentum fraction $z$ dependence of our updated SCM kernel. Unlike the GW and AZZ kernels, even after the substitution $\mu^2 = l_\perp^2/(z(1-z))$, the $z$ integration does not factorize from the $\mu$ and $\xi^-$ integrals. As a result, the full $\bar{K}$ is $z$ dependent. 
As shown in the figure, our medium modification kernel decreases as the quark momentum fraction $z$ increases, or the gluon momentum fraction $(1-z)$ decreases. In other words, softer gluon emission would yield larger difference between our SCM kernel and the earlier GW kernel.

\subsection{Length dependence of energy loss and other facets}

For energetic partons with energy $E$ and virtuality higher than the medium scale ($\mu^2=\sqrt{2\hat{q}E}$), the medium-modified splitting function presented here will become part of a Sudakov form factor that determines the probability of parton splitting inside the medium~\cite{Cao:2017qpx,Cao:2021rpv}. The medium modified splitting function calculated with GW and SCM kernels have similar shapes in terms of their $z$-dependence. However, as plotted in Fig.~\ref{fig:Kernel_Comp}, while the SCM kernel provides a smaller overall magnitude than GW at shorter formation time, it grows faster with propagation time than the GW kernel. 
Assuming that most splits take place within 2-3 formation times, we expect that by applying our updated SCM kernel in jet energy loss calculation, one would expect to extract a somewhat larger jet transport coefficient $\hat{q}$ from the jet quenching data, compared to earlier studies in which the GW kernel was applied. 
However, no other changes in the overall form of jet modification are expected.

\begin{figure}[tb]
    \centering
    \includegraphics[width=0.95\linewidth]{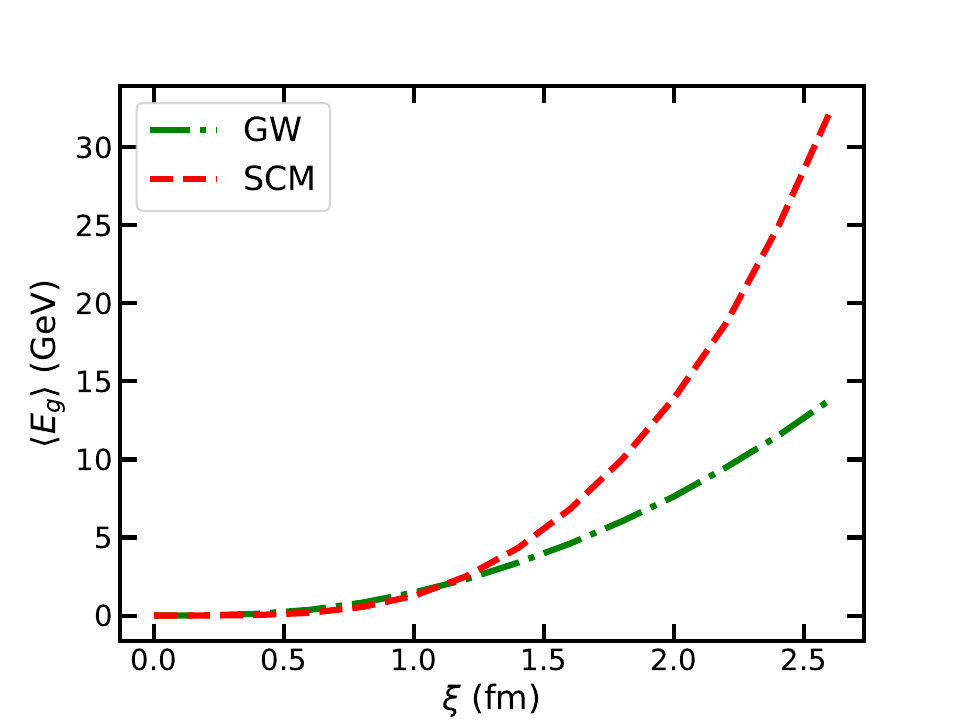}
    \caption{Accumulative energy of emitted gluons as a function of the path-length of a quark with 100~GeV energy through a static medium with $\hat{q}=1$~GeV$^2$/fm [using $n=0$ in Eq.~\eqref{eq:gluonSpectrumSCM}].}
    \label{fig:Eg_vs_t}
\end{figure}

The medium-induced gluon spectrum has also been utilized as the inelastic scattering rate in transport models for high-energy low-virtuality ($\mu^2\lesssim\sqrt{2\hat{q}E}$) partons~\cite{Cao:2017hhk,He:2018xjv}. With the GW calculation, the spectrum of gluons emitted from a hard quark can be written as~\cite{Guo:2000nz,Majumder:2009ge}:
\begin{equation}
\label{eq:gluonSpectrumGW}
\frac{dN_g}{dz' dl_\perp^2 d\xi^-}=\frac{\alpha_S C_A P(z')}{2\pi l_\perp^4}\hat{q}(\xi^-)\left[2-2\cos\left(\frac{\xi^-}{\tau_f}\right)\right],
\end{equation}
where $z'=1-z$ is the fractional energy taken by the emitted gluon, $P(z')=(2-2z'+z'^2)/z'$ and $\tau_f=2Ez'(1-z')/l_\perp^2$. Now with our SCM kernel, the above spectra can be updated as
\begin{align}
\label{eq:gluonSpectrumSCM}
\frac{dN_g}{dz' dl_\perp^2 d\xi^-}&=\frac{\alpha_S C_A P(z')}{2\pi l_\perp^4}\hat{q}(\xi^-)\Bigg[\left(1-\frac{z'}{2}\right)\\
\times \Bigg\{2 -2&\cos\left(\frac{\xi^-}{\tau_f}\right)-2\left(\frac{\xi^-}{\tau_f}\right)\sin\left(\frac{\xi^-}{\tau_f}\right)\Bigg\} \nonumber\\
+ \left( 1 - n \right)& \left(\frac{\xi^-}{\tau_f}\right)^2\Bigg].\nonumber
\end{align}

Within the transport framework, one may evaluate the length dependence of the average accumulative parton energy loss as 
\begin{equation}
\label{eq:energyLoss}
\langle E_g \rangle (\xi^-) = \int_0^{ \xi^- } d\xi^- \int dz'\int dl_\perp^2 \;z'E\frac{dN_g}{dz' dl_\perp^2 d\xi^-}.
\end{equation}
A numerical evaluation is shown in Fig.~\ref{fig:Eg_vs_t} (for $n=0$), compared between using the GW gluon spectrum Eq.~\eqref{eq:gluonSpectrumGW} and the SCM spectrum Eq.~\eqref{eq:gluonSpectrumSCM}. Here, we calculate for a quark with a fixed energy at 100~GeV, and assume a fixed $\hat{q}= 1$~GeV$^2$/fm for jet-medium interaction. A lower cut-off of 1~GeV, around the thermal scale of the medium, is applied for both the energy ($xE$) and the transverse momentum ($l_\perp$) of the emitted gluon, while an upper limit of $\sqrt{2\hat{q}E}$ is enforced for $l_\perp^2$ to constrain the transport description of parton energy loss in the low virtuality regime. 

In Fig.~\ref{fig:Eg_vs_t}, one observes considerable difference in the path-length dependence of parton energy loss between GW and SCM calculations. Here the upper boundary of the SCM band, where $n=0$, is used to illustrate the significance of the energy loss when the SCM kernel is used. Instead of a quadratic length dependence predicted by GW, a cubic dependence could be reached with SCM after the $\vec{k}_\perp$-dependent phase factors combined with the processes allowing pre-emission rescattering are included in our full calculation. This may have a noticeable impact on jet quenching phenomenology that depends on the path-length dependence of parton energy loss, such as the anisotropic flow coefficients of jets and the system size dependence of jet quenching. We note that in this simplified semi-analytical estimation, one has not taken into account the variation of the parton energy while it emits gluons; effects of the dynamically expanding medium have not been included either. A full Monte-Carlo simulation of jet quenching through a realistic hydrodynamic medium is necessary for drawing reliable  conclusions on the phenomenological impacts of our improved medium modification kernel at the next-to-leading twist. This will be investigated in detail in an upcoming study.

\section{\label{sec:level5}Summary and Outlook}

Over the last two decades, there has been a considerable amount of research on jets and jet modification in a dense medium. There are now several formalisms that describe the modification of a hard jet as it propagates through a strongly interacting medium~\cite{Bass:2008rv,Majumder:2007iu}. Of these the higher twist formalism~\cite{Guo:2000nz,Wang:2001ifa,Majumder:2009ge} has been widely used in a variety of applications. 
Within this formalism, the medium induced single gluon emission rate at next-to-leading twist 
forms the underlying kernel of both the LBT~\cite{Cao:2016gvr,He:2015pra} and MATTER~\cite{Majumder:2013re,Cao:2017qpx} event generators. This kernel has also been used in several different 
applications prior to the advent of multi-stage Monte-Carlo generators~\cite{Majumder:2004pt,Deng:2009ncl,Chen:2011vt, Wang:2013cia, Cao:2016gvr, Cao:2017hhk, Majumder:2011uk,Qin:2009gw,Qin:2009uh,Majumder:2004pt,Majumder:2009zu}.

In spite of its wide prevalence, the medium induced single gluon emission rate at next-to-leading twist has 
accumulated a certain amount of dispute regarding the exact form that should be used: the original derivation by Guo and Wang (GW)~\cite{Guo:2000nz,Wang:2001ifa} presents a very simple form [Eq.~\eqref{Eqn:GW_kernel}] by ignoring all contributions that involve scattering prior to emission, and all terms that emanate from a $k_\perp$ expansion of the phase factor. These approximations, though somewhat unjustified, yielded a positive definite kernel, allowing the GW kernel to be reliably used as a probability distribution, which could then be sampled in a transport or Sudakov formalism. These approximations also simplified the number of diagrams that needed to be evaluated. Ignoring contributions from the phase in the $k_\perp$ Taylor expansion, the only non-vanishing topology was that of Fig.~\ref{fig:DoubleggScat1}, i.e., double gluon rescattering.

This interpretation was partially challenged by Aurenche, Zakharov and Zaraket (AZZ)~\cite{Aurenche:2008hm,Aurenche:2008mq}, who questioned the neglect of terms that arose from a $k_\perp$ expansion of the phase. In their reevaluation, AZZ reintroduced the terms from a Taylor expansion 
in terms of $k_\perp$ that arose from the phase and showed that these terms become dominant at lengths larger than one formation length, see Fig.~\ref{fig:Kernel_Comp}. The AZZ kernel [Eq.~\eqref{Eqn:AZZ_kernel}], was however only positive definite up to 1.3 times the formation length $\tau_f = 2 q^-/\mu^2$, and oscillates after that. The plots in Figs.~\ref{fig:Kernel_Comp} and~\ref{fig:Kernel_z_Comp} were evaluated at a fixed light-cone momentum $q^-$, and fixed momentum fraction $z$ and transverse momentum $l_\perp$ of the radiated gluon [with $\mu^2 =  l_\perp^2 / { \{z(1-z)\} }$]. The oscillatory nature of the AZZ kernel prevents it from being used as a probability distribution, and complicates the application of the medium induced single gluon emission rate at next-to-leading twist in any simulation framework~\cite{Majumder:2013re}.

While the assertion of AZZ regarding the dominance of contributions from the phase terms is indeed correct, it is incomplete. In the current paper, we have demonstrated that contributions from phase terms arise from several diagrams with different topology,~Figs.~\ref{fig:DoubleggScat1},~\ref{fig:qgAndggScat1},~\ref{fig:qgAndggScat2},~\ref{fig:qgAndggScat3},~\ref{fig:qgAndggScat4},~\ref{fig:qgAndggScatLeft},~\ref{fig:qgAndggScatRight}. In their analysis, AZZ restricted their terms to only those that arise from Fig.~\ref{fig:DoubleggScat1}, again without justification. 
The evaluation of all these diagrams is presented in 
Sect.~\ref{sec:level3} along with several other central cut contributions that yielded vanishing contributions on Taylor expansion. Most of the diagrams with left and right cuts also yield vanishing contributions, and are evaluated in  Appendix~\ref{sec:appendix1}. The inclusion of all these contributions resurrects the positive definite nature of the final kernel [Eq.~\eqref{Eqn:SCM_kernel}]. We point out yet again that the inclusion of all possible contributions reduces the final result to the square of a complete matrix element which should be positive definite. Increasing the length integration $\xi^-$ in Eq.~\eqref{Eqn:SCM_kernel}, should not turn negative within a few formation lengths, as this would imply that the matrix element is not positive definite (assuming a limit where the second derivative in $k_\perp$ gives a much larger contribution than the fourth derivative and so on).

The inclusion of all diagrams, in the full calculation of the medium induced single gluon emission rate at next-to-leading twist, not only yields a positive definite result, it yields a result that grows swiftly with the 
distance at which the single gluon is emitted. This not only allows one to use such rates in both a Sudakov based formalism, as well as in a transport calculation, but the growing rate clearly outlines a distance beyond which this 
formalism may no longer be used and must be replaced by a multiple scattering induced emission formalism, such as the Baier-Dokshitzer-Mueller-Peigne-Schiff~\cite{Baier:1994bd,Baier:1996kr,Baier:1996sk}, Zakharov~\cite{Zakharov:1996fv,Zakharov:1997uu}, or Arnold-Moore-Yaffe~\cite{Arnold:2001ba,Arnold:2001ba,Arnold:2002ja} formalism. The current paper, allows a future determination of this transition, from few to many scatterings per emission, much more straightforward. 

In spite of the inclusion of several new terms and the positivity of the kernel, the final results are somewhat different from those of GW. This difference is not expected to make much of a difference in a Sudokov like formalism where the rates form GW and the current paper are comparable at a few times of the formation length. However the different functional shapes of the kernels lead to a considerably different gluon emission rate as a function of distance between the current work and that of GW (Fig.~\ref{fig:Eg_vs_t}). The much stronger length dependence of 
the energy loss is expected to introduce updates to many of the transport and earlier non-event-generator based efforts. It is expected that this new kernel will lead to a larger azimuthal anisotropy, consistent with current data 
on jet and leading hadron azimuthal anisotropy. However, the qualitatively different length dependence is expected to complicate the similarity between the LBT and MARTINI generators. Phenomenological calculations with the current kernel will be carried out in future efforts.

\begin{acknowledgements}
The authors thank Guang-You Qin for pointing out an error in an earlier version of this manuscript. This work was supported in part by the U.S. Department of Energy (DOE) under grant number DE-SC001346, in part by the National Science Foundation (NSF) under grant number ACI-1550300 within the framework of the JETSCAPE collaboration, and in part by the National Natural Science Foundation of China (NSFC) under grant numbers 12175122 and 2021-867.
\end{acknowledgements}

\appendix

\section{\label{sec:appendix1}Phase space constraint in the non-central cut diagrams}

In this study, for the first time, the non-vanishing $k_\perp$-dependent phase factors from all Feynman diagrams are taken into account. This included $k_\perp$ dependent terms from both the magnitude and the phase portions of each diagram. Beyond the work of GW and AZZ, an additional six contributing diagrams were identified. While five of these contributing diagrams are central cut diagrams, the other two diagrams are non central cut diagrams.

\begin{figure}[tb]
    \centering
    \includegraphics[width=0.8\linewidth]{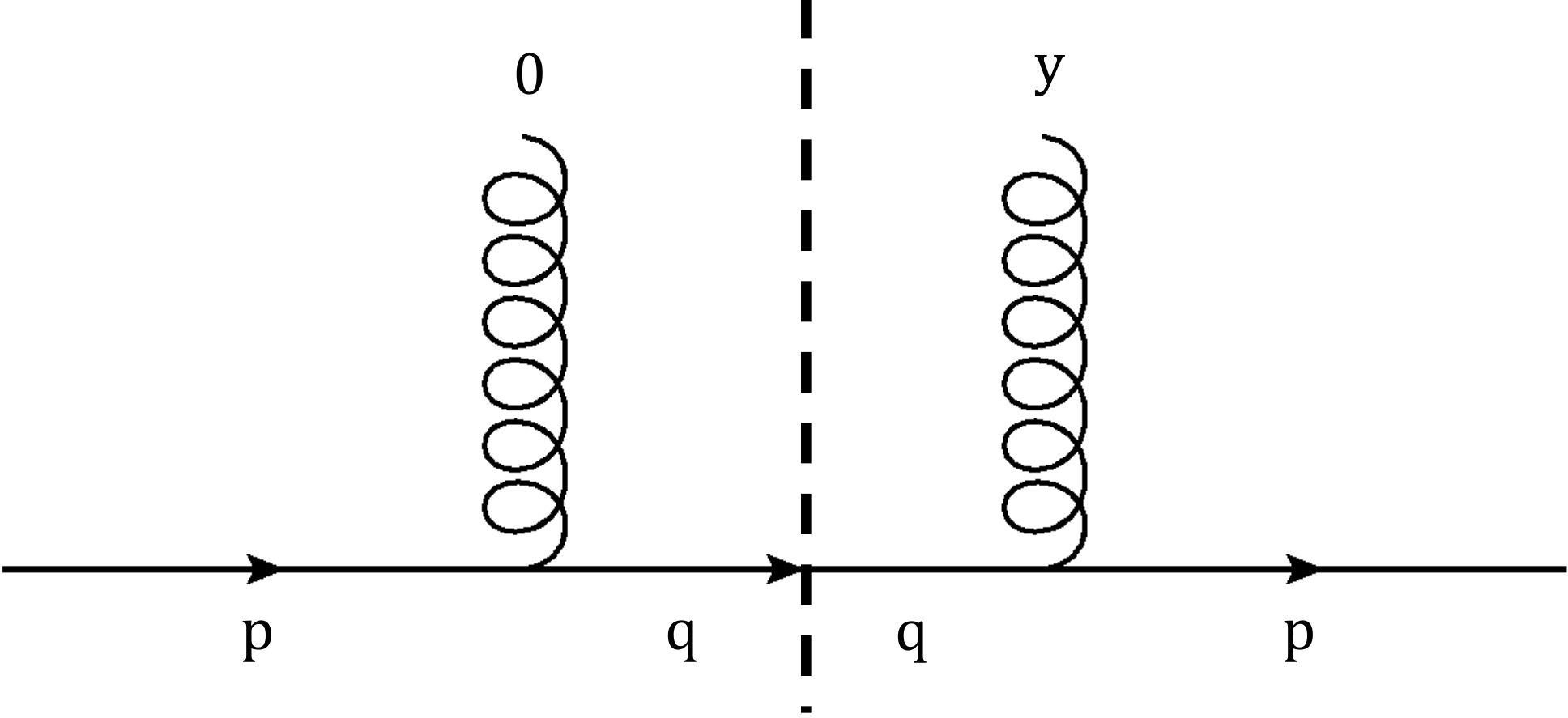}
    \caption{Feynman diagram for quark traveling in positive $z$-direction, scattering by two gluons with the cut line at the center between two scatterings}
    \label{fig:AA_corr_C}
\end{figure}

\begin{figure}[tb]
    \centering
    \includegraphics[width=0.8\linewidth]{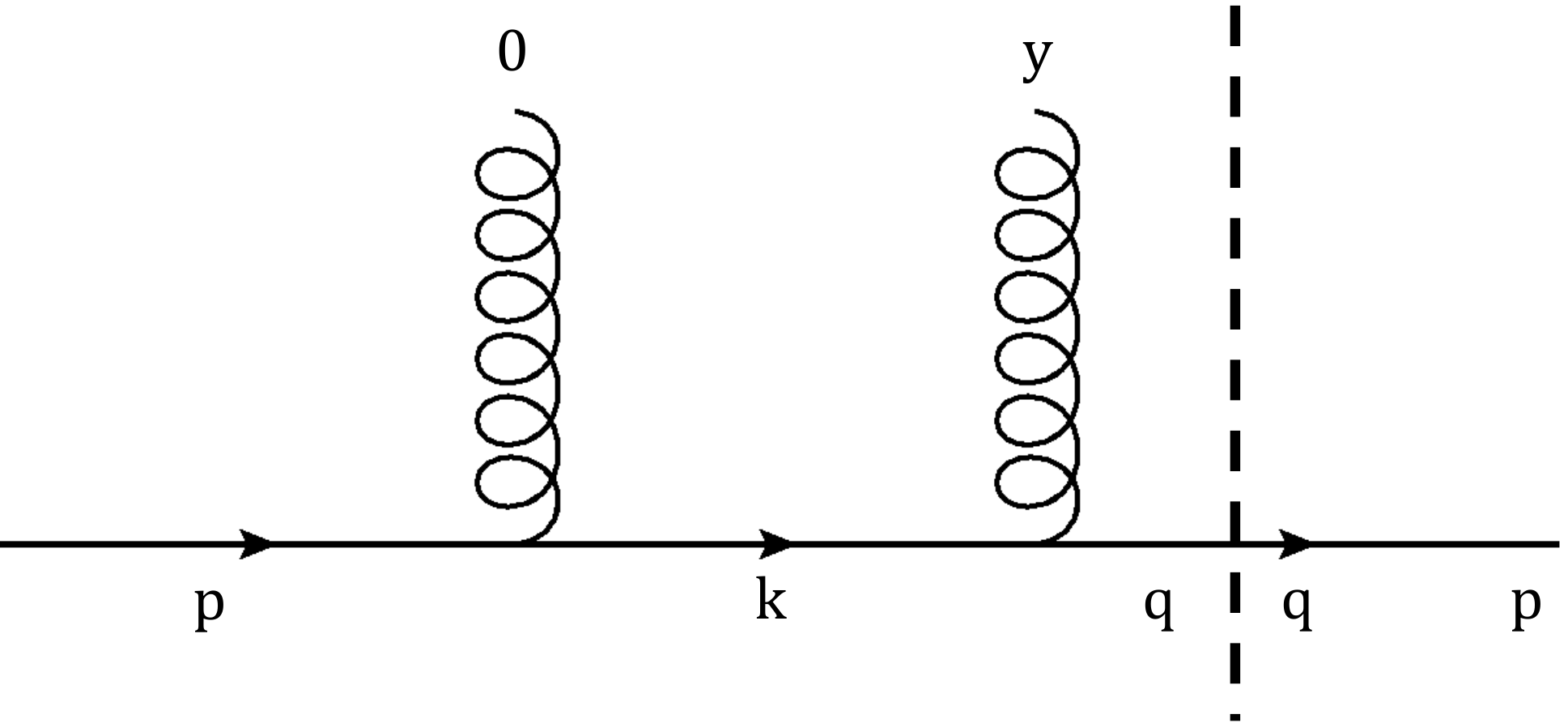}
    \caption{Feynman diagram for quark traveling in positive $z$-direction, scattering by two gluons with the cut line at the external quark line at the right hand side two scatterings}
    \label{fig:AA_corr_R}
\end{figure}

All terms depend on a 4-point matrix element given in Eq.~\eqref{Eqn:MatrixElementFactorization}, where we have assumed a factorization into two 2-point matrix elements. While the matrix element involving quark operators was approximated as the initial parton distribution function for the struck quark, the gluon field-strength$-$field-strength correlator $\langle P | F^{+\mu} (y) F^+_\mu (0) | P \rangle$  represents the rescattering of the quark off the gluon field of nucleons along its path. In all prior efforts, only contributions from central cuts were considered, hence there was no reason to study the difference between the gluon matrix element in central and non-central cut diagrams. These are not the same as illustrated in Figs.~\ref{fig:AA_corr_C},\ref{fig:AA_corr_R}.

The incoming nucleon contains a collection of quarks and gluons. In order to gain a simple understanding of the difference between the two matrix elements, we take the extreme case of considering the gluons emitting off the same quark in the target. 
In the case of a quark travelling in positive $z$-direction, scattering off two in-medium gluons, we can have the cut line either in between two scatterings as shown in Fig. \ref{fig:AA_corr_C} or in one of the external quark line as shown in Fig. \ref{fig:AA_corr_R}. Since field-strength-field-strength correlator $\langle p | F^{+ \mu}(y) F^+_\mu(0) | p \rangle$ is related to the derivatives of the vector-potential-vector-potential correlator $\langle p | A^{+}(y) A^+(0) | p \rangle$, we evaluate only the vector potential correlator in this section. 

For the case of central cut, we can calculate the vector-potential correlator by using the Feynman diagram given in Fig. \ref{fig:AA_corr_C},
\begin{align}
    \label{eq:AA_corr_C}
    \langle P | A^+ (y) A^+ (0) | P \rangle = \int \frac{ d q^+ d^2 q_\perp }{ \left( 2 \pi \right) 2 q^+ } 4 p^+ q^+ \frac{ e^{ i \left( p - q \right) \cdot y } }{ \left( p - q \right)^4 }.
\end{align}

If we only consider light-cone dynamics, i.e., the correlator only depends on $y^+$ and $y^-$, and assume that there is no constraint in the phase space of $y_\perp$, the vector potential correlator for the right cut diagram given in Fig. \ref{fig:AA_corr_R} can be calculated as,
\begin{align}
    \label{eq:AA_corr_R}
    \langle P | A^+ (y) A^+ (0) | P \rangle = \int \frac{ d k^+ d^2 k_\perp }{ \left( 2 \pi \right) 2 k^+ } 4 p^+ k^+ \frac{ e^{ i \left( p - k \right) \cdot y } }{ \left( p - k \right)^4 } \theta \left( y^+ \right).
\end{align}

If we change $k$ to $q$ in Eq. \ref{eq:AA_corr_R}, the only difference between Eq. \ref{eq:AA_corr_C} and Eq. \ref{eq:AA_corr_R} is the Heaviside function, $\theta \left( y^+ \right)$. Recall that in the evaluation of the 4-point functions, one encounters a delta-function in $y^+$, which now yields $\int dy^+ \delta(y^+) \theta(y^+) = 1/2$. The Heaviside function shrinks the phase space of $y^+$ by an overall factor of $1/2$. 

Since the assumption of scattering off a single quark is an extreme case and we ignore the fact that there can be further phase space constraints on $y_\perp$ as well, the factor $1/2$ modulates the maximum value we can get for the non central cut diagrams. Therefore it is convenient to show our result as a band between two extreme cases where the overall factor modulating the left and right cut diagrams lies between $0$ and $1/2$.
Therefore, instead of using exact numerical factor for the non central cut result, we use a variable factor $n$, where $0 \leq n \leq 1/2$.

\section{\label{sec:appendix2} Non-contributing diagrams}

There are ten possible non-central cut diagrams associated with the next-to-leading twist DIS calculation, two of which have been calculated in detail in Sect.~\ref{subsec:level3.3}. In this appendix, we present a summary of calculations for the remaining eight diagrams. 

\begin{figure}[tbp]
    \addtolength{\abovecaptionskip}{-3mm}
%
	\centering
	\includegraphics[width=\linewidth]{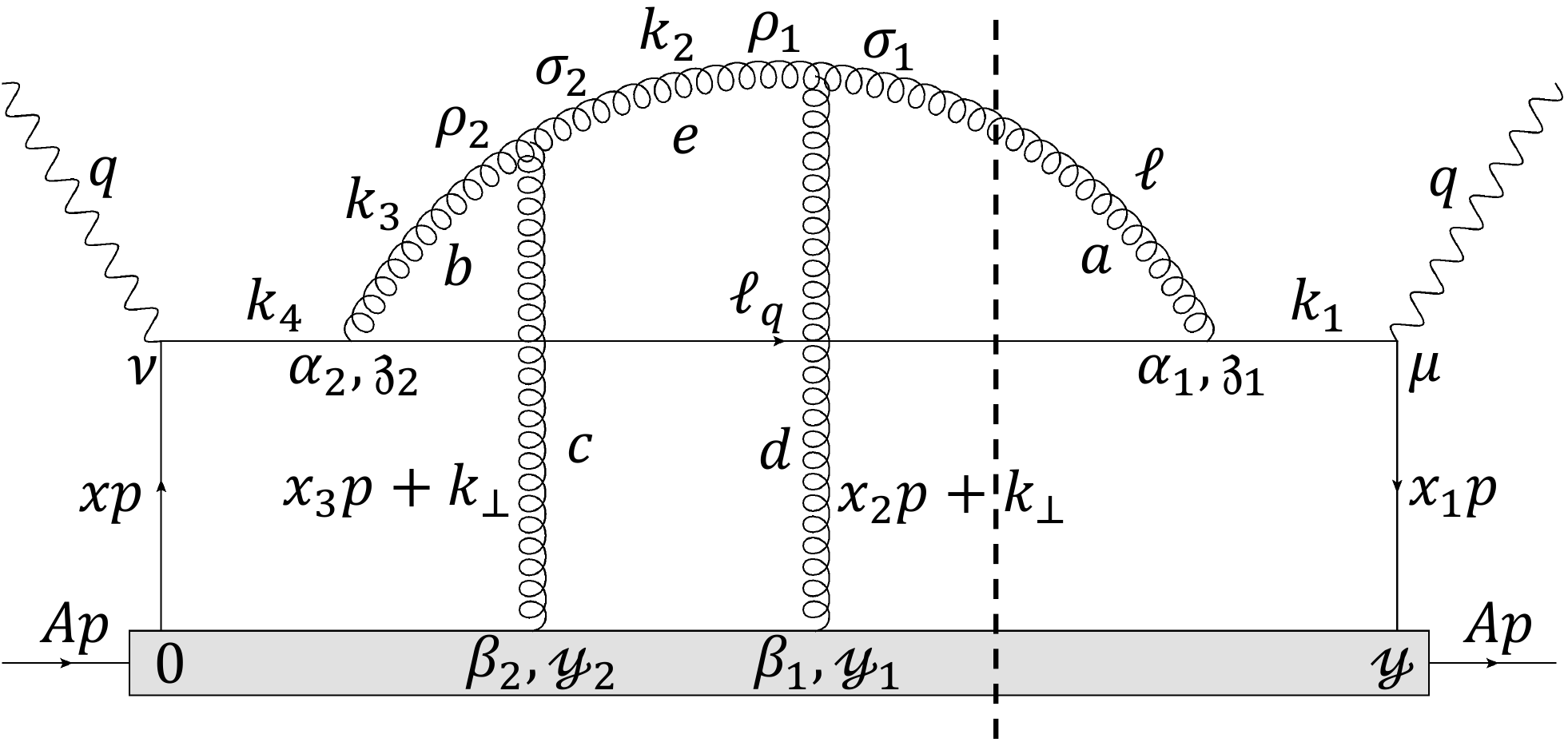}
	\caption{Feynman diagram for the gluon-gluon double rescattering process with right cut.}
	\label{fig:ggDoubleScatRight}
\end{figure}

\begin{figure}[tbp]
    \addtolength{\abovecaptionskip}{-3mm}
%
	\centering
	\includegraphics[width=\linewidth]{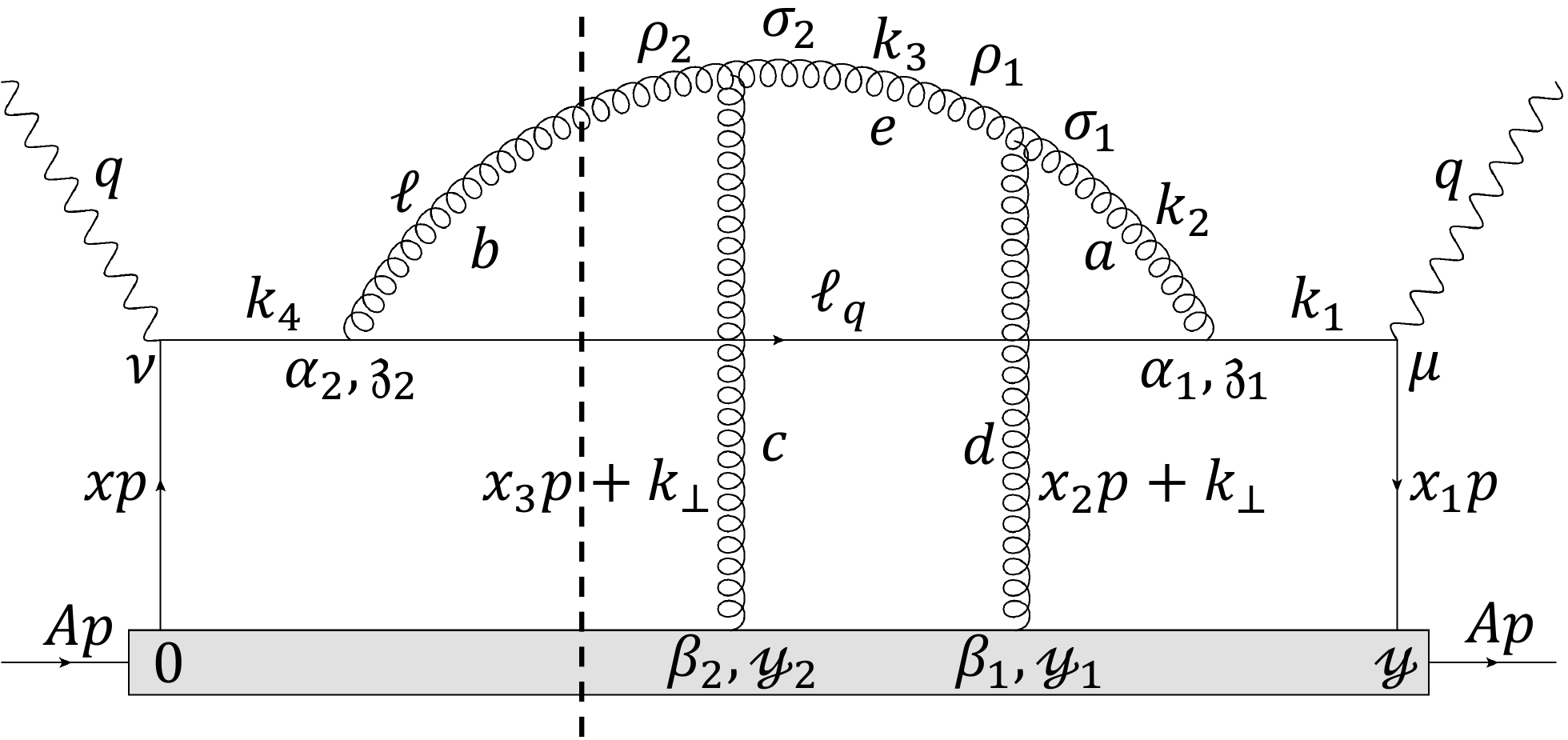}
	\caption{Feynman diagram for the gluon-gluon double rescattering process with left cut.}
	\label{fig:ggDoubleScatLeft}
\end{figure}

We first start with post-emission rescattering diagrams. The central cut diagram for double gluon-gluon rescattering (Fig.~\ref{fig:DoubleggScat1}) has two non-central cut companions, as illustrated in Fig.~\ref{fig:ggDoubleScatRight} (right cut) and Fig.~\ref{fig:ggDoubleScatLeft} (left cut). They are complex conjugate of each other. The hadronic tensor for the former can be written as
%
%
\begin{eqnarray}
\label{Eqn:ggDoubleScatRight_W}
\mathcal{W}^{\mu\nu}_{\ref{fig:ggDoubleScatRight}}  & = & - \frac{1}{2\pi} \int \frac{dy^-}{2\pi} dy_1^- dy_2^- \frac{ d^2y_\perp }{ \left( 2\pi \right)^2 } 
d^2k_\perp \int dz \int dx \nonumber\\
& \times & \left( 2\pi \right) \delta \left[ \left( q + xp \right)^2 \right] \frac{ e_q^2 }{2} 
\Tr \left[ p . \gamma \gamma^\mu \left( q + xp \right) . \gamma \gamma^\nu \right] \frac{ n }{ 2 } \nonumber\\
& \times & \left\langle A \middle| \bar{ \psi } \left( y^- \right) \gamma^+ A^+ \left( y_1^-, \vec{ y }_\perp \right) 
 A^+ \left( y_2^-, \vec{ 0 }_\perp \right) \psi \left( 0 \right) \middle| A \right\rangle  \nonumber\\
& \times & e^{ i \vec{ k }_\perp \cdot \vec{ y }_\perp } \int \frac{ d \ell_\perp^2 }{ \ell_\perp^2 } \frac{ \alpha_s }{ \left( 2\pi \right) } C_F \frac{ 1 + z^2 }{ 1 - z }
\frac{ 2 \pi \alpha_s }{ N_C }  \theta \left(y_2^- \right) \nonumber\\
& \times & \theta \left( y_1^- - y_2^- \right) e^{ -i \left( x_B + x_L \right) p^+ y^- + i  x_D^+ p^+ \left( y_1^- - y_2^- \right)} \nonumber\\
& \times & e^{ -i \frac{ x_D^+ p^+ \left( y_1^- - y_2^- \right) }{ \left( 1 - z \right)} } \left[ 1 - e^{ i x_L p^+ y_2^- } \right],
\end{eqnarray}
while the latter as
%
%
\begin{eqnarray}
\label{Eqn:ggDoubleScatLeft_W}
\mathcal{W}^{\mu\nu}_{\ref{fig:ggDoubleScatLeft}}  & = & - \frac{1}{2\pi} \int \frac{dy^-}{2\pi} dy_1^- dy_2^- \frac{ d^2y_\perp }{ \left( 2\pi \right)^2 } 
d^2k_\perp \int dz \int dx \nonumber\\
& \times & \left( 2\pi \right) \delta \left[ \left( q + xp \right)^2 \right] \frac{ e_q^2 }{2} 
\Tr \left[ p . \gamma \gamma^\mu \left( q + xp \right) . \gamma \gamma^\nu \right] \frac{ n }{ 2 } \nonumber\\
& \times & \left\langle A \middle| \bar{ \psi } \left( y^- \right) \gamma^+ A^+ \left( y_1^-, \vec{ y }_\perp \right) 
 A^+ \left( y_2^-, \vec{ 0 }_\perp \right) \psi \left( 0 \right) \middle| A \right\rangle  \nonumber\\
& \times & e^{ i \vec{ k }_\perp \cdot \vec{ y }_\perp } \int \frac{ d \ell_\perp^2 }{ \ell_\perp^2 } \frac{ \alpha_s }{ \left( 2\pi \right) } C_F \frac{ 1 + z^2 }{ 1 - z }
\frac{ 2 \pi \alpha_s }{ N_C } \theta \left( y_2^- - y_1^- \right)  \nonumber\\
& \times & \theta \left(y_1^- - y^- \right) e^{ -i \left( x_B + x_L \right) p^+ y^- + i  x_D^+ p^+ \left( y_1^- - y_2^- \right)} \nonumber\\
& \times & e^{ -i \frac{ x_D^+ p^+ \left( y_1^- - y_2^- \right) }{ \left( 1 - z \right)} } 
\left[ 1 - e^{ i x_L p^+ \left( y^- - y_1^- \right) } \right] .
\end{eqnarray}
If one sets $ y^- = 0 $ and $ y_1^- = y_2^- = \xi^- $, both Eqs.~\eqref{Eqn:ggDoubleScatRight_W} and~\eqref{Eqn:ggDoubleScatLeft_W} above become independent of $ \vec{k}_\perp $. This leads to zero derivative with respect to $ \vec{k}_\perp $ during collinear expansion. Therefore, there is no next-to-leading twist contribution from these two non-central double gluon-gluon rescattering diagrams.

\begin{figure}[tbp]
    \addtolength{\abovecaptionskip}{-3mm}
%
	\centering
	\includegraphics[width=\linewidth]{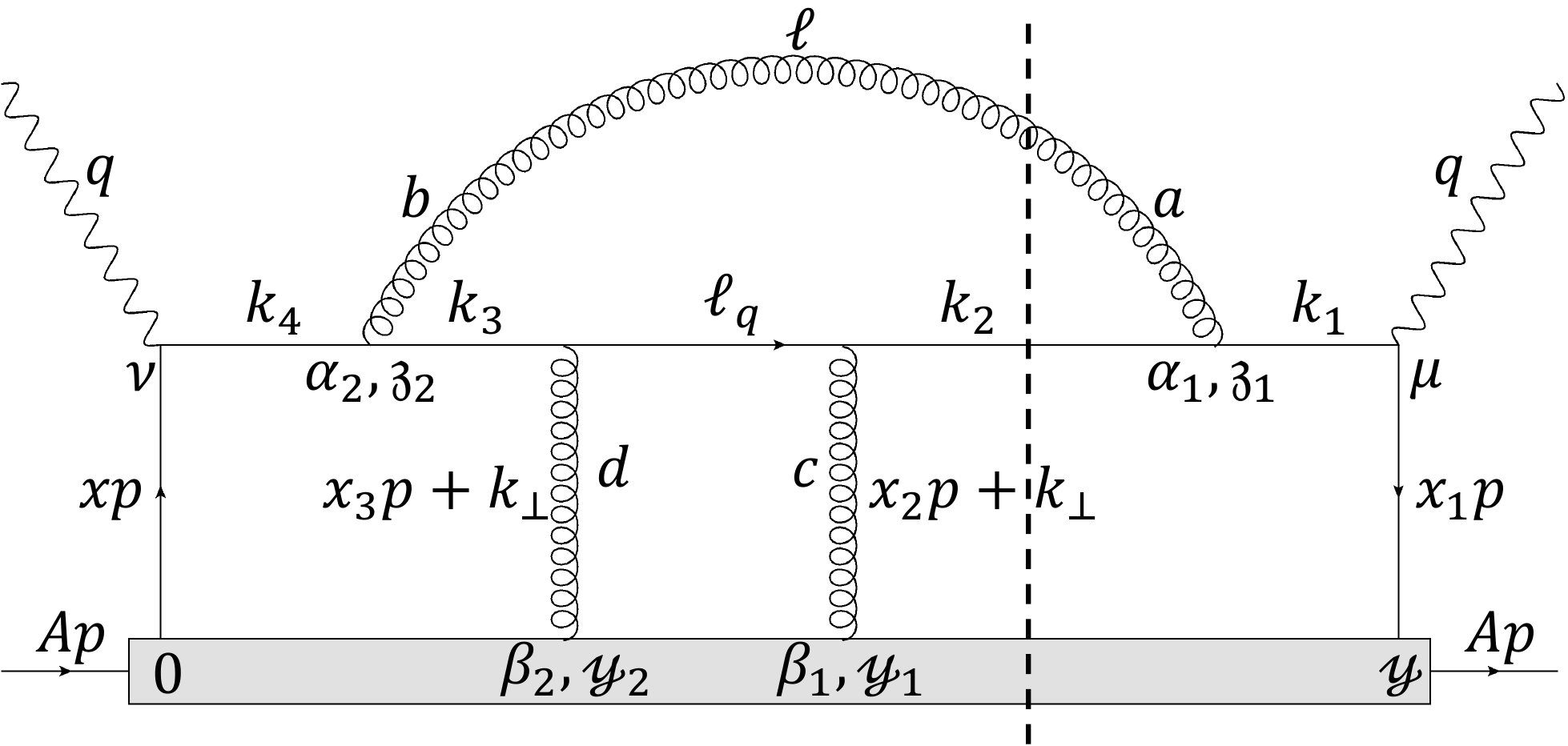}
	\caption{Feynman diagram for the quark-gluon double rescattering process with right cut.}
	\label{fig:qgDoubleScatRight}
\end{figure}

\begin{figure}[tbp]
    \addtolength{\abovecaptionskip}{-3mm}
%
	\centering
	\includegraphics[width=\linewidth]{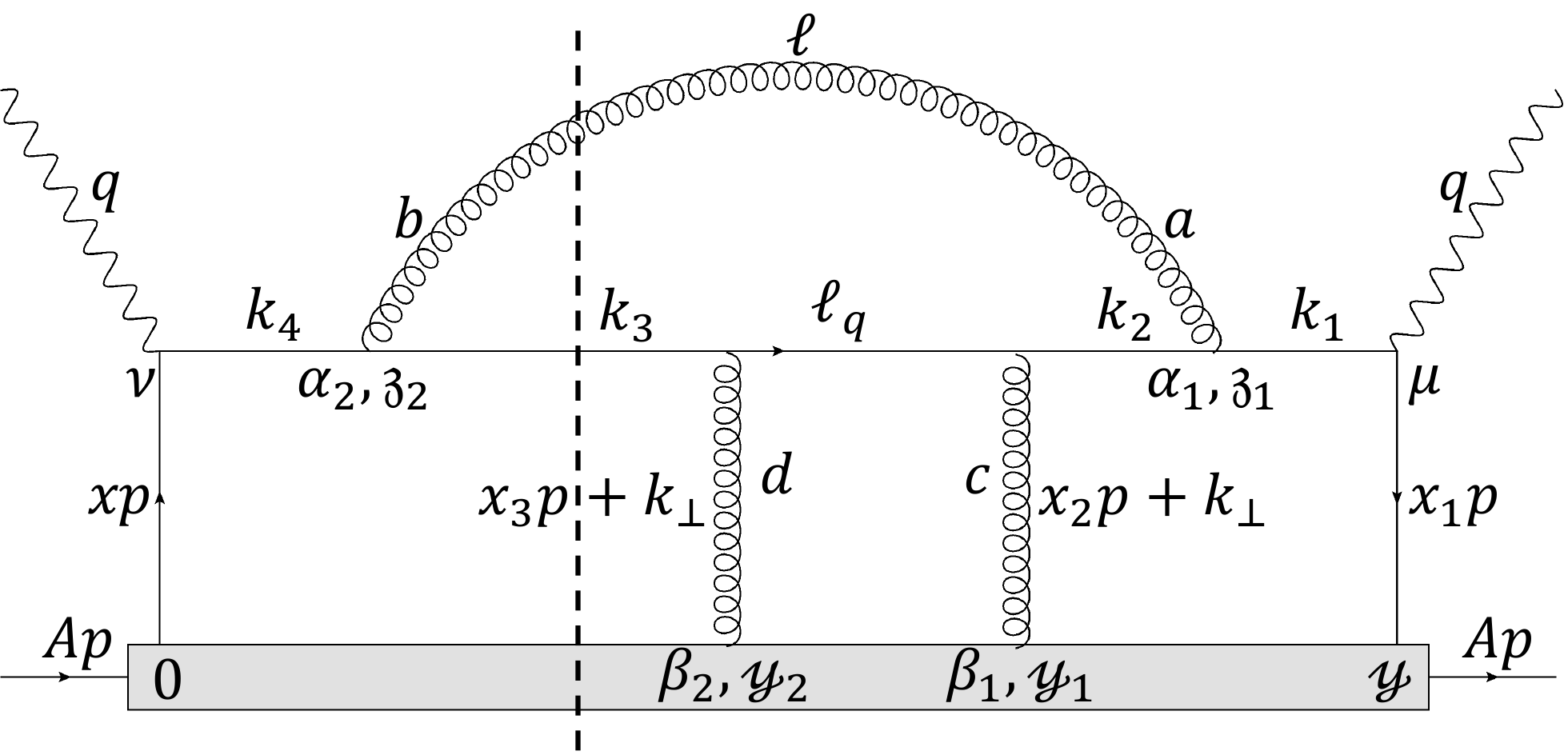}
	\caption{Feynman diagram for the quark-gluon double rescattering process with left cut.}
	\label{fig:qgDoubleScatLeft}
\end{figure}

Similarly, the central cut post-emission double quark-gluon rescattering diagram (Fig.~\ref{fig:qgDoubleScat1}) has two non-central cut companions, as illustrated in Fig.~\ref{fig:qgDoubleScatRight} (right cut) and Fig.~\ref{fig:qgDoubleScatLeft} (left cut), being complex conjugate of each other. The hadronic tensor of the former reads
%
%
\begin{eqnarray}
\label{Eqn:qgDoubleScatRight_W}
\mathcal{W}^{\mu\nu}_{\ref{fig:qgDoubleScatRight}}  & = & - \frac{1}{2\pi} \int \frac{dy^-}{2\pi} dy_1^- dy_2^- \frac{ d^2y_\perp }{ \left( 2\pi \right)^2 } 
d^2k_\perp \int dz \int dx \nonumber\\
& \times & \left( 2\pi \right) \delta \left[ \left( q + xp \right)^2 \right] \frac{ e_q^2 }{2} 
\Tr \left[ p . \gamma \gamma^\mu \left( q + xp \right) . \gamma \gamma^\nu \right] \frac{ n }{ 2 } \nonumber\\
& \times & \left\langle A \middle| \bar{ \psi } \left( y^- \right) \gamma^+ A^+ \left( y_1^-, \vec{ y }_\perp \right) 
 A^+ \left( y_2^-, \vec{ 0 }_\perp \right) \psi \left( 0 \right) \middle| A \right\rangle  \nonumber\\
& \times & e^{ i \vec{ k }_\perp \cdot \vec{ y }_\perp } \int \frac{ d \ell_\perp^2 }{ \ell_\perp^2 } \frac{ \alpha_s }{ \left( 2\pi \right) } C_F \frac{ 1 + z^2 }{ 1 - z } \frac{ 2 \pi \alpha_s }{ N_C }  \nonumber\\
& \times & e^{ -i \left( x_B + x_L \right) p^+ y^- - i  x_D p^+ \left( y_1^- - y_2^- \right)} \theta \left( y_1^- - y_2^- \right) \nonumber\\
& \times & \theta \left(y_2^- \right) \left[ 1 - e^{ - i x_L p^+ y_2^- } \right],
\end{eqnarray}
while the latter reads
%
%
\begin{eqnarray}
\label{Eqn:qgDoubleScatLeft_W}
\mathcal{W}^{\mu\nu}_{\ref{fig:qgDoubleScatLeft}}  & = & - \frac{1}{2\pi} \int \frac{dy^-}{2\pi} dy_1^- dy_2^- \frac{ d^2y_\perp }{ \left( 2\pi \right)^2 } 
d^2k_\perp \int dz \int dx \nonumber\\
& \times & \left( 2\pi \right) \delta \left[ \left( q + xp \right)^2 \right] \frac{ e_q^2 }{2} 
\Tr \left[ p . \gamma \gamma^\mu \left( q + xp \right) . \gamma \gamma^\nu \right] \frac{ n }{ 2 } \nonumber\\
& \times & \left\langle A \middle| \bar{ \psi } \left( y^- \right) \gamma^+ A^+ \left( y_1^-, \vec{ y }_\perp \right) 
 A^+ \left( y_2^-, \vec{ 0 }_\perp \right) \psi \left( 0 \right) \middle| A \right\rangle  \nonumber\\
& \times & e^{ i \vec{ k }_\perp \cdot \vec{ y }_\perp } \int \frac{ d \ell_\perp^2 }{ \ell_\perp^2 } \frac{ \alpha_s }{ \left( 2\pi \right) } C_F \frac{ 1 + z^2 }{ 1 - z } \frac{ 2 \pi \alpha_s }{ N_C }  \nonumber\\
& \times & e^{ -i \left( x_B + x_L \right) p^+ y^- - i  x_D p^+ \left( y_1^- - y_2^- \right)} \theta \left( y_2^- - y_1^- \right) \nonumber\\
& \times & \theta \left(y_1^- - y^- \right) \left[ 1 - e^{ - i x_L p^+ \left( y^- - y_1^- \right) } \right] .
\end{eqnarray}
Again, neither Eq.~\eqref{Eqn:qgDoubleScatRight_W} nor Eq.~\eqref{Eqn:qgDoubleScatLeft_W} depends on $\vec{ k }_\perp$ after we take $ y^- = 0 $ and $ y_1^- = y_2^- = \xi^- $. Therefore, the collinear expansion yields zero at the next-to-leading twist and these two diagrams have no contribution to our final medium modification kernel.

The four diagrams above, together with the two presented in Sect.~\ref{subsec:level3.3}, completes the six possibilities of non-central cut diagrams for post-emission rescattering process. There are four non-central cut diagrams left that allow pre-emission rescattering. Each of them corresponds to one central cut diagram from Fig.~\ref{fig:qgAndggScat3} to Fig.~\ref{fig:qgDoubleScat3}, in which one scattering happens before and the other after the gluon emission. When both scatterings happen before the emission (Fig.~\ref{fig:qgDoubleScat4}), the left or right cut diagram is not associated with the single gluon emission picture, but merely a virtual correction to the leading order diagram with no emission.

\begin{figure}[tbp]
    \addtolength{\abovecaptionskip}{-3mm}
%
	\centering
	\includegraphics[width=\linewidth]{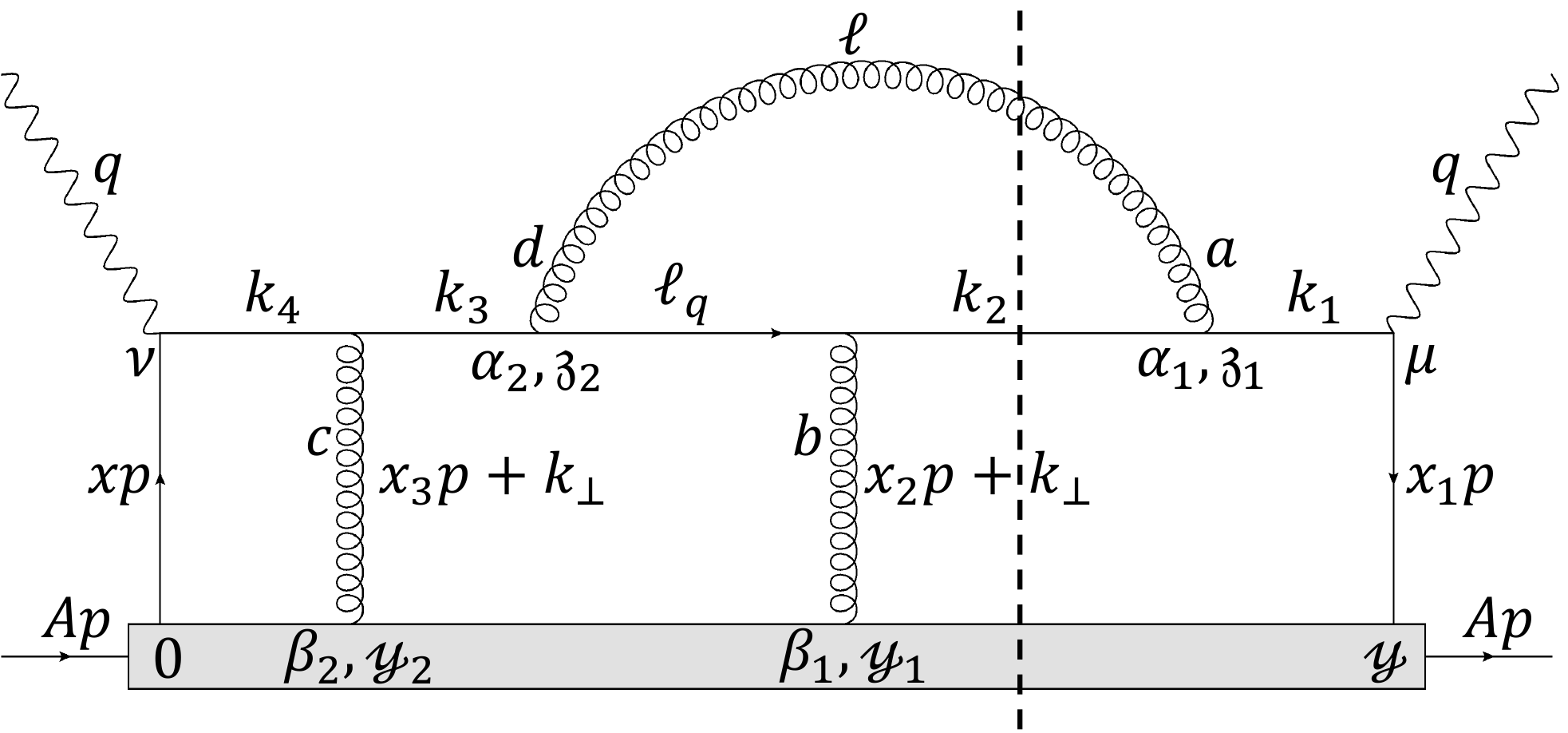}
	\caption{Feynman diagram for the double quark-gluon rescattering process with only one scattering happens before emission with right cut.}
	\label{fig:qgBeforeAfterDoubleScatRight}
\end{figure}

\begin{figure}[tbp]
    \addtolength{\abovecaptionskip}{-3mm}
%
	\centering
	\includegraphics[width=\linewidth]{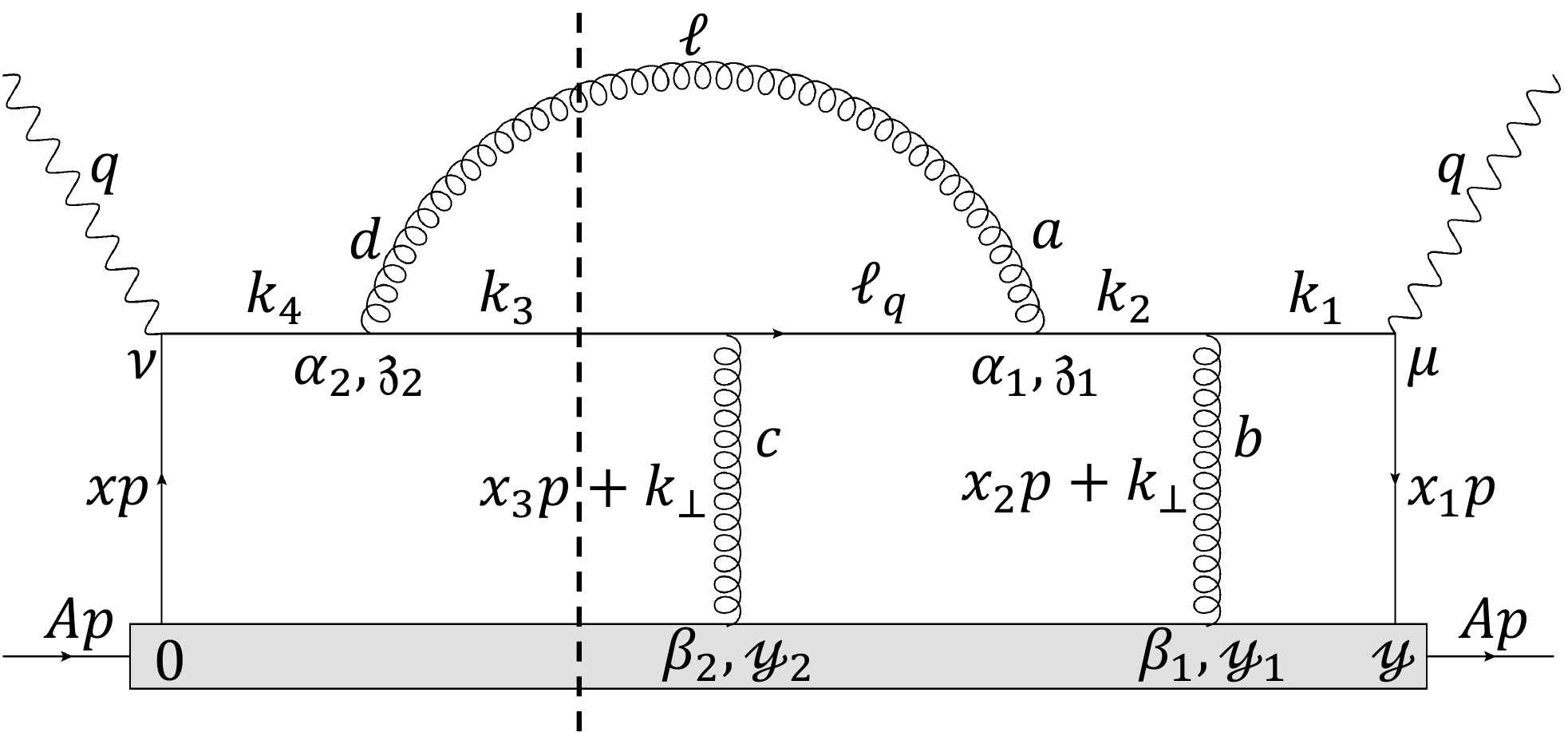}
	\caption{Feynman diagram for the double quark-gluon rescattering process with only one scattering happening before emission with left cut.}
	\label{fig:qgAfterBeforeDoubleScatLeft}
\end{figure}

We now consider the two non-central cut diagrams for double quark-gluon rescatterings with one scattering before the gluon emission, as shown in Figs.~\ref{fig:qgBeforeAfterDoubleScatRight} and~\ref{fig:qgAfterBeforeDoubleScatLeft}. They are complex conjugate of each other.

The hadronic tensor associated with Fig.~\ref{fig:qgBeforeAfterDoubleScatRight} can be expressed as
%
%
\begin{eqnarray}
\label{Eqn:qgBeforeAfterDoubleScatRight_W}
\mathcal{W}^{\mu\nu}_{\ref{fig:qgBeforeAfterDoubleScatRight}}  & = & \frac{1}{2\pi} \int \frac{dy^-}{2\pi} dy_1^- dy_2^- \frac{ d^2y_\perp }{ \left( 2\pi \right)^2 } 
d^2k_\perp \int dz \int dx \nonumber\\
& \times & \left( 2\pi \right) \delta \left[ \left( q + xp \right)^2 \right] \frac{ e_q^2 }{2} 
\Tr \left[ p . \gamma \gamma^\mu \left( q + xp \right) . \gamma \gamma^\nu \right] \frac{ n }{ 2 } \nonumber\\
& \times & \left\langle A \middle| \bar{ \psi } \left( y^- \right) \gamma^+ A^+ \left( y_1^-, \vec{ y }_\perp \right) 
 A^+ \left( y_2^-, \vec{ 0 }_\perp \right) \psi \left( 0 \right) \middle| A \right\rangle  \nonumber\\
& \times & e^{ i \vec{ k }_\perp \cdot \vec{ y }_\perp } \int d \ell_\perp^2 \frac{ \vec{ \ell }_\perp \cdot \left( \vec{ \ell }_\perp - \left( 1 - z \right) \vec{ k }_\perp \right) } 
{ \ell_\perp^2 \left( \vec{ \ell }_\perp - \left( 1 - z \right) \vec{ k }_\perp \right)^2 } \frac{ \alpha_s }{ \left( 2\pi \right) } C_F \nonumber\\
& \times & \frac{ 1 + z^2 }{ 1 - z } \frac{ 2 \pi \alpha_s }{ N_C } e^{ -i \left( x_B + x_L \right) p^+ y^- - i  x_D p^+ \left( y_1^- - y_2^- \right)} \nonumber\\
& \times & \theta \left(y_2^- \right) \theta \left( y_1^- - y_2^- \right) e^{ i x_L p^+ y_2^- } \nonumber\\
& \times & \left[ 1 - e^{ - i \left( x_D^0 - x_D - x_L \right) p^+ \left( y_1^- - y_2^- \right) } \right],
\end{eqnarray}
in which another new symbol of momentum fraction, $ x_D^0 = { k_\perp^2 }/{ (2 p^+ q^-) }$, has been defined to simplify our result. Similarly, the hadronic tensor corresponding to Fig.~\ref{fig:qgAfterBeforeDoubleScatLeft} can be written as
%
%
\begin{eqnarray}
\label{Eqn:qgAfterBeforeDoubleScatLeft_W}
\mathcal{W}^{\mu\nu}_{\ref{fig:qgAfterBeforeDoubleScatLeft}}  & = & \frac{1}{2\pi} \int \frac{dy^-}{2\pi} dy_1^- dy_2^- \frac{ d^2y_\perp }{ \left( 2\pi \right)^2 } 
d^2k_\perp \int dz \int dx \nonumber\\
& \times & \left( 2\pi \right) \delta \left[ \left( q + xp \right)^2 \right] \frac{ e_q^2 }{2} 
\Tr \left[ p . \gamma \gamma^\mu \left( q + xp \right) . \gamma \gamma^\nu \right] \frac{ n }{ 2 } \nonumber\\
& \times & \left\langle A \middle| \bar{ \psi } \left( y^- \right) \gamma^+ A^+ \left( y_1^-, \vec{ y }_\perp \right) 
 A^+ \left( y_2^-, \vec{ 0 }_\perp \right) \psi \left( 0 \right) \middle| A \right\rangle  \nonumber\\
& \times & e^{ i \vec{ k }_\perp \cdot \vec{ y }_\perp } \int d \ell_\perp^2 \frac{ \vec{ \ell }_\perp \cdot \left( \vec{ \ell }_\perp - \left( 1 - z \right) \vec{ k }_\perp \right) } 
{ \ell_\perp^2 \left( \vec{ \ell }_\perp - \left( 1 - z \right) \vec{ k }_\perp \right)^2 } \frac{ \alpha_s }{ \left( 2\pi \right) } C_F \nonumber\\
& \times & \frac{ 1 + z^2 }{ 1 - z } \frac{ 2 \pi \alpha_s }{ N_C } e^{ -i \left( x_B + x_L \right) p^+ y^- - i  x_D p^+ \left( y_1^- - y_2^- \right)} \nonumber\\
& \times & \theta \left( y_1^- - y^- \right) \theta \left( y_2^- - y_1^- \right) e^{ i x_L p^+ \left( y^- - y_1^- \right) } \nonumber\\
& \times & \left[ 1 - e^{ - i \left( x_D^0 - x_D - x_L \right) p^+ \left( y_1^- - y_2^- \right) } \right] .
\end{eqnarray}

\begin{figure}[tbp]
    \addtolength{\abovecaptionskip}{-3mm}
%
	\centering
	\includegraphics[width=\linewidth]{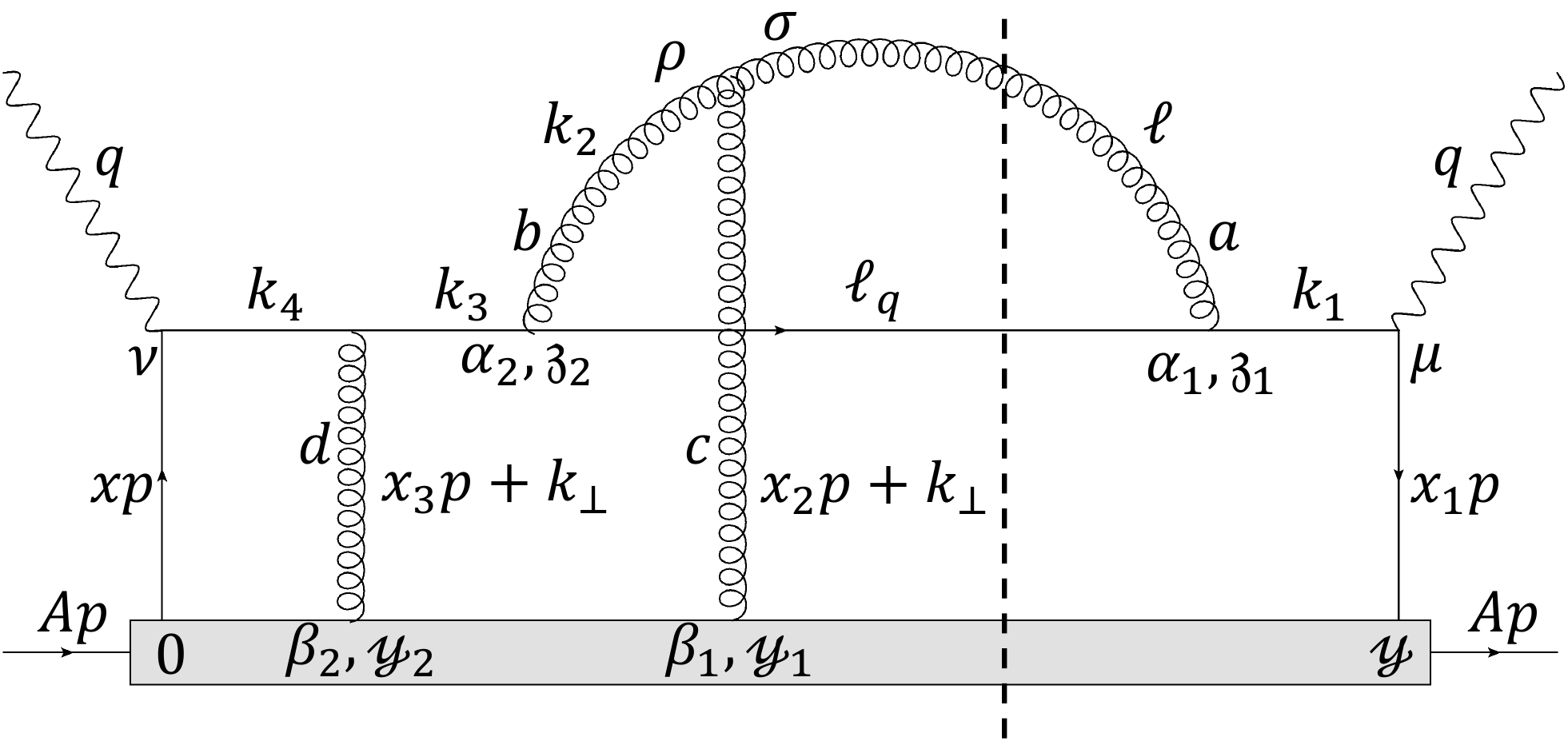}
	\caption{Feynman diagram for the quark-gluon and gluon-gluon rescattering process with scattering on quark happens before the emission with right cut.}
	\label{fig:qgBeforeAndggScatRight}
\end{figure}

\begin{figure}[tbp]
    \addtolength{\abovecaptionskip}{-3mm}
%
	\centering
	\includegraphics[width=\linewidth]{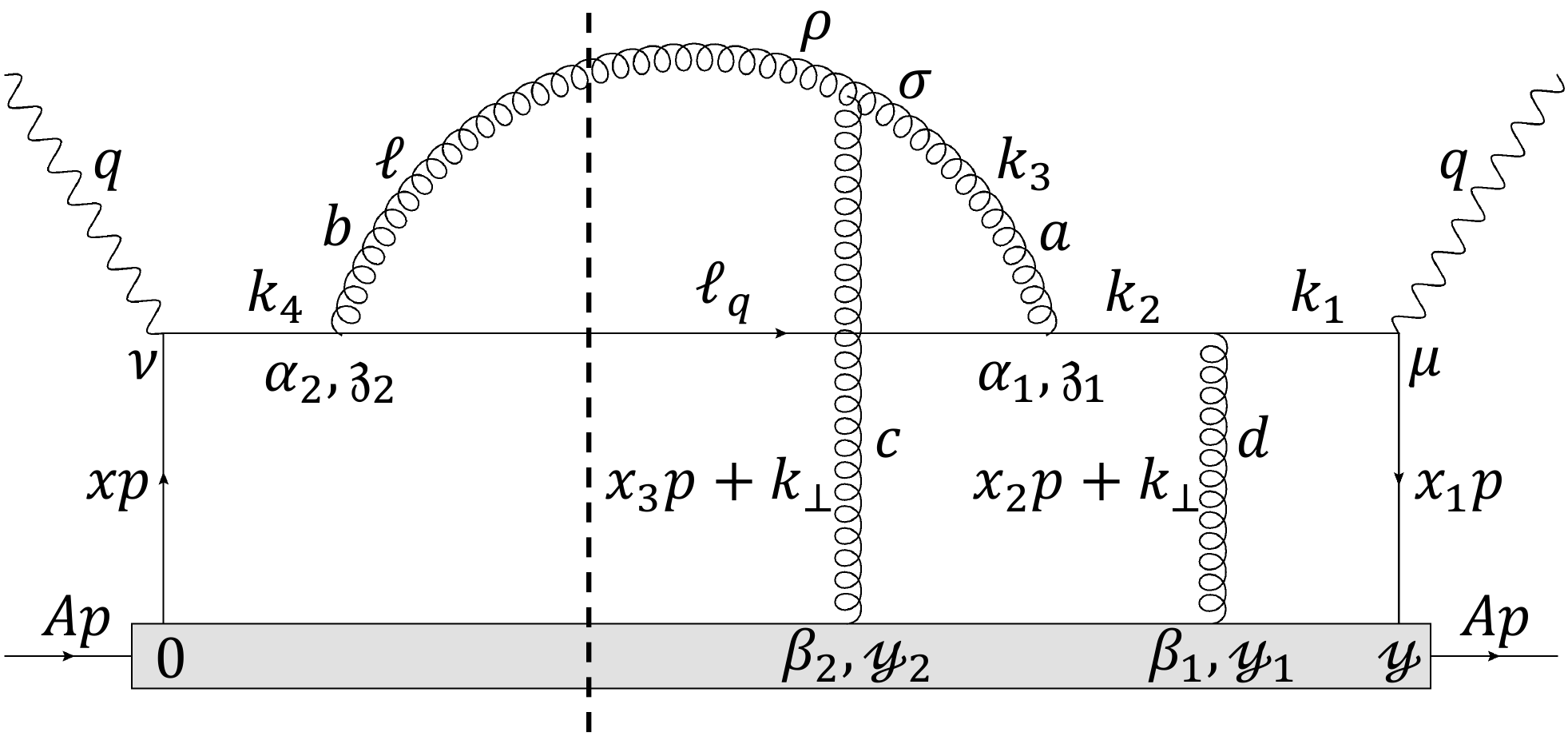}
	\caption{Feynman diagram for the quark-gluon and gluon-gluon rescattering process with scattering on quark happens before the emission with left cut.}
	\label{fig:qgBeforeAndggScatLeft}
\end{figure}

The final two non-central cut diagrams are for quark-gluon and gluon-gluon rescatterings with the quark-gluon scattering happening before the emission, as shown in Figs.~\ref{fig:qgBeforeAndggScatRight} and~\ref{fig:qgBeforeAndggScatLeft}, which are complex conjugate of each other.

The hadronic tensor associated with Fig.~\ref{fig:qgBeforeAndggScatRight} is given by
%
%
\begin{eqnarray}
\label{Eqn:qgBeforeAndggScatRight_W}
\mathcal{W}^{\mu\nu}_{\ref{fig:qgBeforeAndggScatRight}}  & = & - \frac{1}{2\pi} \int \frac{dy^-}{2\pi} dy_1^- dy_2^- \frac{ d^2y_\perp }{ \left( 2\pi \right)^2 } 
d^2k_\perp \int dz \int dx \nonumber\\
& \times & \left( 2\pi \right) \delta \left[ \left( q + xp \right)^2 \right] \frac{ e_q^2 }{2} 
\Tr \left[ p . \gamma \gamma^\mu \left( q + xp \right) . \gamma \gamma^\nu \right] \frac{ n }{ 2 } \nonumber\\
& \times & \left\langle A \middle| \bar{ \psi } \left( y^- \right) \gamma^+ A^+ \left( y_1^-, \vec{ y }_\perp \right) 
 A^+ \left( y_2^-, \vec{ 0 }_\perp \right) \psi \left( 0 \right) \middle| A \right\rangle  \nonumber\\
& \times & e^{ i \vec{ k }_\perp \cdot \vec{ y }_\perp } \int d \ell_\perp^2 \frac{ \vec{ \ell }_\perp \cdot \left( \vec{ \ell }_\perp + z \vec{ k }_\perp \right) } 
{ \ell_\perp^2 \left( \vec{ \ell }_\perp + z \vec{ k }_\perp \right)^2 } \frac{ \alpha_s }{ \left( 2\pi \right) } C_F \frac{ 1 + z^2 }{ 1 - z } \nonumber\\
& \times &  \frac{ 2 \pi \alpha_s }{ N_C } e^{ -i \left( x_B + x_L \right) p^+ y^- + i  x_D^+ p^+ \left( y_1^- - y_2^- \right)} \nonumber\\
& \times & \theta \left( y_1^- - y_2^- \right) \theta \left(y_2^- \right) e^{ i x_L p^+ y_2^- } e^{ - i \frac{ x_D^+ p^+ \left( y_1^- - y_2^- \right) }{ \left( 1 - z \right) } }\nonumber\\
& \times & \left[ 1 - e^{ - i \left( x_D^0 - \frac{ z x_D^+ }{ \left( 1 - z \right) } - x_L \right) p^+ \left( y_1^- - y_2^- \right) } \right] .
\end{eqnarray}
And the hadronic tensor associated with Fig.~\ref{fig:qgBeforeAndggScatLeft} is
%
%
\begin{eqnarray}
\label{Eqn:qgBeforeAndggScatLeft_W}
\mathcal{W}^{\mu\nu}_{\ref{fig:qgBeforeAndggScatLeft}}  & = & - \frac{1}{2\pi} \int \frac{dy^-}{2\pi} dy_1^- dy_2^- \frac{ d^2y_\perp }{ \left( 2\pi \right)^2 } 
d^2k_\perp \int dz \int dx \nonumber\\
& \times & \left( 2\pi \right) \delta \left[ \left( q + xp \right)^2 \right] \frac{ e_q^2 }{2} 
\Tr \left[ p . \gamma \gamma^\mu \left( q + xp \right) . \gamma \gamma^\nu \right] \frac{ n }{ 2 } \nonumber\\
& \times & \left\langle A \middle| \bar{ \psi } \left( y^- \right) \gamma^+ A^+ \left( y_1^-, \vec{ y }_\perp \right) 
 A^+ \left( y_2^-, \vec{ 0 }_\perp \right) \psi \left( 0 \right) \middle| A \right\rangle  \nonumber\\
& \times & e^{ i \vec{ k }_\perp \cdot \vec{ y }_\perp } \int d \ell_\perp^2 \frac{  \vec{ \ell }_\perp \cdot \left( \vec{ \ell }_\perp + z \vec{ k }_\perp \right) } 
{ \ell_\perp^2 \left( \vec{ \ell }_\perp + z \vec{ k }_\perp \right)^2 } \frac{ \alpha_s }{ \left( 2\pi \right) } C_F \frac{ 1 + z^2 }{ 1 - z } \nonumber\\
& \times & \frac{ 2 \pi \alpha_s }{ N_C }  e^{ -i \left( x_B + x_L \right) p^+ y^- + i  x_D^+ p^+ \left( y_1^- - y_2^- \right)} \theta \left(y_1^- - y^- \right) \nonumber\\
& \times &  \theta \left( y_2^- - y_1^- \right) e^{ i x_L p^+ \left( y^- - y_1^- \right) } e^{ - i \frac{ x_D^+ p^+ \left( y_1^- - y_2^- \right) }{ \left( 1 - z \right) }} \nonumber\\
& \times &  \left[ 1 - e^{ - i \left( x_D^0 - \frac{ z x_D^+ }{ \left( 1 - z \right) } - x_L \right) p^+ \left( y_1^- - y_2^- \right) } \right] .
\end{eqnarray}

By setting $ y^- = 0 $ and $ y_1^- = y_2^- = \xi^- $, one observes that the above four equations [\eqref{Eqn:qgBeforeAfterDoubleScatRight_W}, \eqref{Eqn:qgAfterBeforeDoubleScatLeft_W}, \eqref{Eqn:qgBeforeAndggScatRight_W}, and \eqref{Eqn:qgBeforeAndggScatLeft_W}] all contain the factor $ 1 - e^{ - i \left( ... \right) p^+ \left( y_1^- - y_2^- \right) } = 0 $. Therefore, even before the collinear expansion is taken, these last four diagrams are shown to have no contribution to the medium modification kernel at the next-to-leading twist.


\bibliography{FullNextToLeadingTwistDIS,refs}

\providecommand{\noopsort}[1]{}\providecommand{\singleletter}[1]{#1}%
\begin{thebibliography}{55}%
\makeatletter
\providecommand \@ifxundefined [1]{%
 \@ifx{#1\undefined}
}%
\providecommand \@ifnum [1]{%
 \ifnum #1\expandafter \@firstoftwo
 \else \expandafter \@secondoftwo
 \fi
}%
\providecommand \@ifx [1]{%
 \ifx #1\expandafter \@firstoftwo
 \else \expandafter \@secondoftwo
 \fi
}%
\providecommand \natexlab [1]{#1}%
\providecommand \enquote  [1]{``#1''}%
\providecommand \bibnamefont  [1]{#1}%
\providecommand \bibfnamefont [1]{#1}%
\providecommand \citenamefont [1]{#1}%
\providecommand \href@noop [0]{\@secondoftwo}%
\providecommand \href [0]{\begingroup \@sanitize@url \@href}%
\providecommand \@href[1]{\@@startlink{#1}\@@href}%
\providecommand \@@href[1]{\endgroup#1\@@endlink}%
\providecommand \@sanitize@url [0]{\catcode `\\12\catcode `\$12\catcode `\&12\catcode `\#12\catcode `\^12\catcode `\_12\catcode `\%12\relax}%
\providecommand \@@startlink[1]{}%
\providecommand \@@endlink[0]{}%
\providecommand \url  [0]{\begingroup\@sanitize@url \@url }%
\providecommand \@url [1]{\endgroup\@href {#1}{\urlprefix }}%
\providecommand \urlprefix  [0]{URL }%
\providecommand \Eprint [0]{\href }%
\providecommand \doibase [0]{https://doi.org/}%
\providecommand \selectlanguage [0]{\@gobble}%
\providecommand \bibinfo  [0]{\@secondoftwo}%
\providecommand \bibfield  [0]{\@secondoftwo}%
\providecommand \translation [1]{[#1]}%
\providecommand \BibitemOpen [0]{}%
\providecommand \bibitemStop [0]{}%
\providecommand \bibitemNoStop [0]{.\EOS\space}%
\providecommand \EOS [0]{\spacefactor3000\relax}%
\providecommand \BibitemShut  [1]{\csname bibitem#1\endcsname}%
\let\auto@bib@innerbib\@empty
\bibitem [{\citenamefont {Majumder}\ and\ \citenamefont {Van~Leeuwen}(2011)}]{Majumder:2010qh}%
  \BibitemOpen
  \bibfield  {author} {\bibinfo {author} {\bibfnamefont {A.}~\bibnamefont {Majumder}}\ and\ \bibinfo {author} {\bibfnamefont {M.}~\bibnamefont {Van~Leeuwen}},\ }\bibfield  {title} {\bibinfo {title} {{The Theory and Phenomenology of Perturbative QCD Based Jet Quenching}},\ }\href {https://doi.org/10.1016/j.ppnp.2010.09.001} {\bibfield  {journal} {\bibinfo  {journal} {Prog. Part. Nucl. Phys.}\ }\textbf {\bibinfo {volume} {66}},\ \bibinfo {pages} {41} (\bibinfo {year} {2011})},\ \Eprint {https://arxiv.org/abs/1002.2206} {arXiv:1002.2206 [hep-ph]} \BibitemShut {NoStop}%
\bibitem [{\citenamefont {Cao}\ and\ \citenamefont {Wang}(2020)}]{Cao:2020wlm}%
  \BibitemOpen
  \bibfield  {author} {\bibinfo {author} {\bibfnamefont {S.}~\bibnamefont {Cao}}\ and\ \bibinfo {author} {\bibfnamefont {X.-N.}\ \bibnamefont {Wang}},\ }\bibfield  {title} {\bibinfo {title} {{Jet quenching and medium response in high-energy heavy-ion collisions: a review}},\ }\href@noop {} {\  (\bibinfo {year} {2020})},\ \Eprint {https://arxiv.org/abs/2002.04028} {arXiv:2002.04028 [hep-ph]} \BibitemShut {NoStop}%
\bibitem [{\citenamefont {Cao}\ \emph {et~al.}(2017{\natexlab{a}})\citenamefont {Cao} \emph {et~al.}}]{Cao:2017zih}%
  \BibitemOpen
  \bibfield  {author} {\bibinfo {author} {\bibfnamefont {S.}~\bibnamefont {Cao}} \emph {et~al.} (\bibinfo {collaboration} {JETSCAPE}),\ }\bibfield  {title} {\bibinfo {title} {{Multistage Monte-Carlo simulation of jet modification in a static medium}},\ }\href {https://doi.org/10.1103/PhysRevC.96.024909} {\bibfield  {journal} {\bibinfo  {journal} {Phys. Rev.}\ }\textbf {\bibinfo {volume} {C96}},\ \bibinfo {pages} {024909} (\bibinfo {year} {2017}{\natexlab{a}})},\ \Eprint {https://arxiv.org/abs/1705.00050} {arXiv:1705.00050 [nucl-th]} \BibitemShut {NoStop}%
\bibitem [{\citenamefont {Kumar}\ \emph {et~al.}(2021)\citenamefont {Kumar} \emph {et~al.}}]{Kumar:2020vkx}%
  \BibitemOpen
  \bibfield  {author} {\bibinfo {author} {\bibfnamefont {A.}~\bibnamefont {Kumar}} \emph {et~al.} (\bibinfo {collaboration} {JETSCAPE}),\ }\bibfield  {title} {\bibinfo {title} {{Jet quenching in a multi-stage Monte Carlo approach}},\ }\bibfield  {booktitle} {\emph {\bibinfo {booktitle} {{Proceedings, 28th International Conference on Ultrarelativistic Nucleus-Nucleus Collisions (Quark Matter 2019): Wuhan, China}}},\ }\href {https://doi.org/10.1016/j.nuclphysa.2020.122009} {\bibfield  {journal} {\bibinfo  {journal} {Nucl. Phys.}\ }\textbf {\bibinfo {volume} {A1005}},\ \bibinfo {pages} {122009} (\bibinfo {year} {2021})},\ \Eprint {https://arxiv.org/abs/2002.07124} {arXiv:2002.07124 [nucl-th]} \BibitemShut {NoStop}%
\bibitem [{\citenamefont {Tachibana}\ \emph {et~al.}(2018)\citenamefont {Tachibana} \emph {et~al.}}]{Tachibana:2018yae}%
  \BibitemOpen
  \bibfield  {author} {\bibinfo {author} {\bibfnamefont {Y.}~\bibnamefont {Tachibana}} \emph {et~al.} (\bibinfo {collaboration} {JETSCAPE}),\ }\bibfield  {title} {\bibinfo {title} {{Jet substructure modifications in a QGP from multi-scale description of jet evolution with JETSCAPE}},\ }\bibfield  {booktitle} {\emph {\bibinfo {booktitle} {{Proceedings, 9th International Conference on Hard and Electromagnetic Probes of High-Energy Nuclear Collisions: Hard Probes 2018 (HP2018): Aix-Les-Bains, France, October 1-5, 2018}}},\ }\href {https://doi.org/10.22323/1.345.0099} {\bibfield  {journal} {\bibinfo  {journal} {PoS}\ }\textbf {\bibinfo {volume} {HardProbes2018}},\ \bibinfo {pages} {099} (\bibinfo {year} {2018})},\ \Eprint {https://arxiv.org/abs/1812.06366} {arXiv:1812.06366 [nucl-th]} \BibitemShut {NoStop}%
\bibitem [{\citenamefont {Vujanovic}\ \emph {et~al.}(2021)\citenamefont {Vujanovic} \emph {et~al.}}]{Vujanovic:2020wuk}%
  \BibitemOpen
  \bibfield  {author} {\bibinfo {author} {\bibfnamefont {G.}~\bibnamefont {Vujanovic}} \emph {et~al.} (\bibinfo {collaboration} {JETSCAPE}),\ }\bibfield  {title} {\bibinfo {title} {{Multi-stage evolution of heavy quarks in the quark-gluon plasma}},\ }\bibfield  {booktitle} {\emph {\bibinfo {booktitle} {{Proceedings, 28th International Conference on Ultrarelativistic Nucleus-Nucleus Collisions (Quark Matter 2019): Wuhan, China}}},\ }\href {https://doi.org/10.1016/j.nuclphysa.2020.121965} {\bibfield  {journal} {\bibinfo  {journal} {Nucl. Phys.}\ }\textbf {\bibinfo {volume} {A1005}},\ \bibinfo {pages} {121965} (\bibinfo {year} {2021})},\ \Eprint {https://arxiv.org/abs/2002.06643} {arXiv:2002.06643 [nucl-th]} \BibitemShut {NoStop}%
\bibitem [{\citenamefont {Caucal}\ \emph {et~al.}(2018)\citenamefont {Caucal}, \citenamefont {Iancu}, \citenamefont {Mueller},\ and\ \citenamefont {Soyez}}]{Caucal:2018dla}%
  \BibitemOpen
  \bibfield  {author} {\bibinfo {author} {\bibfnamefont {P.}~\bibnamefont {Caucal}}, \bibinfo {author} {\bibfnamefont {E.}~\bibnamefont {Iancu}}, \bibinfo {author} {\bibfnamefont {A.~H.}\ \bibnamefont {Mueller}},\ and\ \bibinfo {author} {\bibfnamefont {G.}~\bibnamefont {Soyez}},\ }\bibfield  {title} {\bibinfo {title} {{Vacuum-like jet fragmentation in a dense QCD medium}},\ }\href {https://doi.org/10.1103/PhysRevLett.120.232001} {\bibfield  {journal} {\bibinfo  {journal} {Phys. Rev. Lett.}\ }\textbf {\bibinfo {volume} {120}},\ \bibinfo {pages} {232001} (\bibinfo {year} {2018})},\ \Eprint {https://arxiv.org/abs/1801.09703} {arXiv:1801.09703 [hep-ph]} \BibitemShut {NoStop}%
\bibitem [{\citenamefont {Cao}\ \emph {et~al.}(2021{\natexlab{a}})\citenamefont {Cao}, \citenamefont {Sirimanna},\ and\ \citenamefont {Majumder}}]{Cao:2021rpv}%
  \BibitemOpen
  \bibfield  {author} {\bibinfo {author} {\bibfnamefont {S.}~\bibnamefont {Cao}}, \bibinfo {author} {\bibfnamefont {C.}~\bibnamefont {Sirimanna}},\ and\ \bibinfo {author} {\bibfnamefont {A.}~\bibnamefont {Majumder}},\ }\bibfield  {title} {\bibinfo {title} {{The medium modification of high-virtuality partons}},\ }\href@noop {} {\  (\bibinfo {year} {2021}{\natexlab{a}})},\ \Eprint {https://arxiv.org/abs/2101.03681} {arXiv:2101.03681 [hep-ph]} \BibitemShut {NoStop}%
\bibitem [{\citenamefont {Baier}(2003)}]{Baier:2002tc}%
  \BibitemOpen
  \bibfield  {author} {\bibinfo {author} {\bibfnamefont {R.}~\bibnamefont {Baier}},\ }\bibfield  {title} {\bibinfo {title} {{Jet quenching}},\ }\href {https://doi.org/10.1016/S0375-9474(02)01429-X} {\bibfield  {journal} {\bibinfo  {journal} {Nucl. Phys.}\ }\textbf {\bibinfo {volume} {A715}},\ \bibinfo {pages} {209} (\bibinfo {year} {2003})},\ \Eprint {https://arxiv.org/abs/hep-ph/0209038} {arXiv:hep-ph/0209038} \BibitemShut {NoStop}%
\bibitem [{\citenamefont {Majumder}(2013{\natexlab{a}})}]{Majumder:2012sh}%
  \BibitemOpen
  \bibfield  {author} {\bibinfo {author} {\bibfnamefont {A.}~\bibnamefont {Majumder}},\ }\bibfield  {title} {\bibinfo {title} {{Calculating the jet quenching parameter $\hat{q}$ in lattice gauge theory}},\ }\href {https://doi.org/10.1103/PhysRevC.87.034905} {\bibfield  {journal} {\bibinfo  {journal} {Phys. Rev.}\ }\textbf {\bibinfo {volume} {C87}},\ \bibinfo {pages} {034905} (\bibinfo {year} {2013}{\natexlab{a}})}\BibitemShut {NoStop}%
\bibitem [{\citenamefont {Cao}\ \emph {et~al.}(2021{\natexlab{b}})\citenamefont {Cao} \emph {et~al.}}]{JETSCAPE:2021ehl}%
  \BibitemOpen
  \bibfield  {author} {\bibinfo {author} {\bibfnamefont {S.}~\bibnamefont {Cao}} \emph {et~al.} (\bibinfo {collaboration} {JETSCAPE}),\ }\bibfield  {title} {\bibinfo {title} {{Determining the jet transport coefficient $\hat{q}$ from inclusive hadron suppression measurements using Bayesian parameter estimation}},\ }\href {https://doi.org/10.1103/PhysRevC.104.024905} {\bibfield  {journal} {\bibinfo  {journal} {Phys. Rev. C}\ }\textbf {\bibinfo {volume} {104}},\ \bibinfo {pages} {024905} (\bibinfo {year} {2021}{\natexlab{b}})},\ \Eprint {https://arxiv.org/abs/2102.11337} {arXiv:2102.11337 [nucl-th]} \BibitemShut {NoStop}%
\bibitem [{\citenamefont {Kumar}\ \emph {et~al.}(2020{\natexlab{a}})\citenamefont {Kumar}, \citenamefont {Majumder},\ and\ \citenamefont {Weber}}]{Kumar:2020wvb}%
  \BibitemOpen
  \bibfield  {author} {\bibinfo {author} {\bibfnamefont {A.}~\bibnamefont {Kumar}}, \bibinfo {author} {\bibfnamefont {A.}~\bibnamefont {Majumder}},\ and\ \bibinfo {author} {\bibfnamefont {J.~H.}\ \bibnamefont {Weber}},\ }\bibfield  {title} {\bibinfo {title} {{Jet transport coefficient $\hat{q}$ in (2+1)-flavor lattice QCD}},\ }\href@noop {} {\  (\bibinfo {year} {2020}{\natexlab{a}})},\ \Eprint {https://arxiv.org/abs/2010.14463} {arXiv:2010.14463 [hep-lat]} \BibitemShut {NoStop}%
\bibitem [{\citenamefont {Putschke}\ \emph {et~al.}(2019)\citenamefont {Putschke} \emph {et~al.}}]{Putschke:2019yrg}%
  \BibitemOpen
  \bibfield  {author} {\bibinfo {author} {\bibfnamefont {J.~H.}\ \bibnamefont {Putschke}} \emph {et~al.},\ }\bibfield  {title} {\bibinfo {title} {{The JETSCAPE framework}},\ }\href@noop {} {\  (\bibinfo {year} {2019})},\ \Eprint {https://arxiv.org/abs/1903.07706} {arXiv:1903.07706 [nucl-th]} \BibitemShut {NoStop}%
\bibitem [{\citenamefont {Cao}\ \emph {et~al.}(2017{\natexlab{b}})\citenamefont {Cao} \emph {et~al.}}]{JETSCAPE:2017eso}%
  \BibitemOpen
  \bibfield  {author} {\bibinfo {author} {\bibfnamefont {S.}~\bibnamefont {Cao}} \emph {et~al.} (\bibinfo {collaboration} {JETSCAPE}),\ }\bibfield  {title} {\bibinfo {title} {{Multistage Monte-Carlo simulation of jet modification in a static medium}},\ }\href {https://doi.org/10.1103/PhysRevC.96.024909} {\bibfield  {journal} {\bibinfo  {journal} {Phys. Rev. C}\ }\textbf {\bibinfo {volume} {96}},\ \bibinfo {pages} {024909} (\bibinfo {year} {2017}{\natexlab{b}})},\ \Eprint {https://arxiv.org/abs/1705.00050} {arXiv:1705.00050 [nucl-th]} \BibitemShut {NoStop}%
\bibitem [{\citenamefont {Cao}\ and\ \citenamefont {Majumder}(2020)}]{Cao:2017qpx}%
  \BibitemOpen
  \bibfield  {author} {\bibinfo {author} {\bibfnamefont {S.}~\bibnamefont {Cao}}\ and\ \bibinfo {author} {\bibfnamefont {A.}~\bibnamefont {Majumder}},\ }\bibfield  {title} {\bibinfo {title} {{Nuclear modification of leading hadrons and jets within a virtuality ordered parton shower}},\ }\href {https://doi.org/10.1103/PhysRevC.101.024903} {\bibfield  {journal} {\bibinfo  {journal} {Phys. Rev.}\ }\textbf {\bibinfo {volume} {C101}},\ \bibinfo {pages} {024903} (\bibinfo {year} {2020})},\ \Eprint {https://arxiv.org/abs/1712.10055} {arXiv:1712.10055 [nucl-th]} \BibitemShut {NoStop}%
\bibitem [{\citenamefont {Majumder}(2013{\natexlab{b}})}]{Majumder:2013re}%
  \BibitemOpen
  \bibfield  {author} {\bibinfo {author} {\bibfnamefont {A.}~\bibnamefont {Majumder}},\ }\bibfield  {title} {\bibinfo {title} {{Incorporating Space-Time Within Medium-Modified Jet Event Generators}},\ }\href {https://doi.org/10.1103/PhysRevC.88.014909} {\bibfield  {journal} {\bibinfo  {journal} {Phys. Rev.}\ }\textbf {\bibinfo {volume} {C88}},\ \bibinfo {pages} {014909} (\bibinfo {year} {2013}{\natexlab{b}})},\ \Eprint {https://arxiv.org/abs/1301.5323} {arXiv:1301.5323 [nucl-th]} \BibitemShut {NoStop}%
\bibitem [{\citenamefont {Cao}\ \emph {et~al.}(2016)\citenamefont {Cao}, \citenamefont {Luo}, \citenamefont {Qin},\ and\ \citenamefont {Wang}}]{Cao:2016gvr}%
  \BibitemOpen
  \bibfield  {author} {\bibinfo {author} {\bibfnamefont {S.}~\bibnamefont {Cao}}, \bibinfo {author} {\bibfnamefont {T.}~\bibnamefont {Luo}}, \bibinfo {author} {\bibfnamefont {G.-Y.}\ \bibnamefont {Qin}},\ and\ \bibinfo {author} {\bibfnamefont {X.-N.}\ \bibnamefont {Wang}},\ }\bibfield  {title} {\bibinfo {title} {{Linearized Boltzmann transport model for jet propagation in the quark-gluon plasma: Heavy quark evolution}},\ }\href {https://doi.org/10.1103/PhysRevC.94.014909} {\bibfield  {journal} {\bibinfo  {journal} {Phys. Rev. C}\ }\textbf {\bibinfo {volume} {94}},\ \bibinfo {pages} {014909} (\bibinfo {year} {2016})},\ \Eprint {https://arxiv.org/abs/1605.06447} {arXiv:1605.06447 [nucl-th]} \BibitemShut {NoStop}%
\bibitem [{\citenamefont {He}\ \emph {et~al.}(2015)\citenamefont {He}, \citenamefont {Luo}, \citenamefont {Wang},\ and\ \citenamefont {Zhu}}]{He:2015pra}%
  \BibitemOpen
  \bibfield  {author} {\bibinfo {author} {\bibfnamefont {Y.}~\bibnamefont {He}}, \bibinfo {author} {\bibfnamefont {T.}~\bibnamefont {Luo}}, \bibinfo {author} {\bibfnamefont {X.-N.}\ \bibnamefont {Wang}},\ and\ \bibinfo {author} {\bibfnamefont {Y.}~\bibnamefont {Zhu}},\ }\bibfield  {title} {\bibinfo {title} {{Linear Boltzmann Transport for Jet Propagation in the Quark-Gluon Plasma: Elastic Processes and Medium Recoil}},\ }\href {https://doi.org/10.1103/PhysRevC.97.019902, 10.1103/PhysRevC.91.054908} {\bibfield  {journal} {\bibinfo  {journal} {Phys. Rev.}\ }\textbf {\bibinfo {volume} {C91}},\ \bibinfo {pages} {054908} (\bibinfo {year} {2015})},\ \bibinfo {note} {[Erratum: Phys. Rev.C97,no.1,019902(2018)]},\ \Eprint {https://arxiv.org/abs/1503.03313} {arXiv:1503.03313 [nucl-th]} \BibitemShut {NoStop}%
\bibitem [{\citenamefont {Schenke}\ \emph {et~al.}(2009)\citenamefont {Schenke}, \citenamefont {Gale},\ and\ \citenamefont {Jeon}}]{Schenke:2009gb}%
  \BibitemOpen
  \bibfield  {author} {\bibinfo {author} {\bibfnamefont {B.}~\bibnamefont {Schenke}}, \bibinfo {author} {\bibfnamefont {C.}~\bibnamefont {Gale}},\ and\ \bibinfo {author} {\bibfnamefont {S.}~\bibnamefont {Jeon}},\ }\bibfield  {title} {\bibinfo {title} {{MARTINI: An Event generator for relativistic heavy-ion collisions}},\ }\href {https://doi.org/10.1103/PhysRevC.80.054913} {\bibfield  {journal} {\bibinfo  {journal} {Phys.Rev.}\ }\textbf {\bibinfo {volume} {C80}},\ \bibinfo {pages} {054913} (\bibinfo {year} {2009})},\ \Eprint {https://arxiv.org/abs/0909.2037} {arXiv:0909.2037 [hep-ph]} \BibitemShut {NoStop}%
\bibitem [{\citenamefont {Guo}\ and\ \citenamefont {Wang}(2000{\natexlab{a}})}]{Guo:2000nz}%
  \BibitemOpen
  \bibfield  {author} {\bibinfo {author} {\bibfnamefont {X.-F.}\ \bibnamefont {Guo}}\ and\ \bibinfo {author} {\bibfnamefont {X.-N.}\ \bibnamefont {Wang}},\ }\bibfield  {title} {\bibinfo {title} {{Multiple scattering, parton energy loss and modified fragmentation functions in deeply inelastic e A scattering}},\ }\href {https://doi.org/10.1103/PhysRevLett.85.3591} {\bibfield  {journal} {\bibinfo  {journal} {Phys. Rev. Lett.}\ }\textbf {\bibinfo {volume} {85}},\ \bibinfo {pages} {3591} (\bibinfo {year} {2000}{\natexlab{a}})},\ \Eprint {https://arxiv.org/abs/hep-ph/0005044} {arXiv:hep-ph/0005044} \BibitemShut {NoStop}%
\bibitem [{\citenamefont {Wang}\ and\ \citenamefont {Guo}(2001{\natexlab{a}})}]{Wang:2001ifa}%
  \BibitemOpen
  \bibfield  {author} {\bibinfo {author} {\bibfnamefont {X.-N.}\ \bibnamefont {Wang}}\ and\ \bibinfo {author} {\bibfnamefont {X.-F.}\ \bibnamefont {Guo}},\ }\bibfield  {title} {\bibinfo {title} {{Multiple parton scattering in nuclei: Parton energy loss}},\ }\href {https://doi.org/10.1016/S0375-9474(01)01130-7} {\bibfield  {journal} {\bibinfo  {journal} {Nucl. Phys.}\ }\textbf {\bibinfo {volume} {A696}},\ \bibinfo {pages} {788} (\bibinfo {year} {2001}{\natexlab{a}})},\ \Eprint {https://arxiv.org/abs/hep-ph/0102230} {arXiv:hep-ph/0102230} \BibitemShut {NoStop}%
\bibitem [{\citenamefont {Majumder}(2012)}]{Majumder:2009ge}%
  \BibitemOpen
  \bibfield  {author} {\bibinfo {author} {\bibfnamefont {A.}~\bibnamefont {Majumder}},\ }\bibfield  {title} {\bibinfo {title} {{Hard collinear gluon radiation and multiple scattering in a medium}},\ }\href {https://doi.org/10.1103/PhysRevD.85.014023} {\bibfield  {journal} {\bibinfo  {journal} {Phys. Rev.}\ }\textbf {\bibinfo {volume} {D85}},\ \bibinfo {pages} {014023} (\bibinfo {year} {2012})}\BibitemShut {NoStop}%
\bibitem [{\citenamefont {Arnold}\ \emph {et~al.}(2001{\natexlab{a}})\citenamefont {Arnold}, \citenamefont {Moore},\ and\ \citenamefont {Yaffe}}]{Arnold:2001ba}%
  \BibitemOpen
  \bibfield  {author} {\bibinfo {author} {\bibfnamefont {P.}~\bibnamefont {Arnold}}, \bibinfo {author} {\bibfnamefont {G.~D.}\ \bibnamefont {Moore}},\ and\ \bibinfo {author} {\bibfnamefont {L.~G.}\ \bibnamefont {Yaffe}},\ }\bibfield  {title} {\bibinfo {title} {Photon emission from ultrarelativistic plasmas},\ }\href@noop {} {\bibfield  {journal} {\bibinfo  {journal} {JHEP}\ }\textbf {\bibinfo {volume} {11}},\ \bibinfo {pages} {057}},\ \Eprint {https://arxiv.org/abs/hep-ph/0109064} {hep-ph/0109064} \BibitemShut {NoStop}%
\bibitem [{\citenamefont {Arnold}\ \emph {et~al.}(2001{\natexlab{b}})\citenamefont {Arnold}, \citenamefont {Moore},\ and\ \citenamefont {Yaffe}}]{Arnold:2001ms}%
  \BibitemOpen
  \bibfield  {author} {\bibinfo {author} {\bibfnamefont {P.}~\bibnamefont {Arnold}}, \bibinfo {author} {\bibfnamefont {G.~D.}\ \bibnamefont {Moore}},\ and\ \bibinfo {author} {\bibfnamefont {L.~G.}\ \bibnamefont {Yaffe}},\ }\bibfield  {title} {\bibinfo {title} {{Photon emission from quark gluon plasma: Complete leading order results}},\ }\href@noop {} {\bibfield  {journal} {\bibinfo  {journal} {JHEP}\ }\textbf {\bibinfo {volume} {12}},\ \bibinfo {pages} {009}},\ \Eprint {https://arxiv.org/abs/hep-ph/0111107} {arXiv:hep-ph/0111107} \BibitemShut {NoStop}%
\bibitem [{\citenamefont {Arnold}\ \emph {et~al.}(2002)\citenamefont {Arnold}, \citenamefont {Moore},\ and\ \citenamefont {Yaffe}}]{Arnold:2002ja}%
  \BibitemOpen
  \bibfield  {author} {\bibinfo {author} {\bibfnamefont {P.}~\bibnamefont {Arnold}}, \bibinfo {author} {\bibfnamefont {G.~D.}\ \bibnamefont {Moore}},\ and\ \bibinfo {author} {\bibfnamefont {L.~G.}\ \bibnamefont {Yaffe}},\ }\bibfield  {title} {\bibinfo {title} {Photon and gluon emission in relativistic plasmas},\ }\href@noop {} {\bibfield  {journal} {\bibinfo  {journal} {JHEP}\ }\textbf {\bibinfo {volume} {06}},\ \bibinfo {pages} {030}},\ \Eprint {https://arxiv.org/abs/hep-ph/0204343} {hep-ph/0204343} \BibitemShut {NoStop}%
\bibitem [{\citenamefont {Majumder}\ \emph {et~al.}(2007)\citenamefont {Majumder}, \citenamefont {Wang},\ and\ \citenamefont {Wang}}]{Majumder:2004pt}%
  \BibitemOpen
  \bibfield  {author} {\bibinfo {author} {\bibfnamefont {A.}~\bibnamefont {Majumder}}, \bibinfo {author} {\bibfnamefont {E.}~\bibnamefont {Wang}},\ and\ \bibinfo {author} {\bibfnamefont {X.-N.}\ \bibnamefont {Wang}},\ }\bibfield  {title} {\bibinfo {title} {{Modified dihadron fragmentation functions in hot and nuclear matter}},\ }\href {https://doi.org/10.1103/PhysRevLett.99.152301} {\bibfield  {journal} {\bibinfo  {journal} {Phys. Rev. Lett.}\ }\textbf {\bibinfo {volume} {99}},\ \bibinfo {pages} {152301} (\bibinfo {year} {2007})},\ \Eprint {https://arxiv.org/abs/nucl-th/0412061} {arXiv:nucl-th/0412061} \BibitemShut {NoStop}%
\bibitem [{\citenamefont {Deng}\ and\ \citenamefont {Wang}(2010)}]{Deng:2009ncl}%
  \BibitemOpen
  \bibfield  {author} {\bibinfo {author} {\bibfnamefont {W.-t.}\ \bibnamefont {Deng}}\ and\ \bibinfo {author} {\bibfnamefont {X.-N.}\ \bibnamefont {Wang}},\ }\bibfield  {title} {\bibinfo {title} {{Multiple Parton Scattering in Nuclei: Modified DGLAP Evolution for Fragmentation Functions}},\ }\href {https://doi.org/10.1103/PhysRevC.81.024902} {\bibfield  {journal} {\bibinfo  {journal} {Phys. Rev. C}\ }\textbf {\bibinfo {volume} {81}},\ \bibinfo {pages} {024902} (\bibinfo {year} {2010})},\ \Eprint {https://arxiv.org/abs/0910.3403} {arXiv:0910.3403 [hep-ph]} \BibitemShut {NoStop}%
\bibitem [{\citenamefont {Chen}\ \emph {et~al.}(2011)\citenamefont {Chen}, \citenamefont {Hirano}, \citenamefont {Wang}, \citenamefont {Wang},\ and\ \citenamefont {Zhang}}]{Chen:2011vt}%
  \BibitemOpen
  \bibfield  {author} {\bibinfo {author} {\bibfnamefont {X.-F.}\ \bibnamefont {Chen}}, \bibinfo {author} {\bibfnamefont {T.}~\bibnamefont {Hirano}}, \bibinfo {author} {\bibfnamefont {E.}~\bibnamefont {Wang}}, \bibinfo {author} {\bibfnamefont {X.-N.}\ \bibnamefont {Wang}},\ and\ \bibinfo {author} {\bibfnamefont {H.}~\bibnamefont {Zhang}},\ }\bibfield  {title} {\bibinfo {title} {{Suppression of high $p_{T}$ hadrons in $Pb+Pb$ Collisions at LHC}},\ }\href {https://doi.org/10.1103/PhysRevC.84.034902} {\bibfield  {journal} {\bibinfo  {journal} {Phys. Rev. C}\ }\textbf {\bibinfo {volume} {84}},\ \bibinfo {pages} {034902} (\bibinfo {year} {2011})},\ \Eprint {https://arxiv.org/abs/1102.5614} {arXiv:1102.5614 [nucl-th]} \BibitemShut {NoStop}%
\bibitem [{\citenamefont {Wang}\ and\ \citenamefont {Zhu}(2013)}]{Wang:2013cia}%
  \BibitemOpen
  \bibfield  {author} {\bibinfo {author} {\bibfnamefont {X.-N.}\ \bibnamefont {Wang}}\ and\ \bibinfo {author} {\bibfnamefont {Y.}~\bibnamefont {Zhu}},\ }\bibfield  {title} {\bibinfo {title} {{Medium Modification of $\gamma$-jets in High-energy Heavy-ion Collisions}},\ }\href {https://doi.org/10.1103/PhysRevLett.111.062301} {\bibfield  {journal} {\bibinfo  {journal} {Phys. Rev. Lett.}\ }\textbf {\bibinfo {volume} {111}},\ \bibinfo {pages} {062301} (\bibinfo {year} {2013})},\ \Eprint {https://arxiv.org/abs/1302.5874} {arXiv:1302.5874 [hep-ph]} \BibitemShut {NoStop}%
\bibitem [{\citenamefont {Cao}\ \emph {et~al.}(2018)\citenamefont {Cao}, \citenamefont {Luo}, \citenamefont {Qin},\ and\ \citenamefont {Wang}}]{Cao:2017hhk}%
  \BibitemOpen
  \bibfield  {author} {\bibinfo {author} {\bibfnamefont {S.}~\bibnamefont {Cao}}, \bibinfo {author} {\bibfnamefont {T.}~\bibnamefont {Luo}}, \bibinfo {author} {\bibfnamefont {G.-Y.}\ \bibnamefont {Qin}},\ and\ \bibinfo {author} {\bibfnamefont {X.-N.}\ \bibnamefont {Wang}},\ }\bibfield  {title} {\bibinfo {title} {{Heavy and light flavor jet quenching at RHIC and LHC energies}},\ }\href {https://doi.org/10.1016/j.physletb.2017.12.023} {\bibfield  {journal} {\bibinfo  {journal} {Phys. Lett. B}\ }\textbf {\bibinfo {volume} {777}},\ \bibinfo {pages} {255} (\bibinfo {year} {2018})},\ \Eprint {https://arxiv.org/abs/1703.00822} {arXiv:1703.00822 [nucl-th]} \BibitemShut {NoStop}%
\bibitem [{\citenamefont {Majumder}\ and\ \citenamefont {Shen}(2012)}]{Majumder:2011uk}%
  \BibitemOpen
  \bibfield  {author} {\bibinfo {author} {\bibfnamefont {A.}~\bibnamefont {Majumder}}\ and\ \bibinfo {author} {\bibfnamefont {C.}~\bibnamefont {Shen}},\ }\bibfield  {title} {\bibinfo {title} {{Suppression of the High $p_T$ Charged Hadron $R_{AA}$ at the LHC}},\ }\href {https://doi.org/10.1103/PhysRevLett.109.202301} {\bibfield  {journal} {\bibinfo  {journal} {Phys.Rev.Lett.}\ }\textbf {\bibinfo {volume} {109}},\ \bibinfo {pages} {202301} (\bibinfo {year} {2012})},\ \Eprint {https://arxiv.org/abs/1103.0809} {arXiv:1103.0809 [hep-ph]} \BibitemShut {NoStop}%
\bibitem [{\citenamefont {Qin}\ and\ \citenamefont {Majumder}(2010)}]{Qin:2009gw}%
  \BibitemOpen
  \bibfield  {author} {\bibinfo {author} {\bibfnamefont {G.-Y.}\ \bibnamefont {Qin}}\ and\ \bibinfo {author} {\bibfnamefont {A.}~\bibnamefont {Majumder}},\ }\bibfield  {title} {\bibinfo {title} {{A pQCD-based description of heavy and light flavor jet quenching}},\ }\href {https://doi.org/10.1103/PhysRevLett.105.262301} {\bibfield  {journal} {\bibinfo  {journal} {Phys.Rev.Lett.}\ }\textbf {\bibinfo {volume} {105}},\ \bibinfo {pages} {262301} (\bibinfo {year} {2010})},\ \Eprint {https://arxiv.org/abs/0910.3016} {arXiv:0910.3016 [hep-ph]} \BibitemShut {NoStop}%
\bibitem [{\citenamefont {Qin}\ \emph {et~al.}(2009)\citenamefont {Qin}, \citenamefont {Majumder}, \citenamefont {Song},\ and\ \citenamefont {Heinz}}]{Qin:2009uh}%
  \BibitemOpen
  \bibfield  {author} {\bibinfo {author} {\bibfnamefont {G.~Y.}\ \bibnamefont {Qin}}, \bibinfo {author} {\bibfnamefont {A.}~\bibnamefont {Majumder}}, \bibinfo {author} {\bibfnamefont {H.}~\bibnamefont {Song}},\ and\ \bibinfo {author} {\bibfnamefont {U.}~\bibnamefont {Heinz}},\ }\bibfield  {title} {\bibinfo {title} {{Energy and momentum deposited into a QCD medium by a jet shower}},\ }\href {https://doi.org/10.1103/PhysRevLett.103.152303} {\bibfield  {journal} {\bibinfo  {journal} {Phys. Rev. Lett.}\ }\textbf {\bibinfo {volume} {103}},\ \bibinfo {pages} {152303} (\bibinfo {year} {2009})},\ \Eprint {https://arxiv.org/abs/0903.2255} {arXiv:0903.2255 [nucl-th]} \BibitemShut {NoStop}%
\bibitem [{\citenamefont {Aurenche}\ \emph {et~al.}(2008{\natexlab{a}})\citenamefont {Aurenche}, \citenamefont {Zakharov},\ and\ \citenamefont {Zaraket}}]{Aurenche:2008hm}%
  \BibitemOpen
  \bibfield  {author} {\bibinfo {author} {\bibfnamefont {P.}~\bibnamefont {Aurenche}}, \bibinfo {author} {\bibfnamefont {B.}~\bibnamefont {Zakharov}},\ and\ \bibinfo {author} {\bibfnamefont {H.}~\bibnamefont {Zaraket}},\ }\bibfield  {title} {\bibinfo {title} {{Failure of the collinear expansion in calculation of the induced gluon emission}},\ }\href {https://doi.org/10.1134/S0021364008110039} {\bibfield  {journal} {\bibinfo  {journal} {JETP Lett.}\ }\textbf {\bibinfo {volume} {87}},\ \bibinfo {pages} {605} (\bibinfo {year} {2008}{\natexlab{a}})},\ \Eprint {https://arxiv.org/abs/0804.4282} {arXiv:0804.4282 [hep-ph]} \BibitemShut {NoStop}%
\bibitem [{\citenamefont {Aurenche}\ \emph {et~al.}(2008{\natexlab{b}})\citenamefont {Aurenche}, \citenamefont {Zakharov},\ and\ \citenamefont {Zaraket}}]{Aurenche:2008mq}%
  \BibitemOpen
  \bibfield  {author} {\bibinfo {author} {\bibfnamefont {P.}~\bibnamefont {Aurenche}}, \bibinfo {author} {\bibfnamefont {B.~G.}\ \bibnamefont {Zakharov}},\ and\ \bibinfo {author} {\bibfnamefont {H.}~\bibnamefont {Zaraket}},\ }\bibfield  {title} {\bibinfo {title} {{Comment on ``Success of collinear expansion in the calculation of induced gluon emission''}},\ }\href@noop {} {\  (\bibinfo {year} {2008}{\natexlab{b}})},\ \Eprint {https://arxiv.org/abs/0806.0160} {arXiv:0806.0160 [hep-ph]} \BibitemShut {NoStop}%
\bibitem [{\citenamefont {Majumder}(2009)}]{Majumder:2009zu}%
  \BibitemOpen
  \bibfield  {author} {\bibinfo {author} {\bibfnamefont {A.}~\bibnamefont {Majumder}},\ }\bibfield  {title} {\bibinfo {title} {{The in-medium scale evolution in jet modification}},\ }\href@noop {} {\  (\bibinfo {year} {2009})},\ \Eprint {https://arxiv.org/abs/0901.4516} {arXiv:0901.4516 [nucl-th]} \BibitemShut {NoStop}%
\bibitem [{\citenamefont {Guo}\ and\ \citenamefont {Wang}(2000{\natexlab{b}})}]{Guo_2000}%
  \BibitemOpen
  \bibfield  {author} {\bibinfo {author} {\bibfnamefont {X.}~\bibnamefont {Guo}}\ and\ \bibinfo {author} {\bibfnamefont {X.-N.}\ \bibnamefont {Wang}},\ }\bibfield  {title} {\bibinfo {title} {Multiple scattering, parton energy loss, and modified fragmentation functions in deeply inelasticeascattering},\ }\href {https://doi.org/10.1103/physrevlett.85.3591} {\bibfield  {journal} {\bibinfo  {journal} {Physical Review Letters}\ }\textbf {\bibinfo {volume} {85}},\ \bibinfo {pages} {3591–3594} (\bibinfo {year} {2000}{\natexlab{b}})}\BibitemShut {NoStop}%
\bibitem [{\citenamefont {Wang}\ and\ \citenamefont {Guo}(2001{\natexlab{b}})}]{Wang_2001}%
  \BibitemOpen
  \bibfield  {author} {\bibinfo {author} {\bibfnamefont {X.-N.}\ \bibnamefont {Wang}}\ and\ \bibinfo {author} {\bibfnamefont {X.}~\bibnamefont {Guo}},\ }\bibfield  {title} {\bibinfo {title} {Multiple parton scattering in nuclei: parton energy loss},\ }\href {https://doi.org/10.1016/s0375-9474(01)01130-7} {\bibfield  {journal} {\bibinfo  {journal} {Nuclear Physics A}\ }\textbf {\bibinfo {volume} {696}},\ \bibinfo {pages} {788–832} (\bibinfo {year} {2001}{\natexlab{b}})}\BibitemShut {NoStop}%
\bibitem [{\citenamefont {Idilbi}\ and\ \citenamefont {Majumder}(2009)}]{Idilbi:2008vm}%
  \BibitemOpen
  \bibfield  {author} {\bibinfo {author} {\bibfnamefont {A.}~\bibnamefont {Idilbi}}\ and\ \bibinfo {author} {\bibfnamefont {A.}~\bibnamefont {Majumder}},\ }\bibfield  {title} {\bibinfo {title} {{Extending Soft-Collinear-Effective-Theory to describe hard jets in dense QCD media}},\ }\href {https://doi.org/10.1103/PhysRevD.80.054022} {\bibfield  {journal} {\bibinfo  {journal} {Phys. Rev.}\ }\textbf {\bibinfo {volume} {D80}},\ \bibinfo {pages} {054022} (\bibinfo {year} {2009})},\ \Eprint {https://arxiv.org/abs/0808.1087} {arXiv:0808.1087 [hep-ph]} \BibitemShut {NoStop}%
\bibitem [{\citenamefont {wei Qiu}\ and\ \citenamefont {Sterman}(1991{\natexlab{a}})}]{Qiu:1990xxa}%
  \BibitemOpen
  \bibfield  {author} {\bibinfo {author} {\bibfnamefont {J.}~\bibnamefont {wei Qiu}}\ and\ \bibinfo {author} {\bibfnamefont {G.}~\bibnamefont {Sterman}},\ }\bibfield  {title} {\bibinfo {title} {{Power corrections in hadronic scattering. 1. Leading 1/Q**2 corrections to the Drell-Yan cross-section}},\ }\href {https://doi.org/10.1016/0550-3213(91)90503-P} {\bibfield  {journal} {\bibinfo  {journal} {Nucl. Phys.}\ }\textbf {\bibinfo {volume} {B353}},\ \bibinfo {pages} {105} (\bibinfo {year} {1991}{\natexlab{a}})}\BibitemShut {NoStop}%
\bibitem [{\citenamefont {wei Qiu}\ and\ \citenamefont {Sterman}(1991{\natexlab{b}})}]{Qiu:1990xy}%
  \BibitemOpen
  \bibfield  {author} {\bibinfo {author} {\bibfnamefont {J.}~\bibnamefont {wei Qiu}}\ and\ \bibinfo {author} {\bibfnamefont {G.}~\bibnamefont {Sterman}},\ }\bibfield  {title} {\bibinfo {title} {{Power corrections to hadronic scattering. 2. Factorization}},\ }\href {https://doi.org/10.1016/0550-3213(91)90504-Q} {\bibfield  {journal} {\bibinfo  {journal} {Nucl. Phys.}\ }\textbf {\bibinfo {volume} {B353}},\ \bibinfo {pages} {137} (\bibinfo {year} {1991}{\natexlab{b}})}\BibitemShut {NoStop}%
\bibitem [{\citenamefont {Kumar}\ \emph {et~al.}(2017)\citenamefont {Kumar}, \citenamefont {Bianchi}, \citenamefont {Elledge}, \citenamefont {Majumder}, \citenamefont {Qin},\ and\ \citenamefont {Shen}}]{Kumar:2017des}%
  \BibitemOpen
  \bibfield  {author} {\bibinfo {author} {\bibfnamefont {A.}~\bibnamefont {Kumar}}, \bibinfo {author} {\bibfnamefont {E.}~\bibnamefont {Bianchi}}, \bibinfo {author} {\bibfnamefont {J.}~\bibnamefont {Elledge}}, \bibinfo {author} {\bibfnamefont {A.}~\bibnamefont {Majumder}}, \bibinfo {author} {\bibfnamefont {G.-Y.}\ \bibnamefont {Qin}},\ and\ \bibinfo {author} {\bibfnamefont {C.}~\bibnamefont {Shen}},\ }\bibfield  {title} {\bibinfo {title} {{Solving the $\hat{q}$ puzzle with $x$ and scale dependence}},\ }\bibfield  {booktitle} {\emph {\bibinfo {booktitle} {{Proceedings, 26th International Conference on Ultra-relativistic Nucleus-Nucleus Collisions (Quark Matter 2017): Chicago, Illinois, USA, February 5-11, 2017}}},\ }\href {https://doi.org/10.1016/j.nuclphysa.2017.05.015} {\bibfield  {journal} {\bibinfo  {journal} {Nucl. Phys.}\ }\textbf {\bibinfo {volume} {A967}},\ \bibinfo {pages} {536} (\bibinfo {year} {2017})},\ \Eprint {https://arxiv.org/abs/1706.07547} {arXiv:1706.07547 [nucl-th]} \BibitemShut {NoStop}%
\bibitem [{\citenamefont {Bianchi}\ \emph {et~al.}(2017)\citenamefont {Bianchi}, \citenamefont {Elledge}, \citenamefont {Kumar}, \citenamefont {Majumder}, \citenamefont {Qin},\ and\ \citenamefont {Shen}}]{Bianchi:2017wpt}%
  \BibitemOpen
  \bibfield  {author} {\bibinfo {author} {\bibfnamefont {E.}~\bibnamefont {Bianchi}}, \bibinfo {author} {\bibfnamefont {J.}~\bibnamefont {Elledge}}, \bibinfo {author} {\bibfnamefont {A.}~\bibnamefont {Kumar}}, \bibinfo {author} {\bibfnamefont {A.}~\bibnamefont {Majumder}}, \bibinfo {author} {\bibfnamefont {G.-Y.}\ \bibnamefont {Qin}},\ and\ \bibinfo {author} {\bibfnamefont {C.}~\bibnamefont {Shen}},\ }\bibfield  {title} {\bibinfo {title} {{The $x$ and $Q^2$ dependence of $\hat{q}$, quasi-particles and the JET puzzle}},\ }\href@noop {} {\  (\bibinfo {year} {2017})},\ \Eprint {https://arxiv.org/abs/1702.00481} {arXiv:1702.00481 [nucl-th]} \BibitemShut {NoStop}%
\bibitem [{\citenamefont {Kumar}\ \emph {et~al.}(2020{\natexlab{b}})\citenamefont {Kumar}, \citenamefont {Majumder},\ and\ \citenamefont {Shen}}]{Kumar:2019uvu}%
  \BibitemOpen
  \bibfield  {author} {\bibinfo {author} {\bibfnamefont {A.}~\bibnamefont {Kumar}}, \bibinfo {author} {\bibfnamefont {A.}~\bibnamefont {Majumder}},\ and\ \bibinfo {author} {\bibfnamefont {C.}~\bibnamefont {Shen}},\ }\bibfield  {title} {\bibinfo {title} {{Energy and scale dependence of $\hat{q}$ and the JET puzzle}},\ }\href {https://doi.org/10.1103/PhysRevC.101.034908} {\bibfield  {journal} {\bibinfo  {journal} {Phys. Rev.}\ }\textbf {\bibinfo {volume} {C101}},\ \bibinfo {pages} {034908} (\bibinfo {year} {2020}{\natexlab{b}})},\ \Eprint {https://arxiv.org/abs/1909.03178} {arXiv:1909.03178 [nucl-th]} \BibitemShut {NoStop}%
\bibitem [{\citenamefont {Landau}\ and\ \citenamefont {Pomeranchuk}(1953{\natexlab{a}})}]{Landau:1953um}%
  \BibitemOpen
  \bibfield  {author} {\bibinfo {author} {\bibfnamefont {L.~D.}\ \bibnamefont {Landau}}\ and\ \bibinfo {author} {\bibfnamefont {I.}~\bibnamefont {Pomeranchuk}},\ }\bibfield  {title} {\bibinfo {title} {{Limits of applicability of the theory of bremsstrahlung electrons and pair production at high-energies}},\ }\href@noop {} {\bibfield  {journal} {\bibinfo  {journal} {Dokl. Akad. Nauk Ser. Fiz.}\ }\textbf {\bibinfo {volume} {92}},\ \bibinfo {pages} {535} (\bibinfo {year} {1953}{\natexlab{a}})}\BibitemShut {NoStop}%
\bibitem [{\citenamefont {Landau}\ and\ \citenamefont {Pomeranchuk}(1953{\natexlab{b}})}]{Landau:1953gr}%
  \BibitemOpen
  \bibfield  {author} {\bibinfo {author} {\bibfnamefont {L.~D.}\ \bibnamefont {Landau}}\ and\ \bibinfo {author} {\bibfnamefont {I.}~\bibnamefont {Pomeranchuk}},\ }\bibfield  {title} {\bibinfo {title} {{Electron cascade process at very high-energies}},\ }\href@noop {} {\bibfield  {journal} {\bibinfo  {journal} {Dokl. Akad. Nauk Ser. Fiz.}\ }\textbf {\bibinfo {volume} {92}},\ \bibinfo {pages} {735} (\bibinfo {year} {1953}{\natexlab{b}})}\BibitemShut {NoStop}%
\bibitem [{\citenamefont {Migdal}(1956)}]{Migdal:1956tc}%
  \BibitemOpen
  \bibfield  {author} {\bibinfo {author} {\bibfnamefont {A.~B.}\ \bibnamefont {Migdal}},\ }\bibfield  {title} {\bibinfo {title} {{Bremsstrahlung and pair production in condensed media at high-energies}},\ }\href {https://doi.org/10.1103/PhysRev.103.1811} {\bibfield  {journal} {\bibinfo  {journal} {Phys. Rev.}\ }\textbf {\bibinfo {volume} {103}},\ \bibinfo {pages} {1811} (\bibinfo {year} {1956})}\BibitemShut {NoStop}%
\bibitem [{\citenamefont {He}\ \emph {et~al.}(2019)\citenamefont {He}, \citenamefont {Cao}, \citenamefont {Chen}, \citenamefont {Luo}, \citenamefont {Pang},\ and\ \citenamefont {Wang}}]{He:2018xjv}%
  \BibitemOpen
  \bibfield  {author} {\bibinfo {author} {\bibfnamefont {Y.}~\bibnamefont {He}}, \bibinfo {author} {\bibfnamefont {S.}~\bibnamefont {Cao}}, \bibinfo {author} {\bibfnamefont {W.}~\bibnamefont {Chen}}, \bibinfo {author} {\bibfnamefont {T.}~\bibnamefont {Luo}}, \bibinfo {author} {\bibfnamefont {L.-G.}\ \bibnamefont {Pang}},\ and\ \bibinfo {author} {\bibfnamefont {X.-N.}\ \bibnamefont {Wang}},\ }\bibfield  {title} {\bibinfo {title} {{Interplaying mechanisms behind single inclusive jet suppression in heavy-ion collisions}},\ }\href {https://doi.org/10.1103/PhysRevC.99.054911} {\bibfield  {journal} {\bibinfo  {journal} {Phys. Rev.}\ }\textbf {\bibinfo {volume} {C99}},\ \bibinfo {pages} {054911} (\bibinfo {year} {2019})},\ \Eprint {https://arxiv.org/abs/1809.02525} {arXiv:1809.02525 [nucl-th]} \BibitemShut {NoStop}%
\bibitem [{\citenamefont {Bass}\ \emph {et~al.}(2009)\citenamefont {Bass}, \citenamefont {Gale}, \citenamefont {Majumder}, \citenamefont {Nonaka}, \citenamefont {Qin} \emph {et~al.}}]{Bass:2008rv}%
  \BibitemOpen
  \bibfield  {author} {\bibinfo {author} {\bibfnamefont {S.~A.}\ \bibnamefont {Bass}}, \bibinfo {author} {\bibfnamefont {C.}~\bibnamefont {Gale}}, \bibinfo {author} {\bibfnamefont {A.}~\bibnamefont {Majumder}}, \bibinfo {author} {\bibfnamefont {C.}~\bibnamefont {Nonaka}}, \bibinfo {author} {\bibfnamefont {G.-Y.}\ \bibnamefont {Qin}}, \emph {et~al.},\ }\bibfield  {title} {\bibinfo {title} {{Systematic Comparison of Jet Energy-Loss Schemes in a realistic hydrodynamic medium}},\ }\href {https://doi.org/10.1103/PhysRevC.79.024901} {\bibfield  {journal} {\bibinfo  {journal} {Phys.Rev.}\ }\textbf {\bibinfo {volume} {C79}},\ \bibinfo {pages} {024901} (\bibinfo {year} {2009})},\ \Eprint {https://arxiv.org/abs/0808.0908} {arXiv:0808.0908 [nucl-th]} \BibitemShut {NoStop}%
\bibitem [{\citenamefont {Majumder}(2007)}]{Majumder:2007iu}%
  \BibitemOpen
  \bibfield  {author} {\bibinfo {author} {\bibfnamefont {A.}~\bibnamefont {Majumder}},\ }\bibfield  {title} {\bibinfo {title} {{A comparative study of jet-quenching schemes}},\ }\href {https://doi.org/10.1088/0954-3899/34/8/S25} {\bibfield  {journal} {\bibinfo  {journal} {J. Phys.}\ }\textbf {\bibinfo {volume} {G34}},\ \bibinfo {pages} {S377} (\bibinfo {year} {2007})},\ \Eprint {https://arxiv.org/abs/nucl-th/0702066} {arXiv:nucl-th/0702066} \BibitemShut {NoStop}%
\bibitem [{\citenamefont {Baier}\ \emph {et~al.}(1995)\citenamefont {Baier}, \citenamefont {Dokshitzer}, \citenamefont {Peigne},\ and\ \citenamefont {Schiff}}]{Baier:1994bd}%
  \BibitemOpen
  \bibfield  {author} {\bibinfo {author} {\bibfnamefont {R.}~\bibnamefont {Baier}}, \bibinfo {author} {\bibfnamefont {Y.~L.}\ \bibnamefont {Dokshitzer}}, \bibinfo {author} {\bibfnamefont {S.}~\bibnamefont {Peigne}},\ and\ \bibinfo {author} {\bibfnamefont {D.}~\bibnamefont {Schiff}},\ }\bibfield  {title} {\bibinfo {title} {{Induced gluon radiation in a QCD medium}},\ }\href {https://doi.org/10.1016/0370-2693(94)01617-L} {\bibfield  {journal} {\bibinfo  {journal} {Phys. Lett.}\ }\textbf {\bibinfo {volume} {B345}},\ \bibinfo {pages} {277} (\bibinfo {year} {1995})},\ \Eprint {https://arxiv.org/abs/hep-ph/9411409} {arXiv:hep-ph/9411409} \BibitemShut {NoStop}%
\bibitem [{\citenamefont {Baier}\ \emph {et~al.}(1997{\natexlab{a}})\citenamefont {Baier}, \citenamefont {Dokshitzer}, \citenamefont {Mueller}, \citenamefont {Peigne},\ and\ \citenamefont {Schiff}}]{Baier:1996kr}%
  \BibitemOpen
  \bibfield  {author} {\bibinfo {author} {\bibfnamefont {R.}~\bibnamefont {Baier}}, \bibinfo {author} {\bibfnamefont {Y.~L.}\ \bibnamefont {Dokshitzer}}, \bibinfo {author} {\bibfnamefont {A.~H.}\ \bibnamefont {Mueller}}, \bibinfo {author} {\bibfnamefont {S.}~\bibnamefont {Peigne}},\ and\ \bibinfo {author} {\bibfnamefont {D.}~\bibnamefont {Schiff}},\ }\bibfield  {title} {\bibinfo {title} {Radiative energy loss of high energy quarks and gluons in a finite-volume quark-gluon plasma},\ }\href@noop {} {\bibfield  {journal} {\bibinfo  {journal} {Nucl. Phys.}\ }\textbf {\bibinfo {volume} {B483}},\ \bibinfo {pages} {291} (\bibinfo {year} {1997}{\natexlab{a}})},\ \Eprint {https://arxiv.org/abs/hep-ph/9607355} {hep-ph/9607355} \BibitemShut {NoStop}%
\bibitem [{\citenamefont {Baier}\ \emph {et~al.}(1997{\natexlab{b}})\citenamefont {Baier}, \citenamefont {Dokshitzer}, \citenamefont {Mueller}, \citenamefont {Peigne},\ and\ \citenamefont {Schiff}}]{Baier:1996sk}%
  \BibitemOpen
  \bibfield  {author} {\bibinfo {author} {\bibfnamefont {R.}~\bibnamefont {Baier}}, \bibinfo {author} {\bibfnamefont {Y.~L.}\ \bibnamefont {Dokshitzer}}, \bibinfo {author} {\bibfnamefont {A.~H.}\ \bibnamefont {Mueller}}, \bibinfo {author} {\bibfnamefont {S.}~\bibnamefont {Peigne}},\ and\ \bibinfo {author} {\bibfnamefont {D.}~\bibnamefont {Schiff}},\ }\bibfield  {title} {\bibinfo {title} {Radiative energy loss and p(t)-broadening of high energy partons in nuclei},\ }\href@noop {} {\bibfield  {journal} {\bibinfo  {journal} {Nucl. Phys.}\ }\textbf {\bibinfo {volume} {B484}},\ \bibinfo {pages} {265} (\bibinfo {year} {1997}{\natexlab{b}})},\ \Eprint {https://arxiv.org/abs/hep-ph/9608322} {hep-ph/9608322} \BibitemShut {NoStop}%
\bibitem [{\citenamefont {Zakharov}(1996)}]{Zakharov:1996fv}%
  \BibitemOpen
  \bibfield  {author} {\bibinfo {author} {\bibfnamefont {B.~G.}\ \bibnamefont {Zakharov}},\ }\bibfield  {title} {\bibinfo {title} {{Fully quantum treatment of the Landau-Pomeranchuk-Migdal effect in QED and QCD}},\ }\href {https://doi.org/10.1134/1.567126} {\bibfield  {journal} {\bibinfo  {journal} {JETP Lett.}\ }\textbf {\bibinfo {volume} {63}},\ \bibinfo {pages} {952} (\bibinfo {year} {1996})},\ \Eprint {https://arxiv.org/abs/hep-ph/9607440} {arXiv:hep-ph/9607440} \BibitemShut {NoStop}%
\bibitem [{\citenamefont {Zakharov}(1997)}]{Zakharov:1997uu}%
  \BibitemOpen
  \bibfield  {author} {\bibinfo {author} {\bibfnamefont {B.~G.}\ \bibnamefont {Zakharov}},\ }\bibfield  {title} {\bibinfo {title} {Radiative energy loss of high energy quarks in finite-size nuclear matter and quark-gluon plasma},\ }\href@noop {} {\bibfield  {journal} {\bibinfo  {journal} {JETP Lett.}\ }\textbf {\bibinfo {volume} {65}},\ \bibinfo {pages} {615} (\bibinfo {year} {1997})},\ \Eprint {https://arxiv.org/abs/hep-ph/9704255} {hep-ph/9704255} \BibitemShut {NoStop}%
\end{thebibliography}%

\end{document}